\documentclass{article}%
\usepackage{amssymb}
\usepackage{amsmath}
\usepackage{amsfonts}
\usepackage{graphicx}%
\newtheorem{theorem}{Theorem}

\newtheorem{definition}[theorem]{Definition}
\newtheorem{example}[theorem]{Example}
\newtheorem{exercise}[theorem]{Exercise}

\newtheorem{proposition}[theorem]{Proposition}
\newtheorem{remark}[theorem]{Remark}

\newenvironment{proof}[1][Proof]{\noindent\textbf{#1.} }{\ \rule{0.5em}{0.5em}}
\begin{document}

\title{Hidden Consequence of Active Local Lorentz Invariance\thanks{ \textit{Int. J.
Geom. Meth. Mod. Phys. }\textbf{2 }(2), 305-357 (2005).}}
\author{{\footnotesize \ W. A. Rodrigues Jr.}$^{1}${\footnotesize \ , R. da
Rocha}$^{2}${\footnotesize , and J. Vaz Jr}$^{1}${\footnotesize .}\\$^{1}\hspace{-0.1cm}${\footnotesize Institute of Mathematics, Statistics and
Scientific Computation}\\{\footnotesize \ IMECC-UNICAMP CP 6065}\\{\footnotesize \ 13083-859 Campinas, SP, Brazil }\\$^{2}${\footnotesize Institute of Physics Gleb Wataghin, UNICAMP CP6165}\\{\footnotesize 13083-970 Campinas, SP, Brazil}\\{\footnotesize e-mails: walrod@ime.unicamp.br; roldao@ifi.unicamp.br;
vaz@ime.unicamp.br}}
\maketitle

\begin{abstract}
In this paper we investigate a hidden consequence of the hypothesis that
Lagrangians and field equations must be invariant under active local Lorentz
transformations. We show that this hypothesis implies in an equivalence
between spacetime structures with several curvature and torsion possibilities.

\end{abstract}

\section{Introduction}

It is now a well established fact that Maxwell, Dirac and Einstein theories
can be formulated in terms of differential forms. The easiest way to see this
is to introduce the Clifford and spin-Clifford bundles \cite{quiro,rod04,moro}
of spacetime. This formalism for the case of Maxwell and Dirac fields will be
briefly recalled in Section 2, since they will play an essential role for
proving the main claim of this paper. For the case of the gravitational field,
see \cite{quiro04} and also \cite{quiro1,thirring,wallner}.

As it is well known \cite{thirring}, any field theory formulated in terms of
differential forms \ is such that the action is invariant under arbitrary
diffeomorphisms. This does not imply that the field equations of the theory
are necessarily invariant under diffeomorphisms, \textit{unless} we are
prepared to accept as equivalent different manifolds equipped with metrics and
connections which may be said diffeomorphically equivalent \footnote{A
thoughtful discussion on this issue will be presented elsewhere.}. Now,
several authors, e.g., \cite{mielke,ramond} insist that the invariance of the
action (and field equations) under arbitrary active Lorentz transformations is
a necessary consequence of the equivalence principle\footnote{We are not going
to enter discussions about the equivalence principle in this paper. One of the
authors' view on the subject is discussed in \cite{rod01}.}. From the
mathematical point of view, in order to have such imposition satisfied it is
necessary to introduce the concept of \textit{generalized gauge covariant
derivatives} (Section 3 and Appendix D). We show that once we accept this
concept we must necessarily also accept the equivalence between spacetime
models with \textit{connections} that have \textit{different} curvature and
torsion tensors. This paper is organized as follows. In Section 2 we introduce
the Clifford and spin-Clifford bundles and discuss the concept of
Dirac-Hestenes spinor fields (which are sections of a spin-Clifford bundle)
and their \textit{representatives} in the Clifford bundle, which are sums of
even nonhomogenous differential forms. In Section 3 we discuss the covariant
derivative of Clifford fields and introduce the curvature and torsion extensor
fields of a given connection. In Section 4 we discuss the covariant derivative
of spinor fields, and how to define an \textit{effective} covariant derivative
for representatives of Dirac-Hestenes spinor fields in the Clifford bundle. In
Section 5 we discuss the many faces of Dirac equation. In Section 6 we discuss
the meaning of active local Lorentz invariance of Maxwell equations. In
Section 7, \ we introduce the Dirac-Hestenes equation in a Riemann-Cartan
spacetime, and in Section 8 we discuss the meaning of active local Lorentz
invariance of the Dirac-Equation. We prove then our main result, namely:
active local Lorentz invariance implies a gauge identification of geometries
with connections that have different torsion and curvature tensors. The
intelligibility of the issues discussed in this paper requires a working
knowledge of the theory of connections, and in particular, the introduction of
the concept of generalized $G$-connections. The main results needed are
presented in Appendices A-D. This paper is dedicated to Ivanenko, that in a
pioneer paper \cite{ivalandau} used sums of nonhomogeneous antisymmetric
tensors to represent spinors. The way in which such an idea becomes
mathematically legitimate is presented below.

\section{Clifford and Spin-Clifford Bundles}

Let $\mathcal{M}=(M,\mathtt{g},\mathbf{\nabla},\tau_{\text{\textbf{g}}%
},\uparrow)$ be an arbitrary Riemann-Cartan spacetime.\ The quadruple
$(M,\mathtt{g},\tau_{\text{\textbf{g}}},\uparrow)$ is a four-dimensional
time-oriented and space-oriented Lorentzian manifold. This means that
$\mathtt{g}\in\sec T_{2}^{0}M$ is a Lorentzian metric of signature (1,3),
$\tau_{\text{\textbf{g}}}\in\sec\bigwedge{}^{4}(T^{\ast}M)$ and $\uparrow$ is
a time-orientation (see details, e.g., in \cite{sawu}). Here, $T^{\ast}M$
[$TM$] is the cotangent [tangent] bundle. $T^{\ast}M=\cup_{x\in M}T_{x}^{\ast
}M$, $TM=\cup_{x\in M}T_{x}M$, and $T_{x}M\simeq T_{x}^{\ast}M\simeq
\mathbb{R}^{1,3}$, where $\mathbb{R}^{1,3}$ is the Minkowski vector
space\footnote{Not to be confused with Minkowski spacetime \cite{sawu}.}.
$\mathbf{\nabla}$ is an arbitrary \ metric compatible connection
$i.e$\textit{.\/}, $\mathbf{\nabla}\mathtt{g}=0$, but in general,
$\mathbf{R}(\mathbf{\nabla})\neq0$, $\mathbf{T}(\mathbf{\nabla})\neq0$,
$\mathbf{R}$ and $\mathbf{T}$ being respectively the curvature and torsion
tensors. \ When $\mathbf{R}(\mathbf{\nabla})\neq0$, $\mathbf{T}(\mathbf{\nabla
})\neq0$, $\mathcal{M}$ is called a \textit{Riemann-Cartan spacetime}. When
$\mathbf{R}(\mathbf{\nabla})\neq0$, $\mathbf{T}(\mathbf{\nabla})=0$,
$\mathcal{M}$ is called a \textit{Lorentzian spacetime}. When $\mathbf{R}%
(\mathbf{\nabla})=0$, $\mathbf{T}(\mathbf{\nabla})\neq0$, $\mathcal{M}$ is
called a \textit{teleparallel spacetime.} Minkowski spacetime is the case
where $\mathbf{R}(\mathbf{\nabla})=0$, $\mathbf{T}(\mathbf{\nabla})=0$, and
$M\simeq\mathbb{R}^{4}$. Let $g\in\sec T_{0}^{2}M$ be the metric of the
\textit{cotangent bundle}. The Clifford bundle of differential forms
$\mathcal{C}\!\ell(M,g)$ is the bundle of algebras, i.e., $\mathcal{C}%
\!\ell(M,g)=\cup_{x\in M}\mathcal{C}\!\ell(T_{x}^{\ast}M,g)$, where $\forall
x\in M$, $\mathcal{C}\!\ell(T_{x}^{\ast}M,g)=\mathbb{R}_{1,3}$, the so called
\emph{spacetime} \emph{algebra }\cite{rod04}. Recall also that $\mathcal{C}%
\!\ell(M,g)$ is a vector bundle associated to the \emph{orthonormal frame
bundle}, i.e., $\mathcal{C}\!\ell(M,g)$ $=P_{\mathrm{SO}_{(1,3)}^{e}}%
(M)\times_{\mathrm{Ad}}\mathcal{C}l_{1,3}$ \cite{lawmi,moro}. For any $x\in
M$, $\mathcal{C}\!\ell(T_{x}^{\ast}M,g)$ as a linear space over the real field
$\mathbb{R}$ is isomorphic to the Cartan algebra $\bigwedge(T_{x}^{\ast}M)$ of
the cotangent space. $\bigwedge(T_{x}^{\ast}M)=\oplus_{k=0}^{4}\bigwedge
^{k}(T_{x}^{\ast}M)$, where $\bigwedge^{k}(T_{x}^{\ast}M)$ is the $\binom
{4}{k}$-dimensional space of $k$-forms. Then, sections of $\mathcal{C}%
\!\ell(M,g)$ can be represented as a sum of non homogeneous differential
forms. Let $\{e_{\mathbf{a}}\}\in\sec P_{\mathrm{SO}_{(1,3)}^{e}}(M)$ (the
frame bundle) be an orthonormal basis for $TU\subset TM$, i.e., $\mathbf{g}%
(e_{\mathbf{a}},e_{\mathbf{a}})=\eta_{\mathbf{ab}}=\mathrm{diag}(1,-1,-1,-1)$.
Let $\theta^{\mathbf{a}}\in\sec\bigwedge^{1}(T^{\ast}M)\hookrightarrow
\sec\mathcal{C}\!\ell(M,g)$ ($\mathbf{a}=0,1,2,3$) be such that the set
$\{\theta^{\mathbf{a}}\}$ is the dual basis of $\{e_{\mathbf{a}}\}$.

\subsection{Clifford Product}

The fundamental \emph{Clifford product} (in what follows to be denoted by
juxtaposition of symbols) is generated by $\theta^{\mathbf{a}}\theta
^{\mathbf{b}}+\theta^{\mathbf{b}}\theta^{\mathbf{a}}=2\eta^{\mathbf{ab}}$ and
if $\mathcal{C}\in\sec\mathcal{C}\!\ell(M,g)$ we have%

\begin{equation}
\mathcal{C}=s+v_{\mathbf{a}}\theta^{\mathbf{a}}+\frac{1}{2!}f_{\mathbf{ab}%
}\theta^{\mathbf{a}}\theta^{\mathbf{b}}+\frac{1}{3!}t_{\mathbf{abc}}%
\theta^{\mathbf{a}}\theta^{\mathbf{b}}\theta^{\mathbf{c}}+p\theta^{5}\;,
\label{3}%
\end{equation}
where $\tau_{\mathtt{g}}=\theta^{5}=\theta^{0}\theta^{1}\theta^{2}\theta^{3}$
is the volume element and $s$, $v_{\mathbf{a}}$, $f_{\mathbf{ab}}$,
$t_{\mathbf{abc}}$, $p\in\sec\bigwedge^{0}(T^{\ast}M)\hookrightarrow
\sec\mathcal{C}\!\ell(M,g)$.

For $A_{r}\in\sec\bigwedge^{r}(T^{\ast}M)\hookrightarrow\sec\mathcal{C}%
\!\ell(M,g),B_{s}\in\sec\bigwedge^{s}(T^{\ast}M)\hookrightarrow\sec
\mathcal{C}\!\ell(M,g)$ we define the \emph{exterior product} in
$\mathcal{C}\!\ell(M,g)$ \ ($\forall r,s=0,1,2,3)$ by
\begin{align}
A_{r}\wedge B_{s}  &  =\langle A_{r}B_{s}\rangle_{r+s},\nonumber\\
A_{r}\wedge B_{s}  &  =(-1)^{rs}B_{s}\wedge A_{r}, \label{5}%
\end{align}
where $\langle\;\;\rangle_{k}$ is the component in $\bigwedge^{k}(T^{\ast}M)$
of the Clifford field. The exterior product is extended by linearity to all
sections of $\mathcal{C}\!\ell(M,g)$.

Let $A_{r}\in\sec\bigwedge^{r}(T^{\ast}M)\hookrightarrow\sec\mathcal{C}%
\!\ell(M,g),B_{s}\in\sec\bigwedge^{s}(T^{\ast}M)\hookrightarrow\sec
\mathcal{C}\!\ell(M,g)$. We define a \emph{scalar product }in\emph{\ }%
$\mathcal{C}\!\ell(M,g)$ (denoted by $\cdot$) as follows:

(i) For $a,b\in\sec\bigwedge^{1}(T^{\ast}M)\hookrightarrow\sec\mathcal{C}%
\!\ell(M,g),$%
\begin{equation}
a\cdot b=\frac{1}{2}(ab+ba)=g(a,b). \label{4}%
\end{equation}

(ii) For $A_{r}=a_{1}\wedge...\wedge a_{r},B_{r}=b_{1}\wedge...\wedge b_{r}
$,
\begin{align}
A_{r}\cdot B_{r}  &  =(a_{1}\wedge...\wedge a_{r})\cdot(b_{1}\wedge...\wedge
b_{r})\nonumber\\
&  =\left\vert
\begin{array}
[c]{lll}%
a_{1}\cdot b_{1} & .... & a_{1}\cdot b_{r}\\
.......... & .... & ..........\\
a_{r}\cdot b_{1} & .... & a_{r}\cdot b_{r}%
\end{array}
\right\vert \label{6}%
\end{align}

We agree that if $r=s=0$, the scalar product is simple the ordinary product in
the real field.

Also, if $r\neq s$, then $A_{r}\cdot B_{s}=0$. Finally, the scalar product is
extended by linearity for all sections of $\mathcal{C}\!\ell(M,g)$.

For $r\leq s,A_{r}=a_{1}\wedge...\wedge a_{r},B_{s}=b_{1}\wedge...\wedge
b_{s\text{ }}$we define the \textit{left contraction} by
\begin{equation}
\lrcorner:(A_{r},B_{s})\mapsto A_{r}\lrcorner B_{s}=%
{\displaystyle\sum\limits_{i_{1}\,<...\,<i_{r}}}
\epsilon^{i_{1}...i_{s}}(a_{1}\wedge...\wedge a_{r})\cdot(b_{_{i_{1}}}%
\wedge...\wedge b_{i_{r}})^{\sim}b_{i_{r}+1}\wedge...\wedge b_{i_{s}}
\label{7}%
\end{equation}
where $\sim$ is the reverse mapping (\emph{reversion}) defined by
\begin{equation}
\sim:\sec%
{\displaystyle\bigwedge\nolimits^{p}}
(T^{\ast}M)\ni a_{1}\wedge...\wedge a_{p}\mapsto a_{p}\wedge...\wedge a_{1}
\label{8}%
\end{equation}
and extended by linearity to all sections of $\mathcal{C}\!\ell(M,g)$. We
agree that for $\alpha,\beta\in\sec\bigwedge^{0}(T^{\ast}M)$ the contraction
is the ordinary (pointwise) product in the real field and that if $\alpha
\in\sec\bigwedge^{0}(T^{\ast}M)$, $A_{r}\in\sec\bigwedge^{r}(T^{\ast}%
M),B_{s}\in\sec\bigwedge^{s}(M)$ then $(\alpha A_{r})\lrcorner B_{s}%
=A_{r}\lrcorner(\alpha B_{s})$. Left contraction is extended by linearity to
all pairs of elements of sections of $\mathcal{C}\!\ell(M,g)$, i.e., for
$A,B\in\sec\mathcal{C}\!\ell(M,g)$%

\begin{equation}
A\lrcorner B=\sum_{r,s}\langle A\rangle_{r}\lrcorner\langle B\rangle_{s},r\leq
s \label{9}%
\end{equation}

It is also necessary to introduce the operator of \emph{right contraction}
denoted by $\llcorner$. The definition is obtained from the one presenting the
left contraction with the imposition that $r\geq s$ and taking into account
that now if $A_{r}\in\sec\bigwedge^{r}(T^{\ast}M),B_{s}\in\sec\bigwedge
^{s}(T^{\ast}M)$ then $A_{r}\llcorner(\alpha B_{s})=(\alpha A_{r})\llcorner
B_{s}$. See also the third formula in Eq.(\ref{10}).

The main formulas used in the Clifford calculus can be obtained from the
following ones (where $a\in\sec\bigwedge^{1}(T^{\ast}M)\hookrightarrow
\sec\mathcal{C}\!\ell(M,g)$):
\begin{align}
aB_{s}  &  =a\lrcorner B_{s}+a\wedge B_{s},\text{ }B_{s}a=B_{s}\llcorner
a+B_{s}\wedge a,\nonumber\\
a\lrcorner B_{s}  &  =\frac{1}{2}(aB_{s}-(-)^{s}B_{s}a),\nonumber\\
A_{r}\lrcorner B_{s}  &  =(-)^{r(s-1)}B_{s}\llcorner A_{r},\nonumber\\
a\wedge B_{s}  &  =\frac{1}{2}(aB_{s}+(-)^{s}B_{s}a),\nonumber\\
A_{r}B_{s}  &  =\langle A_{r}B_{s}\rangle_{|r-s|}+\langle A_{r}\lrcorner
B_{s}\rangle_{|r-s-2|}+...+\langle A_{r}B_{s}\rangle_{|r+s|}\nonumber\\
&  =\sum\limits_{k=0}^{m}\langle A_{r}B_{s}\rangle_{|r-s|+2k},\label{10}\\
A_{r}\cdot B_{r}  &  =B_{r}\cdot A_{r}=\tilde{A}_{r}\lrcorner B_{r}%
=A_{r}\llcorner\tilde{B}_{r}=\langle\tilde{A}_{r}B_{r}\rangle_{0}=\langle
A_{r}\tilde{B}_{r}\rangle_{0}.\nonumber
\end{align}

\subsubsection{Hodge Star Operator}

Let $\star$ be the Hodge star operator, i.e., the mapping
\[
\star:%
{\displaystyle\bigwedge\nolimits^{k}}
(T^{\ast}M)\rightarrow%
{\displaystyle\bigwedge\nolimits^{4-k}}
(T^{\ast}M),\text{ }A_{k}\mapsto\star A_{k}%
\]
where for $A_{k}\in\sec\bigwedge^{k}(T^{\ast}M)\hookrightarrow\sec
\mathcal{C}\!\ell(M,g)$%
\begin{equation}
\lbrack B_{k}\cdot A_{k}]\tau_{g}=B_{k}\wedge\star A_{k},\forall B_{k}\in
\sec\bigwedge\nolimits^{k}(T^{\ast}M)\hookrightarrow\sec\mathcal{C}%
\!\ell(M,g). \label{11a}%
\end{equation}
$\tau_{\mathtt{g}}\in\bigwedge^{4}(M)$ is a \emph{standard} volume element.
Then we can verify that
\begin{equation}
\star A_{k}=\widetilde{A}_{k}\gamma^{5}. \label{11b}%
\end{equation}

\subsubsection{Dirac Operator}

Let $d$ and $\delta$ be respectively the differential and Hodge codifferential
operators acting on sections of $\mathcal{C}\!\ell(M,g)$. If $A_{p}\in
\sec\bigwedge^{p}(T^{\ast}M)\hookrightarrow\sec\mathcal{C}\!\ell(M,g)$, then
$\delta A_{p}=(-)^{p}\star^{-1}d\star A_{p}$, with $\star^{-1}\star
=\mathrm{identity}$.

The Dirac operator acting on sections of $\mathcal{C}\!\ell(M,g)$ is the
invariant first order differential operator
\begin{equation}
{\mbox{\boldmath$\partial$}}=\theta^{\mathbf{a}}\mathbf{\nabla}_{e_{\mathbf{a}%
}}, \label{12}%
\end{equation}
where $\{e_{\mathbf{a}}\}$ is an arbitrary \emph{orthonormal basis} for
$TU\subset TM$ and $\{\theta^{\mathbf{b}}\}$ is a basis for $T^{\ast}U\subset
T^{\ast}M$ dual to the basis $\{e_{\mathbf{a}}\}$, i.e., $\theta^{\mathbf{b}%
}(e_{\mathbf{a}})=\delta_{\mathbf{b}}^{\mathbf{a}}$, $\mathbf{a,b}=0,1,2,3$.
The reciprocal basis of $\{\theta^{\mathbf{b}}\}$ is denoted $\{\theta
_{\mathbf{a}}\}$ and we have $\theta_{\mathbf{a}}\cdot\theta_{\mathbf{b}}%
=\eta_{\mathbf{ab}}$. Also,
\begin{equation}
\mathbf{\nabla}_{e_{\mathbf{a}}}\theta^{\mathbf{b}}=-\omega_{\mathbf{a}%
}^{\mathbf{bc}}\theta_{\mathbf{c}} \label{12n}%
\end{equation}
Defining
\begin{equation}
\mathbf{\omega}_{e_{\mathbf{a}}}=\frac{1}{2}\omega_{\mathbf{a}}^{\mathbf{bc}%
}\theta_{\mathbf{b}}\wedge\theta_{\mathbf{c}}, \label{12nn}%
\end{equation}
we have that for any $A_{p}\in\sec\bigwedge^{p}(T^{\ast}M),$ $p=0,1,2,3,4$
\begin{equation}
\mathbf{\nabla}_{e_{\mathbf{a}}}A_{p}=\partial_{e_{\mathbf{a}}}A_{p}+\frac
{1}{2}[\mathbf{\omega}_{e_{\mathbf{a}}},A_{p}], \label{12nnn}%
\end{equation}
where $\partial_{e_{\mathbf{a}}}$ is the Pfaff derivative, i.e., if
$A_{p}=\frac{1}{p!}A_{\mathbf{i}_{1}...\mathbf{i}_{p}}\theta^{\mathbf{i}%
_{1}...\mathbf{i}_{p}}$,%
\begin{equation}
\partial_{e_{\mathbf{a}}}A_{p}:=\frac{1}{p!}e_{\mathbf{a}}(A_{\mathbf{i}%
_{1}...\mathbf{i}_{p}})\theta^{\mathbf{i}_{1}...\mathbf{i}_{p}}. \label{pfaaf}%
\end{equation}

This important formula (Eq.(\ref{12nnn}) that is valid also for a
nonhomogeneous $A\in\sec\mathcal{C}\!\ell(M,g)$ will be proved below (Section 3).

Using Eq.(\ref{12nnn}) we can show a very important result: The Dirac operator
associated to a Levi-Civita connection satisfies%

\begin{align}
{\mbox{\boldmath$\partial$}}A_{p}  &  ={\mbox{\boldmath$\partial$}}\wedge
A_{p\,}+\,{\mbox{\boldmath$\partial$}}\lrcorner A_{p}=dA_{p}-\delta
A_{p},\nonumber\\
{\mbox{\boldmath$\partial$}}\wedge A_{p}  &  =dA_{p},\hspace{0.1in}%
\,{\mbox{\boldmath$\partial$}}\lrcorner A_{p}=-\delta A_{p}. \label{13}%
\end{align}

With these results, Maxwell equations for $F\in\sec\bigwedge^{2}(T^{\ast
}M)\hookrightarrow\sec\mathcal{C}\!\ell(M,g)$, $J\in\sec\bigwedge^{1}(T^{\ast
}M)\hookrightarrow\sec\mathcal{C}\!\ell(M,g)$ reads%
\begin{equation}
dF=0,\text{ }\delta F=-J, \label{maxwell1}%
\end{equation}
or Maxwell equation\footnote{No misprint here.} reads%
\begin{equation}
{\mbox{\boldmath$\partial$}}F=J{.} \label{maxwell2}%
\end{equation}

\subsection{ Spinor Fields}

How to represent the Dirac spinor fields in this formalism? We can show that
\emph{Dirac-Hestenes} spinor fields do the job\footnote{More details on other
kinds of spinor fields in this formalism can be found in \cite{13}.}. To
introduce this concept in a meaningful way we need to recall several
mathematical concepts.\ First we recall the concept of spin structure of an
oriented (and time-oriented) $(M,\mathtt{g})$. This consists of a principal
fibre bundle $\mathbf{\pi}_{s}:\mathbf{P}_{\mathrm{Spin}_{1,3}^{e}%
}(M)\rightarrow M$ \ (called the \textit{spin frame bundle}) with group
$\mathrm{Spin}_{1,3}^{e}$ and a map
\begin{equation}
s:P_{\mathrm{Spin}_{1,3}^{e}}(M)\rightarrow\mathbf{P}_{\mathrm{SO}_{1,3}^{e}%
}(M) \label{spinor bundle 1}%
\end{equation}
satisfying the following conditions:

(i) $\mathbf{\pi}(s(p))=\mathbf{\pi}_{s}(p)\ \forall p\in\mathbf{P}%
_{\mathrm{Spin}_{1,3}^{e}}(M),\pi$ is the projection map of the bundle
$\mathbf{P}_{\mathrm{SO}_{1,3}^{e}}(M)$.

(ii) $s(pu)=s(p)\mathrm{Ad}_{u}\ ,\forall p\in\mathbf{P}_{\mathrm{Spin}%
_{1,3}^{e}}(M)$ and%
\begin{align*}
\mathrm{Ad}  &  :\mathrm{Spin}_{1,3}^{e}\rightarrow\mathrm{Aut}(\mathbb{R}%
_{1,3}),\\
\mathrm{Ad}_{u}  &  :\mathbb{R}_{1,3}\ni x\mapsto uxu^{-1}\in\mathbb{R}%
_{1,3}.\medskip
\end{align*}

When does a \textit{spin structure} exist on an oriented (and time oriented)
$(M,\mathtt{g})$? For a Lorentzian manifold \ the answer is given by a famous
result due to Geroch \cite{geroch} which says that for an oriented (and
time-oriented) Lorentz manifold $(M,\mathtt{\mathbf{g)}}$, a spin structure
exists if and only if $\mathbf{P}_{\mathrm{SO}_{1,3}^{e}}(M)$ is a trivial bundle.

We call global sections $\xi\in\sec\mathbf{P}_{\mathrm{SO}_{1,3}^{e}}(M)$
Lorentz frames\emph{\ }and global sections $_{s}\Xi\in\sec P_{\mathrm{Spin}%
_{1,3}^{e}}(M)$ \emph{spin frames}. We recall (see, e.g., \cite{moro}) that
each $\xi\in\sec P_{\mathrm{SO}_{1,3}^{e}}(M)$ is a basis for $TM$, which is
completely specified once we give an element of the Lorentz group for each
$x\in M$ and fix a fiducial frame. Each $_{s}\Xi\in\sec P_{\mathrm{Spin}%
_{1,3}^{e}}(M)$ is also a basis for $TM$ and is completely identified once we
give an element of the $\mathrm{Spin}_{1,3}^{e}$ for each $x\in M$ and fix a
fiducial frame. Note that two ordered basis for $TM$ when considered as spin
frames, even if consisting of the same vector fields, but related by a $2\pi$
rotation are considered different. Also, two ordered basis for $TM$ when
considered as spin frames, if consisting of the same vector fields, related by
a $4\pi$ rotation are considered equal. Even if this mathematical construction
seems at first sight impossible of experimental detection, Aharonov and
Susskind \cite{ahsu} warrant that with clever experiments the spinor structure
can be detected.

\label{geroch rem 2}Recall that a principal bundle is trivial if and only if
it admits a global section. Therefore, Geroch's result says that\ a
(non-compact) spacetime admits a spin structure, if and only if, it admits a
(globally defined) Lorentz frame. In fact, it is possible to replace
$\mathbf{P}_{\mathrm{SO}_{1,3}^{e}}(M)$ by $\mathbf{P}_{\mathrm{Spin}%
_{1,3}^{e}}(M)$ in the statement of Geroch theorem (see \cite{geroch},
footnote 25). In this way, when a (non-compact) spacetime admits a spin
structure, the bundle $\mathbf{P}_{\mathrm{Spin}_{1,3}^{e}}(M)$ is trivial
and, therefore, every bundle associated to it is also trivial. This result is
indeed a very important one, because it says to us that the real spacetime of
our universe (that, of course, is inhabited by several different types of
spinor fields) must have a topology that admits a global tetrad field, which
is defined only modulus a local Lorentz transformation. The dual cotetrad have
been associated to the gravitational field in \cite{quiro04}, where we wrote
wave equations for them. In a certain sense that cotetrad field is the
substance of physical spacetime. In what follows we shall use the symbol $\Xi$
to denote a spin coframe dual to a spin frame $_{s}\Xi$. We also write by
\textit{abuse of notation} that $\Xi\in\sec\mathbf{P}_{\mathrm{Spin}_{1,3}%
^{e}}(M)$.

An oriented manifold endowed with a spin structure will be called a
\textit{spin manifold}.

\subsection{Spinor Bundles and Spinor Fields}

We now present the most usual definitions of spinor bundles appearing in the
literature\footnote{We recall that there are some other (equivalent)
definitions of spinor bundles that we are not going to introduce in this paper
as, e.g., the one given in \cite{bleecker} in terms of mappings from
\ $\mathbf{P}_{\mathrm{Spin}_{1,3}^{e}}$ to some appropriate vector space.}
and next we find appropriate vector bundles such that particular sections are
\textit{Dirac-Hestenes spinor fields. }

\label{LRSB}A \textit{real spinor bundle} for $M$ is a vector bundle
\begin{equation}
S(M,g)=\mathbf{P}_{\mathrm{Spin}_{1,3}^{e}}(M)\times_{\mu_{l}}\mathbf{M}
\label{1.7}%
\end{equation}
where $\mathbf{M}$ is a left module for $\mathbb{R}_{1,3}$ and $\mu_{l}$ is a
representation of $\mathrm{Spin}_{1,3}^{e}$ on $\mathrm{End}(\mathbf{M)}$
given by left multiplication by elements of $\mathrm{Spin}_{1,3}^{e}$.

The \textit{dual bundle} $S^{\star}(M)$ is a real spinor bundle
\begin{equation}
S^{\star}(M,g)=\mathbf{P}_{\mathrm{Spin}_{1,3}^{e}}(M)\times_{\mu_{r}%
}\mathbf{M}^{\star} \label{1.7bis}%
\end{equation}
where $\mathbf{M}^{\star}$ is a right module for $\mathbb{R}_{1,3}$ and
$\mu_{r}$ is a representation $\mathrm{Spin}_{1,3}^{e}$ in \textrm{End}%
$(\mathbf{M)}$ given by right multiplication by (inverse) elements of
$\mathrm{Spin}_{1,3}^{e}$. By \textit{right multiplication} we mean that given
$a\in\mathbf{M}^{\ast},$ $\mu_{r}(u)a=au^{-1}$, then
\begin{equation}
\mu_{r}(uu^{\prime})a=a(uu^{\prime})^{-1}=au^{\prime-1}u^{-1}=\mu_{r}%
(u)\mu_{r}(u^{\prime})a. \label{1.7biss}%
\end{equation}

A \textit{complex spinor bundle} for $M$ is a vector bundle
\begin{equation}
S_{c}(M,g)=\mathbf{P}_{\mathrm{Spin}_{1,3}^{e}}(M)\times_{\mu_{c}}%
\mathbf{M}_{c} \label{4.7}%
\end{equation}
where $\mathbf{M}_{c}$ is a complex left module for $\mathbb{C}\otimes
\mathbb{R}_{1,3}\simeq\mathbb{R}_{4,1}\simeq\mathbb{C(}4\mathbb{)}$, and where
$\mu_{c}$ is a representation of $\mathrm{Spin}_{1,3}^{e}$ in $\mathrm{End}%
(\mathbf{M}_{c})$ given by left multiplication by elements of $\mathrm{Spin}%
_{1,3}^{e}$.

The \textit{dual complex spinor bundle} for $M$ is a vector bundle
\begin{equation}
S_{c}^{\star}(M,g)=\mathbf{P}_{\mathrm{Spin}_{1,3}^{e}}(M)\times_{\mu_{c}%
}\mathbf{M}_{c}^{\star} \label{4.7bis}%
\end{equation}
where $\mathbf{M}_{c}^{\star}$ is a complex right module for $\mathbb{C}%
\otimes\mathbb{R}_{1,3}\simeq\mathbb{R}_{4,1}\simeq\mathbb{C(}4\mathbb{)}$,
and where $\mu_{c}$ is a representation of $\mathrm{Spin}_{1,3}^{e}$ in
\textrm{End}$(\mathbf{M}_{c})$ given by right multiplication by (inverse)
elements of $\mathrm{Spin}_{1,3}^{e}$.

Taking, e.g., $\mathbf{M}_{c}=\mathbb{C}^{4}$ and $\mu_{c}$ the $D^{(1/2,0)}%
\oplus D^{(0,1/2)}$ representation of $\mathrm{Spin}_{1,3}^{e}\cong
Sl(2,\mathbb{C})$ in $\mathrm{End}(\mathbb{C}^{4})$, we immediately recognize
the usual definition of the (Dirac) covariant spinor bundle of $M$, as given,
e.g., in~\cite{choquet}.

Let $\{\mathbf{E}_{a}\}$ be an orthonormal basis $\mathbb{R}_{1,3}$. The ideal
$I=\mathbb{R}_{1,3}\frac{1}{2}(1+\mathbf{E}_{0})$ is a minimal left ideal of
$\mathbb{R}_{1,3}$. Besides $I$, other ideals exist in $\mathbb{R}_{1,3}$,
that are only \textit{algebraically} equivalent to this one \cite{rod04}. In
order to capture all possibilities we recall that $\mathbb{R}_{1,3}$ can be
considered as a module over itself by left (or right) multiplication. We have:

\label{LsCbundle}The \textit{left real spin-Clifford bundle} of $M$ is the
vector bundle
\begin{equation}
\mathcal{C}\ell_{\mathrm{Spin}_{1,3}^{e}}^{l}(M,g)=\mathbf{P}_{\mathrm{Spin}%
_{1,3}^{e}}(M)\times_{l}\mathbb{R}_{1,3}%
\end{equation}
where $l$ is the representation of $\mathrm{Spin}_{1,3}^{e}$ on $\mathbb{R}%
_{1,3}$ given by $l(a)x=ax$. Sections of $\mathcal{C}\ell_{\mathrm{Spin}%
_{1,3}^{e}}^{l}(M,g)$ are\textit{\ called left spin-Clifford fields}.

\label{''principalbunde''}$\mathcal{C}\ell_{\mathrm{Spin}_{1,3}^{e}}^{l}(M)$
is a \ `principal $\mathbb{R}_{1,3}$-bundle', i.e., it admits a free action of
$\mathbb{R}_{1,3}$ on the right \cite{lawmi}, which is denoted by $R_{g}$,
$g\in\mathbb{R}_{1,3}$.

There is a \textit{natural} embedding $\mathbf{P}_{\mathrm{Spin}_{1,3}^{e}%
}(M)\hookrightarrow\mathcal{C}\ell_{\mathrm{Spin}_{1,3}^{e}}^{l}(M,g)$ which
comes from the embedding $\mathrm{Spin}_{1,3}^{e}\hookrightarrow
\mathbb{R}_{1,3}^{0}$. Hence (as we shall see in more details below), every
real left spinor bundle for $M$ can be captured from $\mathcal{C}%
\ell_{\mathrm{Spin}_{1,3}^{e}}^{l}(M,g)$, which is a vector bundle very
different from $\mathcal{C}\ell(M,g)$. Their relation is presented below, but
before, we introduce ideal left algebraic spinor fields.

Let $I(M,g)$ be a subbundle of $\mathcal{C}\ell_{\mathrm{Spin}_{1,3}^{e}}%
^{l}(M,g)$ such that there exists a primitive idempotent $\mathbf{e}$ of
$\mathbb{R}_{1,3}$ with
\begin{equation}
R_{\mathbf{e}}\Psi=\Psi
\end{equation}
for all $\Psi\in\sec I(M,g)\hookrightarrow\sec\mathcal{C}\ell_{\mathrm{Spin}%
_{1,3}^{e}}^{l}(M,g).$ Then, $I(M,g)$ is called a subbundle of \textit{ideal
left algebraic spinor fields. }Any $\Psi\in\sec I(M,g)\hookrightarrow
\sec\mathcal{C}\ell_{\mathrm{Spin}_{1,3}^{e}}^{l}(M,g)$ is called a
\textit{left ideal algebraic spinor field }(\textit{LIASF)\label{LIASF}}.

$I(M,g)$ can be thought of as being a real spinor bundle for $M$ such that
$\mathbf{M}$ in Eq.(\ref{4.7}) is a minimal left ideal of $\mathbb{R}_{1,3}.$

We say that two subbundles $I(M,g)$ and$\ \ I^{\prime}(M,g)$ of \textit{LIASF
are geometrically equivalent if the idempotents }$\mathbf{e},\mathbf{e}%
^{\prime}\in\mathbb{R}_{1,3}$ (appearing in the previous definition) are
related by an element $u\in$ $\mathrm{Spin}_{1,3}^{e}$, i.e., $\mathbf{e}%
^{\prime}=u\mathbf{e}u^{-1}$.

\label{RsCbundle}The \textit{right\ real spin-Clifford bundle }of $M$ is the
vector bundle
\begin{equation}
\mathcal{C}\ell_{\mathrm{Spin}_{1,3}^{e}}^{r}(M,g)=\mathbf{P}_{\mathrm{Spin}%
_{1,3}^{e}}(M)\times_{r}\mathbb{R}_{1,3}. \label{RsCbundle'}%
\end{equation}
Sections of $\mathcal{C}\ell_{\mathrm{Spin}_{1,3}^{e}}^{r}(M,\mathtt{g})$ are
called \textit{right spin-Clifford fields}

In Eq.(\ref{RsCbundle'}) $r$ refers to a representation of $\mathrm{Spin}%
_{1,3}^{e}$ on $\mathbb{R}_{1,3}$, given by $r(a)x=xa^{-1}$. As in the case
for the left real spin-Clifford bundle, there is a \emph{natural} embedding
$\mathbf{P}_{\mathrm{Spin}_{1,3}^{e}}(M)\hookrightarrow\mathcal{C}%
\ell_{\mathrm{Spin}_{1,3}^{e}}^{r}(M,g)$ which comes from the embedding
$\mathrm{Spin}_{1,3}^{e}\hookrightarrow\mathbb{R}_{1,3}^{0}$. There exists
also a natural left $L_{a}$ action of $a\in\mathbb{R}_{1,3}$ on $\mathcal{C}%
\ell_{\mathrm{Spin}_{1,3}^{e}}^{r}(M,g)$. This will be proved below.

Let $I^{\star}(M,g)$ be a subbundle of $\mathcal{C}\ell_{\mathrm{Spin}%
_{1,3}^{e}}^{r}(M,g)$ such that there exists a primitive idempotent element
$\mathbf{e}$ of $\mathbb{R}_{1,3}$ with%
\begin{equation}
L_{\mathbf{e}}\Psi=\Psi\label{IM1}%
\end{equation}
for any $\Psi$\ $\in\sec I^{\star}(M,g)\hookrightarrow\sec\mathcal{C}%
\ell_{\mathrm{Spin}_{1,3}^{e}}^{r}(M,g)$. Then, $I^{\star}(M,g)$ is called a
subbundle of right\textit{\ ideal algebraic spinor fields}. Any $\Psi\in\sec
I^{\star}(M,g)\hookrightarrow\sec\mathcal{C}\ell_{\mathrm{Spin}_{1,3}^{e}}%
^{r}(M,g)$ is called a RIASF\label{RIASF}. $I^{\star}(M,g)$ can be thought of
as being a real spinor bundle for $M$ such that $\mathbf{M}^{\star}$ in
Eq.(\ref{4.7bis}) is a minimal right ideal of $\mathbb{R}_{1,3}.$\ \ 

We say that two subbundles $I^{\ast}(M,g)$ and$\ I^{\ast\prime}(M,g)$ of
\textit{RIASF }are\textit{\ geometrically equivalent }if the
idempotents\textit{\ }$\mathbf{e},\mathbf{e}^{\prime}\in\mathbb{R}_{1,3}$
(appearing in the previous definition) are related by an element $u\in$
$\mathrm{Spin}_{1,3}^{e}$, i.e., $\mathbf{e}^{\prime}=u\mathbf{e}u^{-1}$.

The following result proved in \cite{moro} is crucial: in a spin manifold, we
have
\begin{equation}
\mathcal{C}\ell(M,g)=\mathbf{P}_{\mathrm{Spin}_{1,3}^{e}}(M)\times
_{\mathrm{Ad}}\mathbb{R}_{1,3}. \label{clifalt}%
\end{equation}

We recall also that $S(M,g)$ (or $\mathcal{C}\ell_{\mathrm{Spin}_{1,3}^{e}%
}^{l}(M,g)$) is a bundle of (left) \emph{modules} over the bundle of algebras
$\mathcal{C}\ell(M,g)$. In particular, the sections of the spinor bundle
($S(M,g)$ or $\mathcal{C}\ell_{\mathrm{Spin}_{1,3}^{e}}^{l}(M,g)$) are a
module over the sections of the Clifford bundle \cite{lawmi}. Then,
\label{colorbom}if $\Phi,\Psi\in\sec C\ell_{\mathrm{Spin}_{1,3}^{e}}^{l}(M,g)$
and $\Psi\neq0$, there exists $\mathit{\psi}\in\sec\mathcal{C}\ell(M,g)$
\ such that
\begin{equation}
\Psi=\mathit{\psi}\Phi. \label{3.20b}%
\end{equation}

This allows to identify a \emph{correspondence} between some sections of
$\mathcal{C}\ell(M,g)$ and some sections of $I(M)$ or $\mathcal{C}%
\ell_{\mathrm{Spin}_{1,3}^{e}}^{l}(M,g)$ once we fix a section on
$\mathcal{C}\ell_{\mathrm{Spin}_{1,3}^{e}}^{l}(M,g)$. This and other
correspondences are essential for the theory of Dirac-Hestenes spinor fields
(more details are given in \cite{moro}).\ Once we clarified which is the
meaning of a bundle of modules $S(M,g)$ over a bundle of algebras
$\mathcal{C}\ell(M,g),$ we can say that:

Two real left spinor bundles are equivalent, if and only if, they are
equivalent as bundles of $\mathcal{C}\ell(M,g)$ modules.\medskip

Of course, geometrically equivalently\ real left spinor bundles are equivalent.

\begin{remark}
\label{CSCBUNDLE}In what follows we denote the complexified left spin Clifford
bundle and the complexified right spin-Clifford bundle by
\begin{align}
\mathbb{C}\ell_{\mathrm{Spin}_{1,3}^{e}}^{l}(M)  &  =\mathbf{P}_{\mathrm{Spin}%
_{1,3}^{e}}(M)\times_{l}\mathbb{C\otimes R}_{1,3}\simeq\mathbf{P}%
_{\mathrm{Spin}_{1,3}^{e}}(M)\times_{l}\mathbb{R}_{4,1},\nonumber\\
\mathbb{C}\ell_{\mathrm{Spin}_{1,3}^{e}}^{r}(M)  &  =\mathbf{P}_{\mathrm{Spin}%
_{1,3}^{e}}(M)\times_{r}\mathbb{C\otimes R}_{1,3}\simeq\mathbf{P}%
_{\mathrm{Spin}_{1,3}^{e}}(M)\times_{r}\mathbb{R}_{4,1}.
\end{align}

\end{remark}

\subsection{Dirac-Hestenes Spinor Fields}

Let $\mathbf{E}^{\mu}$, $\mu=0,1,2,3$ be the canonical basis of $\mathbb{R}%
^{1,3}\hookrightarrow\mathbb{R}_{1,3}$ which generates the algebra
$\mathbb{R}_{1,3}$. They satisfy the basic relation $\mathbf{E}^{\mu
}\mathbf{E}^{\nu}+\mathbf{E}^{\nu}\mathbf{E}^{\mu}=2\eta^{\mu\nu}$. We recall
that \cite{rod04}%
\begin{equation}
\mathbf{e=}\frac{1}{2}(1+\mathbf{E}^{0})\in\mathbb{R}_{1,3} \label{dh1}%
\end{equation}
is a primitive idempotent of $\mathbb{R}_{1,3}$ and
\begin{equation}
\mathbf{f=}\frac{1}{2}(1+\mathbf{E}^{0})\frac{1}{2}(1+i\mathbf{E}%
^{2}\mathbf{E}^{1})\in\mathbb{C\otimes R}_{1,3} \label{dh2}%
\end{equation}
is a primitive idempotent of $\mathbb{C\otimes R}_{1,3}$. Now, let
$\mathbf{I=}\mathbb{R}_{1,3}\mathbf{e}$ and $\mathbf{I}_{\mathbb{C}%
}=\mathbb{C\otimes R}_{1,3}\mathbf{f}$ be respectively the minimal left ideals
of $\mathbb{R}_{1,3}$ and $\mathbb{C\otimes R}_{1,3}$ generated by
$\mathbf{e}$ and $\mathbf{f}$. Let $\mathbf{\phi=\phi e\in I}$ and
$\mathbf{\Psi=\Psi f\in I}_{\mathbb{C}}$. Then, any $\mathbf{\phi\in I}$ can
be written as
\begin{equation}
\mathbf{\phi=\psi e} \label{dh3}%
\end{equation}
with $\mathbf{\psi}\in\mathbb{R}_{1,3}^{0}$. Analogously, any $\mathbf{\Psi\in
I}_{\mathbb{C}}$ can be written as
\begin{equation}
\mathbf{\Psi=\psi e}\frac{1}{2}(1+i\mathbf{E}^{2}\mathbf{E}^{1}), \label{dh4}%
\end{equation}
with $\mathbf{\psi}\in\mathbb{R}_{1,3}^{0}$.

Recall moreover that $\mathbb{C\otimes R}_{1,3}\simeq\mathbb{R}_{4,1}$
$\simeq\mathbb{C(}4)$, where $\mathbb{C(}4)$ is the algebra of the $4\times4 $
complex matrices. We can verify that
\begin{equation}
\left(
\begin{array}
[c]{cccc}%
1 & 0 & 0 & 0\\
0 & 0 & 0 & 0\\
0 & 0 & 0 & 0\\
0 & 0 & 0 & 0
\end{array}
\right)  \label{dh5}%
\end{equation}
is a primitive idempotent of $\mathbb{C(}4)$ which is a matrix representation
of $\mathbf{f}$. In that way (see details in \cite{rod04}) there is a
bijection between column spinors, i.e., elements of $\mathbb{C}^{4}$ (the
complex $4$-dimensional vector space) and the elements of $\mathbf{I}%
_{\mathbb{C}}$. All that, plus the definitions of the left real and complex
spin bundles and the subbundle $I(M,g)$ suggests the following definition
\cite{moro}:

Let $\Phi\in\sec I(M,\mathtt{g})\hookrightarrow\sec\mathcal{C}\ell
_{\mathrm{Spin}_{1,3}^{e}}^{l}(M,\mathtt{g})$, i.e.,%
\begin{equation}
R_{\mathbf{e}}\Phi=\Phi\mathbf{e}=\Phi,\text{ }\mathbf{e}^{2}=\mathbf{e=\frac
{1}{2}(1+\mathbf{E}^{0})}\in\mathbb{R}_{1,3}. \label{dh5'}%
\end{equation}
A \textit{Dirac-Hestenes spinor field }(\textit{DHSF}) associated with $\Phi$
is an \emph{even} section $\mathbf{\psi}$ of $\mathcal{C}\ell_{\mathrm{Spin}%
_{1,3}^{e}}^{l}(M,g)$ such that
\begin{equation}
\Phi=\mathbf{\psi e}. \label{dh6}%
\end{equation}

An equivalent definition of a\textit{\ DHSF} is the following. Let $\Psi
\in\sec\mathbb{C}\ell_{\mathrm{Spin}_{1,3}^{e}}^{l}(M,g)$ be such that
\begin{equation}
R_{\mathbf{f}}\Psi=\Psi\mathbf{f=}\Psi,\mathbf{f}^{2}=\mathbf{f=\frac{1}%
{2}(1+\mathbf{E}^{0})\frac{1}{2}(1+i\mathbf{E}^{2}\mathbf{E}^{1})}%
\in\mathbb{C}\mathbf{\mathbb{\otimes}}\mathbb{R}_{1,3}.
\end{equation}
Then, a \textit{DHSF} associated to $\Psi$ is an even section $\mathbf{\psi}$
of $\mathcal{C}\ell_{\mathrm{Spin}_{1,3}^{e}}^{l}(M,\mathtt{g})\hookrightarrow
\mathbb{C}\ell_{\mathrm{Spin}_{1,3}^{e}}^{l}(M,\mathtt{g})$ such that
\begin{equation}
\Psi=\mathbf{\psi f.} \label{dh7}%
\end{equation}

In what follows, when we refer to a \textit{DHSF} $\mathbf{\psi}$ we omit for
simplicity the wording associated with $\Phi$ (or $\Psi$). It is very
important to observe that \textit{DHSF} are \textit{not }sums of even
multivector fields although, under a local trivialization, $\mathbf{\psi}$
$\in\sec C\ell_{\mathrm{Spin}_{1,3}^{e}}^{l}(M,\mathtt{g})$ is mapped on an
even element\footnote{Note that it is meaningful to speak about even (or odd)
elements in $C\ell_{\mathrm{Spin}_{1,3}^{e}}^{l}(M,g)$ since $\mathrm{Spin}%
_{1,3}^{e}\hookrightarrow\mathbb{R}_{1,3}^{0}.$
\par
{}} of \ $\mathbb{R}_{1,3}$. We emphasize that\textit{\ DHSF} are particular
sections of a spinor bundle, not of the Clifford bundle. However, we show
below how these objects have \textit{representatives} in the Clifford bundle.
To understand how this happens, recall that, when $M$ is a spin manifold:

(i) The elements of $\mathcal{C}\ell(M,\mathtt{g})=\mathbf{P}_{\mathrm{Spin}%
_{1,3}^{e}}(M)\times_{\mathrm{Ad}}\mathbb{R}_{1,3}$ are equivalence classes
$\left[  (p,a)\right]  $ of pairs $(p,a),$ where $p\in\mathbf{P}%
_{\mathrm{Spin}_{1,3}^{e}}(M)$, $a\in\mathbb{R}_{1,3}$ and $(p,a)\sim
(p^{\prime},a^{\prime})$ $\Leftrightarrow p^{\prime}=pu^{-1},$ $a^{\prime
}=uau^{-1}$, for some $u\in\mathrm{Spin}_{1,3}^{e}$;

(ii) The elements of $\mathcal{C}\ell_{\mathrm{Spin}_{1,3}^{e}}^{l}%
(M,\mathtt{g})$ are equivalence classes of pairs $(p,a),$ where $p\in
\mathbf{P}_{\mathrm{Spin}_{1,3}^{e}}(M)$, $a\in\mathbb{R}_{1,3}$ and
$(p,a)\sim(p^{\prime},a^{\prime})$ $\Leftrightarrow p^{\prime}=pu^{-1},$
$a^{\prime}=ua$, for some $u\in\mathrm{Spin}_{1,3}^{e}$;

(iii) The elements of $\mathcal{C}\ell_{\mathrm{Spin}_{1,3}^{e}}^{r}(M,g)$ are
equivalence classes of pairs $(p,a),$ where $p\in\mathbf{P}_{\mathrm{Spin}%
_{1,3}^{e}}(M)$, $a\in\mathbb{R}_{1,3}$ and $(p,a)\sim(p^{\prime},a^{\prime})$
$\Leftrightarrow p^{\prime}=pu^{-1},$ $a^{\prime}=au^{-1}$, for some
$u\in\mathrm{Spin}_{1,3}^{e}$.

In this way, it is possible to define the following natural actions on these
associated bundles \cite{moro}:

\textbf{1. }There is a natural right action of $\mathbb{R}_{1,3}$ on
$\mathcal{C}\ell_{\mathrm{Spin}_{1,3}^{e}}^{l}(M,\mathtt{g})$ and a natural
left action of $\mathbb{R}_{1,3}$ on $\mathcal{C}\ell_{\mathrm{Spin}_{1,3}%
^{e}}^{r}(M,\mathtt{g})$.\label{R_13 on Cl_left}

Indeed, given $b\in\mathbb{R}_{1,3}$ and $\alpha\in\mathcal{C}\ell
_{\mathrm{Spin}_{1,3}^{e}}^{l}(M,g),$ select a representative $(p,a)$ for
$\alpha$ and define $\alpha b:=\left[  (p,ab)\right]  .$ If another
representative $(pu^{-1},ua)$ is chosen for $\alpha,$ we have $(pu^{-1}%
,uab)\sim(p,ab)$ and thus $\alpha b$ is a well-defined element of
$\mathcal{C}\ell_{\mathrm{Spin}_{1,3}^{e}}^{l}(M,g)$.

Let us denote the space of $\mathbb{R}_{1,3}$-valued smooth functions on $M$
by $\mathcal{F}(M,\mathbb{R}_{1,3})$. Then, the above proposition immediately
yields the following.

\textbf{2. }There is a natural right action of $\mathcal{F}(M,\mathbb{R}%
_{1,3})$ on the sections of $\mathcal{C}\ell_{\mathrm{Spin}_{1,3}^{e}}%
^{l}(M,\mathtt{g})$ and a natural left action of $\mathcal{F}(M,\mathbb{R}%
_{1,3})$ on the sections of $\mathcal{C}\ell_{\mathrm{Spin}_{1,3}^{e}}%
^{r}(M,g)$.\label{C(R_13) on sec(Cl_left)}

\textbf{3. }There is a natural left action of $\sec\mathcal{C}\ell
(M,\mathtt{g})$ on the sections of $\mathcal{C}\ell_{\mathrm{Spin}_{1,3}^{e}%
}^{l}(M,\mathtt{g})$ and a natural right action of $\sec\mathcal{C}%
\ell(M,\mathtt{g})$ on sections of $\mathcal{C}\ell_{\mathrm{Spin}_{1,3}^{e}%
}^{r}(M,g)$.\label{sec(Cl) on sec(Cl_left)}

Indeed, given $\alpha\in\sec\mathcal{C}\ell(M,\mathtt{g})$ and $\beta\in
\sec\mathcal{C}\ell_{\mathrm{Spin}_{1,3}^{e}}^{l}(M,\mathtt{g}),$ select
representatives $(p,a)$ for $\alpha(x)$ and $(p,b)$ for $\beta(x)$ (with
$p\in\pi^{-1}(x)$) and define $(\alpha\beta)(x):=\left[  (p,ab)\right]
\in\mathcal{C}\ell_{\mathrm{Spin}_{1,3}^{e}}^{l}(M,\mathtt{g}).$ If
alternative representatives $(pu^{-1},uau^{-1})$ and $(pu^{-1},ub)$ are chosen
for $\alpha(x)$ and $\beta(x),$ we have%
\[
(pu^{-1},uau^{-1}ub)=(pu^{-1},uab)\sim(p,ab)
\]
and thus $(\alpha\beta)(x)$ is a well-defined element of $\mathcal{C}%
\ell_{\mathrm{Spin}_{1,3}^{e}}^{l}(M,g)$.

\textbf{4. }\label{sec(Cl_left) x sec(Cl_right)}There is a natural pairing%
\[
\sec\mathcal{C}\ell_{\mathrm{Spin}_{1,3}^{e}}^{l}(M,g)\times\sec
\mathcal{C}\ell_{\mathrm{Spin}_{1,3}^{e}}^{r}(M,g)\rightarrow\sec
\mathcal{C}\ell(M,g).
\]

Indeed, given $\alpha\in\sec\mathcal{C}\ell_{\mathrm{Spin}_{1,3}^{e}}%
^{l}(M,g)$ and $\beta\in\sec\mathcal{C}\ell_{\mathrm{Spin}_{1,3}^{e}}%
^{r}(M,g),$ select representatives $(p,a)$ for $\alpha(x)$ and $(p,b)$ for
$\beta(x)$ (with $p\in\pi^{-1}(x)$) and define $(\alpha\beta)(x):=\left[
(p,ab)\right]  \in\mathcal{C}\ell(M,\mathtt{g})$. If alternative
representatives $(pu^{-1},ua)$ and $(pu^{-1},bu^{-1})$ are chosen for
$\alpha(x)$ and $\beta(x),$ we have $(pu^{-1},uabu^{-1})\sim(p,ab)$ and thus
$(\alpha\beta)(x)$ is a well-defined element of $\mathcal{C}\ell(M,g)$.

\textbf{5. }\label{sec(CL_right) on sec(Cl_left)}There is a natural pairing%
\[
\sec\mathcal{C}\ell_{\mathrm{Spin}_{1,3}^{e}}^{r}(M,\mathtt{g})\times
\sec\mathcal{C}\ell_{\mathrm{Spin}_{1,3}^{e}}^{l}(M,\mathtt{g})\rightarrow
\mathcal{F}(M,\mathbb{R}_{1,3}).
\]

Indeed, given $\alpha\in\sec\mathcal{C}\ell_{\mathrm{Spin}_{1,3}^{e}}%
^{r}(M,\mathtt{g})$ and $\beta\in\sec\mathcal{C}\ell_{\mathrm{Spin}_{1,3}^{e}%
}^{l}(M,\mathtt{g}),$ select representatives $(p,a)$ for $\alpha(x)$ and
$(p,b)$ for $\beta(x)$ (with $p\in\pi^{-1}(x)$) and define $(\alpha
\beta)(x):=ab\in\mathbb{R}_{1,3}$. If alternative representatives
$(pu^{-1},au^{-1})$ and $(pu^{-1},ub)$ are chosen for $\alpha(x)$ and
$\beta(x),$ we have $au^{-1}ub=ab$ and thus $(\alpha\beta)(x)$ is a
well-defined element of $\mathbb{R}_{1,3}$.

\subsubsection{\textquotedblleft Unit Sections\textquotedblright}

We now show how to define \textquotedblleft unit sections\textquotedblright%
\ on the various vector bundles associated to the principal bundle
$\mathbf{P}_{\mathrm{Spin}_{1,3}^{e}}(M)$.

Let%
\begin{equation}
\Phi_{i}:\mathbf{\pi}^{-1}(U_{i})\rightarrow U_{i}\times\mathrm{Spin}%
_{1,3}^{e},\quad\Phi_{j}:\mathbf{\pi}^{-1}(U_{j})\rightarrow U_{j}%
\times\mathrm{Spin}_{1,3}^{e}\nonumber
\end{equation}
be two local trivializations for $\mathbf{P}_{\mathrm{Spin}_{1,3}^{e}}(M),$
with%
\[
\Phi_{i}(u)=(\pi(u)=x,\phi_{i,x}(u)),\quad\Phi_{j}(u)=(\pi(u)=x,\phi
_{j,x}(u)).
\]
Recall that the transition function on $g_{ij}:U_{i}\cap U_{j}\rightarrow
\mathrm{Spin}_{1,3}^{e}$ is then given by%
\[
g_{ij}(x)=\phi_{i,x}(u)\circ\phi_{j,x}(u)^{-1},
\]
which does not depend on $u$.

\textbf{6}. $\mathcal{C}\ell(M,g)$ has a naturally defined global unit
section.\textbf{\ }\ Indeed,\textbf{\ }for the associated bundle
$\mathcal{C}\ell(M,g)$, the transition functions corresponding to local
trivializations
\begin{equation}
\Psi_{i}:\mathbf{\pi}_{c}^{-1}(U_{i})\rightarrow U_{i}\times\mathbb{R}%
_{1,3}\text{,\quad}\Psi_{j}:\mathbf{\pi}_{c}^{-1}(U_{j})\rightarrow
U_{j}\times\mathbb{R}_{1,3},
\end{equation}
are given by $h_{ij}(x)=\mathrm{Ad}_{g_{ij}(x)}$. Define the local sections
\begin{equation}
\mathbf{1}_{i}(x)=\Psi_{i}^{-1}(x,1),\quad\mathbf{1}_{j}(x)=\Psi_{j}%
^{-1}(x,1), \label{?1}%
\end{equation}
where $1$ is the unit element of $\mathbb{R}_{1,3}$. Since
\[
h_{ij}(x)\cdot1=\mathrm{Ad}_{g_{ij}(x)}(1)=g_{ij}(x)1g_{ij}(x)^{-1}=1,
\]
we see that the expressions above uniquely define a global section
$\mathbf{1\in}$ $\mathcal{C}\ell(M,g)$ with $\mathbf{1}|_{U_{i}}%
=\mathbf{1}_{i}$.

\textbf{7. }\label{IDENT?}There exists a unit section on $\mathcal{C}%
\ell_{\mathrm{Spin}_{p,q}^{e}}^{r}(M,g)$ (and also on $\mathcal{C}%
\ell_{\mathrm{Spin}_{p,q}^{e}}^{l}(M,g)$), if and only if, $\mathbf{P}%
_{\mathrm{Spin}_{p,q}^{e}}(M)$ is trivial.

Indeed, we show the necessity for the case of $\mathcal{C}\ell_{\mathrm{Spin}%
_{p,q}^{e}}^{r}(M,g)$,\footnote{The proof for the case of $\mathcal{C}%
\ell_{\mathrm{Spin}_{p,q}^{e}}^{l}(M,g)$ is analogous.} the sufficiency is
trivial. For $\mathcal{C}\ell_{\mathrm{Spin}_{p,q}^{e}}^{r}(M,g)$, the
transition functions corresponding to local trivializations
\begin{equation}
\Omega_{i}:\mathbf{\pi}_{sc}^{-1}(U_{i})\rightarrow U_{i}\times\mathbb{R}%
_{p,q}\text{,\quad}\Omega_{j}:\mathbf{\pi}_{sc}^{-1}(U_{j})\rightarrow
U_{j}\times\mathbb{R}_{p,q},
\end{equation}
are given by $k_{ij}(x)=R_{g_{ij}(x)}$, with $R_{a}:$ $\mathbb{R}%
_{p,q}\rightarrow\mathbb{R}_{p,q},x\mapsto xa^{-1}$. A unit section in
$\mathcal{C}\ell_{\mathrm{Spin}_{p,q}^{e}}^{r}(M,\mathtt{g})$ -- if it exists
-- is written in terms of these two local trivializations as
\begin{equation}
\mathbf{1}_{i}^{r}(x)=\Omega_{i}^{-1}(x,1),\quad\mathbf{1}_{j}^{r}%
(x)=\Omega_{j}^{-1}(x,1),
\end{equation}
and we must have $\mathbf{1}_{i}^{r}(x)=\mathbf{1}_{j}^{r}(x)$ $\forall x\in
U_{i}\cap U_{j}$. As $\Omega_{i}(\mathbf{1}_{i}^{r}(x))=(x,1)=$ $\Omega
_{j}(\mathbf{1}_{j}^{r}(x))$, we have $\mathbf{1}_{i}^{r}(x)=\mathbf{1}%
_{j}^{r}(x)$ $\Leftrightarrow1=k_{ij}(x)\cdot1\Leftrightarrow1=1g_{ij}%
(x)^{-1}\Leftrightarrow g_{ij}(x)=1$.\footnote{For general spin manifolds, the
bundle $\mathbf{P}_{\mathrm{Spin}_{p,q}^{e}}(M)$ is not necessarily trivial
for arbitrary $(p,q)$, but Geroch's theorem warrants that, for the special
case $(p,q)=(1,3)$ with $M$ non-compact, $\mathbf{P}_{\mathrm{Spin}_{1,3}^{e}%
}(M)$ is trivial. By the above proposition, we then see that $\mathcal{C}%
\ell_{\mathrm{Spin}_{1,3}^{e}}^{r}(M,g)$ and also $\mathcal{C}\ell
_{\mathrm{Spin}_{p,q}^{e}}^{l}(M,g)$ have\ global \textquotedblleft unit
sections\textquotedblright. It is most important to note, however, that each
different choice of a (global) trivialization $\Omega_{i}$ on $\mathcal{C}%
\ell_{\mathrm{Spin}_{1,3}^{e}}^{r}(M,\mathtt{g})$ (respectively $\mathcal{C}%
\ell_{\mathrm{Spin}_{p,q}^{e}}^{l}(M,g)$) induces a different global unit
section $\mathbf{1}_{i}^{r}$ (respectively $\mathbf{1}_{i}^{l}$). Therefore,
even in this case there is no canonical unit section on $\mathcal{C}%
\ell_{\mathrm{Spin}_{1,3}^{e}}^{r}(M,g)$ (respectively on $\mathcal{C}%
\ell_{\mathrm{Spin}_{1,3}^{e}}^{l}(M,g)$).}

We already recalled that when the (non-compact) spacetime $M$ is a spin
manifold, the bundle $\mathbf{P}_{\mathrm{Spin}_{1,3}^{e}}(M)$ admits global
sections. With this in mind, let us fix a \textit{spin frame }$_{s}\Xi
$\textit{\ }and its dual spin\textit{\ coframe }$\Xi$ for $M$. This induces a
global trivialization for $\mathbf{P}_{\mathrm{Spin}_{1,3}^{e}}(M),$ which we
denote by $\Phi_{\Xi}:\mathbf{P}_{\mathrm{Spin}_{1,3}^{e}}(M)\rightarrow
M\times\mathrm{Spin}_{1,3}^{e},$ with $\Phi_{\Xi}^{-1}(x,1)=\Xi(x)$. As we
show in the following, the spin coframe $\Xi$ can also be used to induce
certain fiducial global sections on the various vector bundles associated to
$\mathbf{P}_{\mathrm{Spin}_{1,3}^{e}}(M)$:

\textbf{(a)} $\mathcal{C}\ell(M,g)$

Let $\{\mathbf{E}^{\mathbf{a}}\}$ be a fixed orthonormal basis of
$\mathbb{R}^{1,3}\hookrightarrow\mathbb{R}_{1,3}$ (which can be thought of as
being the \textit{canonical} basis of $\mathbb{R}^{1,3}$). We define basis
sections in $\mathcal{C}\ell(M,g)=\mathbf{P}_{\mathrm{Spin}_{1,3}^{e}%
}(M)\times_{\mathrm{Ad}}\mathbb{R}_{1,3}$ by $\theta_{\mathbf{a}}(x)=\left[
(\Xi(x),\mathbf{E}_{a})\right]  $. Of course, this induces a
\textit{multiform} basis $\{%
\mbox{\boldmath{$\theta$}}%
_{I}(x)\}$ for each $x\in M$. Note that a more precise notation for
$\theta_{\mathbf{a}}$ would be, for instance, $\theta_{\mathbf{a}}^{(\Xi)}$.

\textbf{(b)} $\mathcal{C}\ell_{\mathrm{Spin}_{1,3}^{e}}^{l}(M,g)$

Let $\mathbf{1}_{\Xi}^{l}\in\sec\mathcal{C}\ell_{\mathrm{Spin}_{1,3}^{e}}%
^{l}(M,g)$ be defined by $\mathbf{1}_{\Xi}^{l}(x)=\left[  (\Xi(x),1)\right]
$. Then the natural right action of $\mathbb{R}_{1,3}$ on $\mathcal{C}%
\ell_{\mathrm{Spin}_{1,3}^{e}}^{l}(M,g)$ leads to $\mathbf{1}_{\Xi}%
^{l}(x)a=\left[  (\Xi(x),a)\right]  $ for all $a\in\mathbb{R}_{1,3}$. It
follows from property \textbf{2 }above that an arbitrary section $\alpha
\in\sec\mathcal{C}\ell_{\mathrm{Spin}_{1,3}^{e}}^{l}(M,g)$ can be written as
$\alpha=\mathbf{1}_{\Xi}^{l}f$, with $f\in\mathcal{F}(M,\mathbb{R}_{1,3})$.

\textbf{(c) }$\mathcal{C}\ell_{\mathrm{Spin}_{1,3}^{e}}^{r}(M,g)$

Let $\mathbf{1}_{\Xi}^{r}\in\sec\mathcal{C}\ell_{\mathrm{Spin}_{1,3}^{e}}%
^{r}(M,g)$ be defined by $\mathbf{1}_{\Xi}^{r}(x)=\left[  (\Xi(x),1)\right]
$. Then the natural left action of $\mathbb{R}_{1,3}$ on $\mathcal{C}%
\ell_{\mathrm{Spin}_{1,3}^{e}}^{r}(M,g)$ leads to $a\mathbf{1}_{\Xi}%
^{r}(x)=\left[  (\Xi(x),a)\right]  $ for all $a\in\mathbb{R}_{1,3}$. It
follows from property \textbf{2} that an arbitrary section $\alpha\in
\sec\mathcal{C}\ell_{\mathrm{Spin}_{1,3}^{e}}^{r}(M,g)$ can be written as
$\alpha=f\mathbf{1}_{\Xi}^{r}$, with $f\in\mathcal{F}(M,\mathbb{R}_{1,3}%
)$.\bigskip

Now recall again that a spin structure on $M$ is a 2-1 bundle map
$s:\mathbf{P}_{\mathrm{Spin}_{1,3}^{e}}(M)\rightarrow\mathbf{P}_{\mathrm{SO}%
_{1,3}^{e}}(M)$ such that $s(pu)=s(p)\mathrm{Ad}_{u},\ \forall p\in
P_{\mathrm{Spin}_{1,3}^{e}}(M),$ $u\in\mathrm{Spin}_{1,3}^{e}$, where
$\mathrm{Ad:Spin}_{1,3}^{e}\rightarrow\mathrm{SO}_{1,3}^{e},$ $\mathrm{Ad}%
_{u}:x\mapsto uxu^{-1}$. We see that the specification of the global section
in the case (\textbf{a}) above is compatible with the Lorentz coframe
$\{\theta_{a}\}=s(\Xi)$ assigned by $s$. More precisely, for each $x\in M$,
the element $s(\Xi(x))\in\mathbf{P}_{\mathrm{SO}_{1,3}^{e}}(M)$ is to be
regarded as a proper isometry $s(\Xi(x)):\mathbb{R}^{1,3}\rightarrow
T_{x}^{\ast}M$, so that $\theta_{\mathbf{a}}(x):=s(p)\cdot\mathbf{E}%
_{\mathbf{a}}$ yields a Lorentz coframe $\{\theta_{\mathbf{a}}\}$ on $M$,
which we denoted by $s(\Xi)$. On the other hand, $\mathcal{C}\ell(M,g)$ is
isomorphic to $\mathbf{P}_{\mathrm{Spin}_{1,3}^{e}}(M)\times_{\mathrm{Ad}%
}\mathbb{R}_{1,3}$, and we can always arrange things so that $\theta
_{\mathbf{a}}(x)$ is represented in this bundle as $\theta_{\mathbf{a}%
}(x)=\left[  (\Xi(x),\mathbf{E}_{\mathbf{a}})\right]  .$ In fact, all we have
to do is to verify that this identification is covariant under a change of
coframes. To see that, let $\Xi^{\prime}\in\sec(\mathbf{P}_{\mathrm{Spin}%
_{1,3}^{e}}(M))$ be another spin coframe on $M$. From the principal bundle
structure of $\mathbf{P}_{\mathrm{Spin}_{1,3}^{e}}(M)$, we know that, for each
$x\in M$, there exists (a unique) $u(x)\in\mathrm{Spin}_{1,3}^{e}$ such that
$\Xi^{\prime}(x)=\Xi(x)u(x)$. If we define, as above, $\theta_{\mathbf{a}%
}^{\prime}(x)=s(\Xi^{\prime}(x))\cdot\mathbf{E}_{\mathbf{a}}$, then%
\begin{align*}
\theta_{\mathbf{a}}^{\prime}(x)  &  =s(\Xi(x)u(x))\cdot\mathbf{E}_{\mathbf{a}%
}=s(\Xi(x))\mathrm{Ad}_{u(x)}\cdot\mathbf{E}_{\mathbf{a}}\\
&  =\left[  (\Xi(x),\mathrm{Ad}_{u(x)}\cdot\mathbf{E}_{\mathbf{a}})\right]
=\left[  (\Xi(x)u(x),\mathbf{E}_{\mathbf{a}})\right]  =\left[  (\Xi^{\prime
}(x),\mathbf{E}_{a})\right]  .
\end{align*}
This proves our claim. The following results proved in \cite{moro} are also
needed for what follows.%

\begin{align}
\text{(i) }\mathbf{E}_{\mathbf{a}}  &  =\mathbf{1}_{\Xi}^{r}(x)\theta
_{\mathbf{a}}(x)\mathbf{1}_{\Xi}^{l}(x)\text{, }\forall x\in M,\nonumber\\
\text{(ii) }\mathbf{1}_{\Xi}^{l}\mathbf{1}_{\Xi}^{r}  &  =1\in\mathcal{C}%
\ell(M,g),\label{moro1}\\
\text{(iii) }\mathbf{1}_{\Xi}^{r}\mathbf{1}_{\Xi}^{l}  &  =1\in\mathbb{R}%
_{1,3}.\nonumber
\end{align}

Consider now how the various global sections defined above transform when the
spin coframe $\Xi$ is changed. Let $\Xi^{\prime}\in\sec\mathbf{P}%
_{\mathrm{Spin}_{1,3}^{e}}(M)$ be another spin coframe with $\Xi^{\prime
}(x)=\Xi(x)u(x)$, where $u(x)\in\mathrm{Spin}_{1,3}^{e}$. Let $\theta
_{\mathbf{a}}$, $\mathbf{1}_{\Xi}^{r}$, $\mathbf{1}_{\Xi}^{l}$ and
$\theta_{\mathbf{a}}^{\prime}$, $\mathbf{1}_{\Xi^{\prime}}^{r}$,
$\mathbf{1}_{\Xi^{\prime}}^{l}$ be the global sections respectively defined by
$\Xi$ and $\Xi^{\prime}$ (as above). We then have (see proof in \cite{moro})
that if $\Xi,\Xi^{\prime}$ are two spin coframes related by $\Xi^{\prime}=\Xi
u$, where $u:M\rightarrow\mathrm{Spin}_{1,3}^{e}$, then%
\begin{align}
\text{(i) }\theta_{\mathbf{a}}^{\prime}  &  =U\theta_{\mathbf{a}}%
U^{-1}\nonumber\\
\text{(ii) }\mathbf{1}_{\Xi^{\prime}}^{l}  &  =\mathbf{1}_{\Xi}^{l}%
u=U\mathbf{1}_{\Xi}^{l},\nonumber\\
\text{(iii) }\mathbf{1}_{\Xi^{\prime}}^{r}  &  =u^{-1}\mathbf{1}_{\Xi}%
^{r}=\mathbf{1}_{\Xi}^{r}U^{-1}, \label{transforms}%
\end{align}
where $U\in\sec\mathcal{C}\ell(M,g)$ is the Clifford field associated to $u$
by $U(x)=[(\Xi(x),u(x))]$. Also, in (ii) and (iii), $u$ and $u^{-1}$
respectively act on $\mathbf{1}_{\Xi}^{l}\in\sec\mathcal{C}\ell_{\mathrm{Spin}%
_{1,3}^{e}}^{l}(M,g)$ and $\mathbf{1}_{\Xi}^{r}\in\sec\mathcal{C}%
\ell_{\mathrm{Spin}_{1,3}^{e}}^{r}(M,g).$

\section{ Covariant Derivatives of Clifford Fields}

Since the Clifford bundle of differential forms is $\mathcal{C}\ell
(M,g)=\mathcal{T}M/J_{\mathtt{g}},$ it is clear that any linear connection
$\nabla$ on $\mathcal{T}M$ which is metric compatible ($\mathbf{\nabla
}\mathtt{\mathbf{g}}=0$) passes to the quotient $\mathcal{T}M/J_{\mathbf{g}}$,
and thus define an algebra bundle connection \cite{crume}. In this way, the
covariant derivative of a Clifford field $A\in\sec\mathcal{C}\ell(M,g)$ is
completely determined.

We now derive \cite{moro} \ formulas for the covariant derivative of Clifford
fields (and for Dirac-Hestenes spinor fields)\emph{\ }using\emph{\ }the
general theory of connections in principal bundles and covariant derivatives
in associate vector bundles as developed in the Appendices (see also, e.g.,
\cite{choquet,koni,palais})

\label{DERCLIFFORD}The covariant derivative of a Clifford field $A\in
\sec\mathcal{C}\ell(M,g)$ (in a given gauge), in the direction of the vector
field $V\in\sec TM$ is given by
\begin{equation}
\mathbf{\nabla}_{V}A=\partial_{V}(A)+{\frac{1}{2}[}\omega_{V},A],
\label{N4.00}%
\end{equation}
where $\omega_{V}$ is the usual ($\bigwedge^{2}(T^{\ast}M)$-valued) connection
1-form written in the basis $\{\theta_{\mathbf{a}}\}$ . We recall that
$\partial_{V}$ is the Pfaff derivative operator (Eq.(\ref{pfaaf})), that is,
if $A=A^{I}\theta_{I},$ then $\partial_{V}(A):=V(A^{I})\theta_{I}%
$.\footnote{$I$ denotes collective indices of a basis of $\mathcal{C\ell
}(M,\mathtt{g})$, e.g., $\theta_{12}=\theta_{1}\theta_{2}$.}

Indeed, writing $A(t)=A(\sigma(t))$ in terms of the multiform basis
$\{\varepsilon_{I}\}$ of sections associated to a given spin coframe, we have
$A(t)=A^{I}(t)\varepsilon_{I}(t)=A^{I}(t)[(\Xi(t),\mathbf{E}_{I}%
)]=[(\Xi(t),A^{I}(t)\mathbf{E}_{I})]=[(\Xi(t),a(t))],$ with $a(t):=A^{I}%
(t)\mathbf{E}_{I}\in\mathbb{R}_{1,3}$. Then the parallel transported field is
given by \cite{choquet}%
\begin{equation}
A_{||t}^{0}=[(\Xi(0),g(t)a(t)g(t)^{-1})] \label{A^0_||t}%
\end{equation}
for some $g(t)\in\mathrm{Spin}_{1,3}^{e}$, with $g(0)=1$. Thus%
\begin{align*}
\lim_{t\rightarrow0}\frac{1}{t}\left[  g(t)a(t)g(t)^{-1}-a(0)\right]   &
=\left[  \frac{dg}{dt}ag^{-1}+g\frac{da}{dt}g^{-1}+ga\frac{dg_{t}^{-1}}%
{dt}\right]  _{t=0}=\\
&  =\dot{a}(0)+\dot{g}(0)a(0)-a(0)\dot{g}(0)=\\
&  =V(A^{I})\mathbf{E}_{I}+[\dot{g}(0),a(0)],
\end{align*}
where $\dot{g}(0)\in\mathrm{spin}_{1,3}^{e}=\bigwedge\nolimits^{2}%
(\mathbb{R}^{1,3})$. Therefore%
\[
\mathbf{\nabla}_{V}A=V(A^{I})\theta_{I}+\frac{1}{2}[\omega_{V},A],
\]
where%
\begin{align}
\omega_{V}  &  =[(p,\dot{g}(0))]{=\frac{1}{2}}\omega_{V\mathbf{b}}%
^{\mathbf{a}}\theta_{\mathbf{a}}\wedge\theta^{\mathbf{b}}=\frac{1}{2}%
\omega_{\text{ }V}^{\mathbf{a}}{}^{\mathbf{b}}\theta_{\mathbf{a}}\wedge
\theta_{\mathbf{b}}\label{def of w}\\
&  =\frac{1}{2}V^{\mathbf{c}}\omega_{\mathbf{c}\text{ }}^{\mathbf{ab}}%
\theta_{\mathbf{a}}\wedge\theta_{\mathbf{b}}={\frac{1}{2}}V^{\mathbf{c}}%
\omega_{\mathbf{acb}}\theta^{\mathbf{a}}\wedge\theta^{\mathbf{b}}.
\end{align}

In particular, calculating the covariant derivative of the basis 1-covector
fields $\varepsilon_{a}$ yields $\frac{1}{2}[\omega_{\mathbf{e}_{\mathbf{c}}%
},\theta_{\mathbf{a}}]=\omega_{\mathbf{ca}}^{\mathbf{b}}{}{}\theta
_{\mathbf{b}}$. Note that
\begin{equation}
\omega_{\mathbf{acb}}=\eta_{\mathbf{ad}}\omega_{\mathbf{c}}^{\mathbf{d}}%
{}_{\mathbf{b}}=-\omega_{\mathbf{bca}}%
\end{equation}
and
\begin{equation}
\omega_{\mathbf{c}}^{\mathbf{ab}}=\eta^{\mathbf{ka}}\omega_{\mathbf{kcl}}%
\eta^{\mathbf{lb}}=-\omega_{\mathbf{c}}^{\mathbf{ba}}%
\end{equation}
In this way, $\omega:V\mapsto\omega_{V}$ is the usual ($\bigwedge^{2}(T^{\ast
}M)$-valued) connection 1-form on $M$ written in a given gauge (i.e., relative
to a spin frame and associated orthonormal (vector) coframe).

Eq.(\ref{N4.00}) shows that the covariant derivative preserves the degree of a
homogeneous Clifford field, as can be easily verified.

The general formula given by Eq.(\ref{N4.00}) and the associative law in the
Clifford algebra immediately yields the result that the covariant derivative
$\mathbf{\nabla}_{V}$ on $\mathcal{C}\ell(M,g)$ acts as a derivation on the
algebra of sections, i.e., for $A,B\in\sec\mathcal{C}\ell(M,g)$ and $V\in\sec
TM$, it holds
\begin{equation}
\mathbf{\nabla}_{V}(AB)=(\mathbf{\nabla}_{V}A)B+A(\mathbf{\nabla}_{V}B).
\label{Leibnitz-Clifford}%
\end{equation}

Under a change of gauge (local Lorentz transformation), $\omega_{V}$
transforms as
\begin{equation}
\frac{1}{2}\omega_{V}\mapsto U\frac{1}{2}\omega_{V}U^{-1}+(\mathbf{\nabla}%
_{V}U)U^{-1}, \label{connection transf}%
\end{equation}

Indeed, using Eq.(\ref{N4.00}) we can calculate $\mathbf{\nabla}_{V}A$ in two
different gauges as%
\begin{equation}
\mathbf{\nabla}_{V}A=\partial_{V}(A)+{\frac{1}{2}[}\omega_{V},A], \label{dn1}%
\end{equation}
or%
\begin{equation}
\mathbf{\nabla}_{V}A=\partial_{V}^{\prime}(A)+{\frac{1}{2}[}\omega_{V}%
^{\prime},A], \label{dn2}%
\end{equation}
where by definition $\partial_{V}(A)=V(A_{I})\theta^{I}$ and $\partial
_{V}^{\prime}(A)=V(A_{I}^{\prime})\theta^{\prime I}$. Now, we observe that
since $\theta^{\prime I}=U\theta^{I}U^{-1}$, we can write%
\begin{align*}
U\partial_{V}(U^{-1}A)  &  =U\partial_{V}(U^{-1}A_{I}^{\prime}\theta^{\prime
I})=V(A_{I}^{\prime})\theta^{\prime I}+A_{I}^{\prime}\theta^{\prime
I}U\partial_{V}(U^{-1})\\
&  =\partial_{V}^{\prime}(A)+AU\partial_{V}(U^{-1}).
\end{align*}
Now, $U\partial_{V}(U^{-1}A)=\partial_{V}A+U(\partial_{V}U^{-1})A$ and it
follows that%
\[
\partial_{V}^{\prime}(A)=\partial_{V}(A)-[\left(  \partial_{V}U\right)
U^{-1},A]
\]
Then, we see comparing the second members of Eq.(\ref{dn1}) and Eq.(\ref{dn2})
that%
\begin{align}
{\frac{1}{2}}\omega_{V}  &  ={\frac{1}{2}}\omega_{V}^{\prime}+U\left(
\partial_{V}U^{-1}\right)  ,\nonumber\\
{\frac{1}{2}}\omega_{V}^{\prime}  &  ={\frac{1}{2}}\omega_{V}+\left(
\partial_{V}U\right)  U^{-1}, \label{dn3}%
\end{align}

and
\begin{align}
{\frac{1}{2}}\omega_{V}^{\prime}  &  ={\frac{1}{2}}\omega_{V}+\left[
\mathbf{\nabla}_{V}U-{\frac{1}{2}}\omega_{V}U+{\frac{1}{2}}U\omega_{V}\right]
U^{-1}\nonumber\\
&  ={\frac{1}{2}}U\omega_{V}U^{-1}+(\mathbf{\nabla}_{V}U)U^{-1}. \label{dn4}%
\end{align}

\subsection{Curvature and Torsion Extensors of a Riemann-Cartan Connection}

Let $\mathbf{u,v,t,z\in}\sec TM$ and $u,v,t,z$ $\in\bigwedge^{1}(T^{\ast
}M)\hookrightarrow\mathcal{C\ell}(M,g)$ the physically equivalent $1$-forms,
i.e., $u=$ \texttt{\textbf{g}}$(\mathbf{u,}$ $)$, etc\texttt{\textbf{.}}Let
moreover, as usual $\{e_{\mathbf{a}}\}$ be an orthonormal basis for $TM$ and
$\{\theta^{\mathbf{a}}\}$, $\theta^{\mathbf{a}}\in\sec\bigwedge\nolimits^{1}%
(T^{\ast}M)\hookrightarrow\mathcal{C\ell}(M,g)$ the corresponding dual basis
and consider the Riemann-Cartan spacetime $(M,\mathtt{\mathbf{g}}%
,\tau_{\mathtt{\mathbf{g}}},\uparrow,\nabla)$. Call $\mathbf{\partial}%
=\theta^{\mathbf{a}}e_{\mathbf{a}}$.

The \textit{torsion operator }of the connection is the $(2,1)$-extensor
field\footnote{For a detailed theory of \ multivector and extensor fields, see
\cite{fmr1,fmr2,fmr3,fmr4,fmr5,fmr6,fmr7}.
\par
{}}
\begin{align}
\mathbf{\tau:}\sec &  \bigwedge\nolimits^{2}(T^{\ast}M)\rightarrow
\sec\bigwedge\nolimits^{1}(T^{\ast}M),\nonumber\\
\mathbf{\tau(}u\wedge v)  &  =(u\cdot{\mbox{\boldmath$\partial$}}%
)v-(v\cdot{\mbox{\boldmath$\partial$}})u-\mathbf{[}u,v\mathbf{]}, \label{ten1}%
\end{align}
where the Lie bracket of form fields is\footnote{$\mathbf{e}_{\mathbf{a}%
}\theta_{\mathbf{b}}:=h_{\mathbf{a}}^{\mu}\partial_{\mu}(\eta_{\mathbf{bc}%
}\theta^{\mathbf{c}})=h_{\mathbf{a}}^{\mu}\eta_{\mathbf{bc}}\partial_{\mu
}(h_{\nu}^{\mathbf{c}}dx^{\nu})=h_{\mathbf{a}}^{\mu}\partial_{\mu
}(h_{\mathbf{b}\nu})dx^{\nu}$}
\begin{equation}
\lbrack u,v]=(u\cdot\partial)v-(v\cdot\partial)u,
\end{equation}

The \textit{torsion} of the connection is the extensor field%
\begin{align}
\mathbf{T}  &  \mathbf{:}\sec\mathbf{(}\bigwedge\nolimits^{1}(T^{\ast}%
M)\times\bigwedge\nolimits^{2}(T^{\ast}M))\rightarrow\sec\bigwedge
\nolimits^{0}(T^{\ast}M),\label{ten2}\\
\hspace{1.6cm}\mathbf{T(}z,u\wedge v)  &  =z\cdot\mathbf{\tau(}u\wedge v).
\end{align}

The \textit{connection} $(1,2)$-extensor field $%
\mbox{\boldmath{$\omega$}}%
$ is given by
\begin{align}%
\mbox{\boldmath{$\omega$}}%
&  :\sec\bigwedge\nolimits^{1}(T^{\ast}M)\rightarrow\sec\bigwedge
\nolimits^{2}(T^{\ast}M),\nonumber\\
v  &  \mapsto%
\mbox{\boldmath{$\omega$}}%
_{v}=\omega_{\mathbf{v}},\text{ }v=\mathtt{\mathbf{g}}(\mathbf{v,}\text{ }),
\end{align}
where $\omega_{\mathbf{v}}$ is the ($\bigwedge^{2}(T^{\ast}M)$-valued)
connection 1-form introduced in Eq.(\ref{def of w}).

The \textit{curvature biform} of the connection is the ($2$,$2)$-extensor
field\footnote{Note that $[\mathbf{u,v}]$ is the standard Lie bracket of
vector fields.} $\mathfrak{R}(u\wedge v)$%
\begin{align}
\mathfrak{R}  &  :\sec\bigwedge\nolimits^{2}(T^{\ast}M)\rightarrow
\sec\bigwedge\nolimits^{2}(T^{\ast}M)\nonumber\\
\mathfrak{R}(u\wedge v)  &  =u\cdot\mathbf{\partial}%
\mbox{\boldmath{$\omega$}}%
(v)-v\cdot\mathbf{\partial}%
\mbox{\boldmath{$\omega$}}%
(u)+%
\mbox{\boldmath{$\omega$}}%
(u)\times%
\mbox{\boldmath{$\omega$}}%
(v). \label{rie1}%
\end{align}

We now show that the curvature biform can be written as%
\begin{equation}
\mathfrak{R}(u\wedge v)=u\cdot{\mbox{\boldmath$\partial$}\omega}_{\mathbf{v}%
}-v\cdot{\mbox{\boldmath$\partial$}\omega}_{\mathbf{u}}-\frac{1}{2}[{\omega
}_{\mathbf{u}},{\omega}_{\mathbf{v}}]-\omega_{\lbrack\mathbf{u,v}]}.
\label{curvature}%
\end{equation}

Indeed, recall that by definition
\begin{equation}
u\cdot{\mbox{\boldmath$\partial$}(%
\mbox{\boldmath{$\omega$}}%
(v))}=u\cdot{\partial(%
\mbox{\boldmath{$\omega$}}%
(v))+[%
\mbox{\boldmath{$\omega$}}%
(u),%
\mbox{\boldmath{$\omega$}}%
(v)].}%
\end{equation}
Also,%
\begin{equation}
\left(  {u\cdot{\partial}}%
\mbox{\boldmath{$\omega$}}%
\right)  {{(v)\equiv u\cdot{\partial%
\mbox{\boldmath{$\omega$}}%
(v)}}=u\cdot{\partial(%
\mbox{\boldmath{$\omega$}}%
(v))}-%
\mbox{\boldmath{$\omega$}}%
(u\cdot{\partial v}).}%
\end{equation}
So,%
\begin{align}
u\cdot{\partial(%
\mbox{\boldmath{$\omega$}}%
(v))-v\cdot{\partial(%
\mbox{\boldmath{$\omega$}}%
(u))}}  &  {=u\cdot{\partial%
\mbox{\boldmath{$\omega$}}%
(v)}-}v\cdot{\partial%
\mbox{\boldmath{$\omega$}}%
(u)+%
\mbox{\boldmath{$\omega$}}%
(u\cdot{\partial v})-%
\mbox{\boldmath{$\omega$}}%
(v\cdot{\partial}u)}\nonumber\\
&  ={u\cdot{\partial%
\mbox{\boldmath{$\omega$}}%
(v)}-}v\cdot{\partial%
\mbox{\boldmath{$\omega$}}%
(u)+%
\mbox{\boldmath{$\omega$}}%
([u,v])}\nonumber\\
&  ={u\cdot{\partial\omega}_{\mathbf{v}}}-v\cdot{\partial\omega}_{\mathbf{u}%
}+{\omega}_{[\mathbf{u},\mathbf{v}]},
\end{align}
and using the above equations in Eq.(\ref{rie1}) we derives
Eq.(\ref{curvature}).

We can also easily show that
\begin{equation}
\lbrack\mathbf{\nabla}_{\mathbf{u}},\mathbf{\nabla}_{\mathbf{v}}%
]t-\mathbf{\nabla}_{[\mathbf{u,v}]}t=\frac{1}{2}[\mathfrak{R}(u\wedge v),t]
\label{exercise}%
\end{equation}

It is important to have in mind that the curvature operator of the theory of
covariant derivatives is given by (see, e.g., \cite{choquet})%
\begin{equation}
\rho(\mathbf{u,v})=[\mathbf{\nabla}_{\mathbf{u}},\mathbf{\nabla}_{\mathbf{v}%
}]-\mathbf{\nabla}_{[\mathbf{u,v}]} \label{cur oper}%
\end{equation}

The \textit{Riemann curvature extensor} of the connection is
\begin{align}
\mathbf{R}:\sec &  \bigwedge\nolimits^{2}(T^{\ast}M)\times\bigwedge
\nolimits^{2}(T^{\ast}M)\rightarrow\sec\bigwedge\nolimits^{0}(T^{\ast
}M),\nonumber\\
\mathbf{R(}t\wedge z,u\wedge v\mathbf{)}\mathbf{=(}  &  t\wedge z)\cdot
\mathfrak{R}(u\wedge v) \label{rie2}%
\end{align}

We can show that if $[e_{\mathbf{a}},e_{\mathbf{b}}]=c_{\mathbf{ab}%
}^{\mathbf{d}}e_{\mathbf{d}}$, then $[\hspace{-0.1cm}[\mathbf{\theta
}_{\mathbf{a}},\mathbf{\theta}_{\mathbf{a}}]\hspace{-0.1cm}]=c_{\mathbf{ab}%
}^{\mathbf{d}}\theta_{\mathbf{d}}$. Then, using the above formulas we have
immediately that%
\begin{align}
\mathbf{T(}z,u\wedge v)  &  =z_{\mathbf{d}}u^{\mathbf{a}}v^{\mathbf{b}%
}T_{\mathbf{ab}}^{\mathbf{d}},\nonumber\\
T_{\mathbf{ab}}^{\mathbf{c}}  &  =\omega_{\mathbf{ab}}^{\mathbf{c}}%
-\omega_{\mathbf{ba}}^{\mathbf{c}}-c_{\mathbf{ab}}^{\mathbf{c}}
\label{TORSION rie}%
\end{align}
and%
\begin{align}
\mathbf{R(}t\wedge z,u\wedge v\mathbf{)}  &  =t^{\mathbf{c}}z_{\mathbf{d}%
}u^{\mathbf{a}}v^{\mathbf{b}}R_{\mathbf{dab}}^{\mathbf{c}},\nonumber\\
R_{\mathbf{dab}}^{\mathbf{c}}  &  =e_{\mathbf{a}}(\omega_{\mathbf{bc}%
}^{\mathbf{d}})-e_{\mathbf{b}}(\omega_{\mathbf{ac}}^{\mathbf{d}}%
)+\omega_{\mathbf{ak}}^{\mathbf{d}}\omega_{\mathbf{bc}}^{\mathbf{k}}%
-\omega_{\mathbf{bk}}^{\mathbf{d}}\omega_{\mathbf{ac}}^{\mathbf{k}%
}-c_{\mathbf{ab}}^{\mathbf{k}}\omega_{\mathbf{kc}}^{\mathbf{d}}.
\label{riemann impo}%
\end{align}

Also, we have the very important result. For any $t\in\sec\bigwedge
\nolimits^{1}(T^{\ast}M)\hookrightarrow\sec\mathcal{C}\ell(M,g)$
\begin{equation}
\lbrack\mathbf{\nabla}_{e_{\mathbf{a}}},\mathbf{\nabla}_{e_{\mathbf{b}}%
}]t=\mathfrak{R}(\mathbf{\theta}_{\mathbf{a}}\wedge\mathbf{\theta}%
_{\mathbf{b}})\llcorner t-(T_{\mathbf{ab}}^{\mathbf{c}}-\omega_{\mathbf{ab}%
}^{\mathbf{c}}+\omega_{\mathbf{ba}}^{\mathbf{c}})\nabla_{e_{\mathbf{c}}}t.
\label{commut ident 1}%
\end{equation}

Indeed, using the previous results, we can write%
\begin{align}
\lbrack\mathbf{\nabla}_{e_{\mathbf{a}}},\mathbf{\nabla}_{e_{\mathbf{b}}}]t  &
=\frac{1}{2}[\mathfrak{R}(\mathbf{\theta}_{\mathbf{a}}\wedge\mathbf{\theta
}_{\mathbf{b}}),t]+\mathbf{\nabla}_{[e_{\mathbf{a}}\mathbf{,}e_{\mathbf{b}}%
]}t\nonumber\\
&  =\mathfrak{R}(\mathbf{\theta}_{\mathbf{a}}\wedge\mathbf{\theta}%
_{\mathbf{b}})\llcorner t+\mathbf{\nabla}_{[e_{\mathbf{a}}\mathbf{,}%
e_{\mathbf{b}}]-\nabla_{\mathbf{e}_{\mathbf{a}}}e_{\mathbf{b}}+\nabla
_{e_{\mathbf{b}}}e_{\mathbf{a}}}t+\mathbf{\nabla}_{\nabla_{e_{\mathbf{a}}%
}e_{\mathbf{b}}}t-\mathbf{\nabla}_{\nabla_{e_{\mathbf{b}}}e_{\mathbf{a}}%
}t\nonumber\\
&  =\mathfrak{R}(\mathbf{\theta}_{\mathbf{a}}\wedge\mathbf{\theta}%
_{\mathbf{b}})\llcorner t+\mathbf{\nabla}_{-T_{\mathbf{ab}}^{c}e_{\mathbf{c}}%
}t+\mathbf{\nabla}_{\omega_{\mathbf{ab}}^{\mathbf{c}}e_{\mathbf{c}}%
}t-\mathbf{\nabla}_{\omega_{\mathbf{ab}}^{\mathbf{c}}e_{\mathbf{c}}%
}t\nonumber\\
&  =\mathfrak{R}(\mathbf{\theta}_{\mathbf{a}}\wedge\mathbf{\theta}%
_{\mathbf{b}})\llcorner t-T_{\mathbf{ab}}^{\mathbf{c}}\nabla_{e_{\mathbf{c}}%
}t+\left(  \omega_{\mathbf{ab}}^{\mathbf{c}}-\omega_{\mathbf{ab}}^{\mathbf{c}%
}\right)  \nabla_{e_{\mathbf{c}}}t.
\end{align}

\section{Covariant Derivative of Spinor Fields}

The spinor bundles introduced in Sections 2.2, 2.3 and 2.4, like
$I(M,g)=\mathbf{P}_{\mathrm{Spin}_{1,3}^{e}}(M)\times_{\ell}I$, $I=\mathbb{R}%
_{1,3}\frac{1}{2}(1+E_{0})$, and $\mathcal{C}\ell_{\mathrm{Spin}_{1,3}^{e}%
}^{l}(M,g)$, $\mathcal{C}\ell_{\mathrm{Spin}_{1,3}^{e}}^{r}(M,g)$ (and
subbundles) are vector bundles. Thus, as in the case of Clifford fields we can
use the general theory of covariant derivative operators on associate vector
bundles (described in the Appendices) to obtain formulas for the covariant
derivatives of sections of these bundles. Given $\Psi\in\sec\mathcal{C}%
\ell_{\mathrm{Spin}_{1,3}^{e}}^{l}(M,g)$ and $\Phi\in\sec\mathcal{C}%
\ell_{\mathrm{Spin}_{1,3}^{e}}^{r}(M,g)$, we denote the corresponding
covariant derivatives by $\mathbf{\nabla}_{V}^{s}\Psi$ and $\mathbf{\nabla
}_{V}^{s}\Phi$ \footnote{Recall that $I^{l}(M)\hookrightarrow C\ell
_{\mathrm{Spin}_{1,3}^{e}}^{l}(M,\mathtt{g})$ and $I^{r}(M)\hookrightarrow
C\ell_{\mathrm{Spin}_{1,3}^{e}}^{r}(M,\mathtt{g})$.}. We have that for any
$\Psi\in\sec\mathcal{C}\ell_{\mathrm{Spin}_{1,3}^{e}}^{l}(M,g)$ and $\Phi
\in\sec\mathcal{C}\ell_{\mathrm{Spin}_{1,3}^{e}}^{r}(M,g)$
\begin{align}
\mathbf{\nabla}_{V}^{s}\Psi &  =\partial_{V}(\Psi)+{\frac{1}{2}}\omega_{V}%
\Psi,\label{NCS00}\\
\mathbf{\nabla}_{V}^{s}\Phi &  =\partial_{V}(\Phi)-{\frac{1}{2}}\Phi\omega
_{V}. \label{NCS000}%
\end{align}
The proof is analogous to the case of the covaraint derivative of Clifford
fields, with the difference that Eq.(\ref{A^0_||t}) should be substituted by%
\[
\Psi_{||t}^{0}=[(\Xi(0),g(t)a(t))]\text{ and }\Phi_{||t}^{0}=[(\Xi
(0),a(t)g(t)^{-1})].
\]

We observe that in the case of (covariant) spinor fields, the matrix
representation of the $\omega_{e_{\mathbf{i}}}$ are called
\textit{Fock-Ivanenko coefficients.}

Another important result (see \cite{moro}) is the following:

\label{Leib 1} Let $\mathbf{\nabla}$ be the connection on $\mathcal{C}%
\ell(M,g)$ to which $\mathbf{\nabla}^{s}$ is related. Then, for any $V\in\sec
TM$, $A\in\sec\mathcal{C}\ell(M,g)$, $\Psi\in\sec\mathcal{C}\ell
_{\mathrm{Spin}_{1,3}^{e}}^{l}(M,g)$ and $\Phi\in\sec\mathcal{C}%
\ell_{\mathrm{Spin}_{1,3}^{e}}^{r}(M,g),$%
\begin{align}
\mathbf{\nabla}_{V}^{s}(A\Psi)  &  =A(\mathbf{\nabla}_{V}^{s}\Psi
)+(\mathbf{\nabla}_{V}A)\Psi,\label{SCD1}\\
\mathbf{\nabla}_{V}^{s}(\Phi A)  &  =\Phi(\mathbf{\nabla}_{V}A)\mathbf{+}%
(\mathbf{\nabla}_{V}^{s}\Phi)A. \label{SCD11}%
\end{align}

Using the above results we can prove that the right unit sections associated
with spin coframes are \textit{not} constant in any covariant way. Indeed, if
$\mathbf{1}_{\Xi}^{r}\in\sec\mathcal{C}\ell_{\mathrm{Spin}_{1,3}^{e}}%
^{r}(M,g)$ is the right unit section associated to the spin coframe $\Xi$,
then%
\begin{equation}
\mathbf{\nabla}_{e_{\mathbf{a}}}^{s}\mathbf{1}_{\Xi}^{r}=-{\frac{1}{2}%
}\mathbf{1}_{\Xi}^{r}\omega_{\mathbf{e}_{\mathbf{a}}}\text{.}
\label{deriv of 1a}%
\end{equation}

We now calculate the covariant derivative of a spinor field $\Psi\in
\sec\mathcal{C}\ell_{\mathrm{Spin}_{1,3}^{e}}^{l}(M,g)$ in the direction of
the vector field $V\in\sec TM$ to confirm the validity of
Eq.(\ref{connection transf}).\footnote{Authors are grateful to Dr. R. A. Mosna
for discussions on this issue.}

Let $u:M\rightarrow\mathrm{Spin}_{1,3}^{e}\hookrightarrow\mathbb{R}_{1,3}$ be
such that in the spin gauge $\Xi\in\sec\mathbf{P}_{\mathrm{Spin}_{1,3}^{e}}$
$(M)$ we have for $U\in\sec\mathcal{C}\ell^{(0)}(M,g)$, $UU^{-1}=1$,
$U(x)=[(\Xi(x),u(x))]$. Then, we can write the covariant derivative of
$\Psi\in\sec\mathcal{C}\ell_{\mathrm{Spin}_{1,3}^{e}}^{l}(M,g)$ in two
different gauges $\Xi,\Xi^{\prime}\in\sec\mathbf{P}_{\mathrm{Spin}_{1,3}^{e}}$
$(M)$ as
\begin{align}
\mathbf{\nabla}_{V}^{s}\Psi &  =\partial_{V}(\Psi)+{\frac{1}{2}}\omega_{V}%
\Psi,\label{spin deriv 1}\\
\mathbf{\nabla}_{V}^{s}\Psi &  =\partial_{V}^{\prime}(\Psi)+{\frac{1}{2}%
}\omega_{V}^{\prime}\Psi. \label{spin deriv 2}%
\end{align}
Now,
\[
\Psi=\Psi_{I}s^{I}=\Psi_{I}^{\prime}s^{\prime I},
\]
where $s^{I},s^{\prime I}$ are the following equivalence classes in
$\sec\mathcal{C}\ell_{\mathrm{Spin}_{1,3}^{e}}^{l}(M,g)$
\[
s^{I}=\left[  \left(  \Xi(x),S^{I}\right)  \right]  _{\ell},\text{ }s^{\prime
I}=\left[  \left(  \Xi^{\prime}(x),S^{I}\right)  \right]  _{\ell},
\]
with $S^{I}$ a spinor basis in a minimal left ideal in $\mathbb{R}_{1,3}$.
Now, if \ $\Xi^{\prime}=\Xi u$ we can write, using the fact that
$\mathcal{C}\ell_{\mathrm{Spin}_{1,3}^{e}}^{l}(M,g)$ is a module over
$\mathcal{C}\ell(M,g)$ that
\begin{align*}
\text{ }s^{\prime I}  &  =\left[  \left(  \Xi^{\prime}(x),S^{I}\right)
\right]  _{\ell}=\left[  \left(  \Xi(x)u(x),S^{I}\right)  \right]  _{\ell
}=\left[  \left(  \Xi(x),uS^{I}\right)  \right]  _{\ell}\\
&  =\left[  \left(  \Xi(x),u(x)\right)  \right]  _{\mathcal{C}\ell}\left[
\left(  \Xi(x),S^{I}\right)  \right]  _{\ell}%
\end{align*}
Now, recalling that $U=\left[  \left(  \Xi(x),u(x)\right)  \right]
_{\mathcal{C}\ell}$ $\in\sec\mathcal{C}\ell(M,g)$ we can write%
\begin{equation}
\text{ }s^{\prime I}=Us^{I}. \label{spin deriv 4}%
\end{equation}
Recalling that
\begin{align}
\partial_{V}(\Psi)  &  =V(\Psi_{I})s^{I},\nonumber\\
\partial_{V}^{\prime}(\Psi)  &  =\partial_{V}(\Psi_{I}^{\prime})s^{\prime I},
\label{spin deriv 5}%
\end{align}
we get%
\begin{equation}
\partial_{V}^{\prime}(\Psi)=\partial_{V}(\Psi)+U^{-1}(\partial_{V}U).
\label{spin deriv 6}%
\end{equation}

Using Eq.(\ref{spin deriv 6}) in Eq.(\ref{spin deriv 1}) and
Eq.(\ref{spin deriv 2}) we immediately confirm the validity of
Eq.(\ref{connection transf}). We explore a little bit more some details in the
calculation of \ $\mathbf{\nabla}_{V}^{s}\Psi$.

Recall that since \ $\Psi=\left[  \left(  \Xi,\Psi_{\Xi}\right)  \right]
_{\ell}=\left[  \left(  \Xi^{\prime},\Psi_{\Xi}^{\prime}\right)  \right]
_{\ell}$ we can write%
\[
\left[  \left(  \Xi^{\prime},\Psi_{\Xi}^{\prime}\right)  \right]  _{\ell
}=\left[  \left(  \Xi u,\Psi_{\Xi}^{\prime}\right)  \right]  _{\ell}=\left[
\left(  \Xi,u\Psi_{\Xi}^{\prime}\right)  \right]  _{\ell},
\]

i.e.,
\begin{equation}
\Psi_{\Xi}^{\prime}=u^{-1}\Psi_{\Xi} \label{spin deriv 7}%
\end{equation}
Then,%

\begin{align}
\partial_{V}(\Psi)  &  =V(\Psi_{I})\left[  \left(  \Xi,S^{I}\right)  \right]
_{\ell}=\left[  \left(  \Xi,V(\Psi_{I})S^{I}\right)  \right]  _{\ell}=\left[
\left(  \Xi,\partial_{V}(\Psi_{\Xi}\right)  \right] \label{spin deriv 7a}\\
\partial_{V}^{\prime}(\Psi)  &  =V(\Psi_{I}^{\prime})\left[  \left(  \Xi
,S^{I}\right)  \right]  _{\ell}=\left[  \left(  \Xi,V(\Psi_{I}^{\prime}%
)S^{I}\right)  \right]  _{\ell}=\left[  \left(  \Xi,\partial_{V}(\Psi_{\Xi
}^{\prime}\right)  \right]  \label{spin deriv 7b}%
\end{align}
and
\begin{align}
\mathbf{\nabla}_{V}^{s}\Psi &  =\partial_{V}(\Psi)+{\frac{1}{2}}\omega_{V}%
\Psi=\left[  \left(  \Xi,\partial_{V}(\Psi_{\Xi})+{\frac{1}{2}}w_{V}\Psi_{\Xi
}\right)  \right]  _{\ell},\label{spin deriv 8}\\
\mathbf{\nabla}_{V}^{s}\Psi &  =\partial_{V}^{\prime}(\Psi)+{\frac{1}{2}%
}\omega_{V}^{\prime}\Psi=\left[  \left(  \Xi^{\prime},\partial_{V}(\Psi
_{\Xi^{\prime}})+{\frac{1}{2}}w_{V}^{\prime}\Psi_{\Xi}^{\prime}\right)
\right]  _{\ell}, \label{spin deriv 9}%
\end{align}
with%
\begin{align}
\omega_{V}  &  =\left[  \left(  \Xi,w_{V}\right)  \right]  _{\mathcal{C\ell}%
}\text{ },\label{spin deriv 9a}\\
\omega_{V}^{\prime}  &  =\left[  \left(  \Xi^{\prime},w_{V}^{\prime}\right)
\right]  _{\mathcal{C\ell}}. \label{spin deriv9aa}%
\end{align}
Then we can write using Eq.(\ref{spin deriv 9}),
\begin{align}
\mathbf{\nabla}_{V}^{s}\Psi &  =\left[  \left(  \Xi u,\partial_{V}(u^{-1}%
\Psi_{\Xi})+{\frac{1}{2}}w_{V}^{\prime}u^{-1}\Psi_{\Xi}\right)  \right]
_{\ell}\nonumber\\
&  =\left[  \left(  \Xi,u\partial_{V}(u^{-1}\Psi_{\Xi})+{\frac{1}{2}}%
uw_{V}^{\prime}u^{-1}\Psi_{\Xi}\right)  \right]  _{\ell}\nonumber\\
&  =\left[  \left(  \Xi,\partial_{V}\Psi_{\Xi}+u\partial_{V}(u^{-1})\Psi_{\Xi
}+{\frac{1}{2}}uw_{V}^{\prime}u^{-1}\Psi_{\Xi}\right)  \right]  _{\ell}
\label{spin deriv 10}%
\end{align}

Comparing Eqs.(\ref{spin deriv 8}) and (\ref{spin deriv 9}) we get%
\begin{equation}
{\frac{1}{2}}w_{V}^{\prime}=u^{-1}w_{V}u-\partial_{V}(u^{-1})u.
\label{spin deriv 11}%
\end{equation}

Next, we must verify that Eqs.(\ref{spin deriv 11}) and
(\ref{connection transf}) are compatible. To do this, we use
Eq.(\ref{spin deriv9aa}) to write
\begin{align}
\frac{1}{2}\omega_{V}^{\prime}  &  =\frac{1}{2}\left[  \left(  \Xi^{\prime
},w_{V}^{\prime}\right)  \right]  _{\mathcal{C\ell}}=U\frac{1}{2}\omega
_{V}U^{-1}+(\mathbf{\nabla}_{V}U)U^{-1}\nonumber\\
&  =\frac{1}{2}\omega_{V}+(\partial_{V}U)U^{-1}\nonumber\\
&  =\left[  \left(  \Xi,{\frac{1}{2}}w_{V}+\partial_{V}(u)u^{-1}\right)
\right]  _{\mathcal{C}\ell}\nonumber\\
&  =\left[  \left(  \Xi^{\prime}u^{-1},{\frac{1}{2}}w_{V}+\partial
_{V}(u)u^{-1}\right)  \right]  _{\mathcal{C}\ell}\nonumber\\
&  =\left[  \left(  \Xi^{\prime},{\frac{1}{2}}u^{-1}w_{V}u+u^{-1}\partial
_{V}(u)\right)  \right]  _{\mathcal{C}\ell}. \label{spin deriv 14}%
\end{align}

Comparing Eqs.(\ref{spin deriv 14}) and (\ref{spin deriv 11}), we see that
Eqs.(\ref{spin deriv 11}) and (\ref{connection transf}) are indeed compatible.

\section{ Many Faces of the Dirac Equation}

As well known \cite{choquet,rod04,moro}, a \textit{covariant} Dirac spinor
field is a section $\mathbf{\Psi}\in\sec S_{c}(M,g)=\mathbf{P}_{\mathrm{Spin}%
_{1,3}^{e}}(M)\times_{\mu_{l}}\mathbb{C}^{4}$. Let $(U=M,\Phi),\Phi
(\mathbf{\Psi})=(x,|\Psi(x)\rangle)$ be a \textit{global} trivialization
corresponding to a spin coframe $\Xi$, such that for $\{e_{\mathbf{a}}%
\}\in\mathbf{P}_{\mathrm{SO}_{1,3}^{e}}(M),$%
\begin{align}
s(\Xi)  &  =\{\theta^{\mathbf{a}}\}\text{, }\theta^{\mathbf{a}}\in
\sec\mathcal{C}\ell(M,g),\text{ }\theta^{\mathbf{a}}(e_{\mathbf{b}}%
)=\delta_{\mathbf{b}}^{\mathbf{a}}\nonumber\\
\theta^{\mathbf{a}}\theta^{\mathbf{b}}+\theta^{\mathbf{b}}\theta^{\mathbf{a}}
&  =2\eta^{\mathbf{ab}}\text{, }\mathbf{a,b}=0,1,2,3. \label{DE0}%
\end{align}
The usual Dirac equation in a Lorentzian \textit{spacetime} for the spinor
field $\mathbf{\Psi}$ --- in interaction with an electromagnetic field
$A\in\sec\bigwedge\nolimits^{1}(M)\hookrightarrow\sec\mathcal{C}\ell(M,g)$ ---
is then%
\begin{equation}
i\mathbf{\gamma}^{\mathbf{a}}(\mathbf{\nabla}_{e_{\mathbf{a}}}^{s}%
+iqA_{\mathbf{a}})|\Psi(x)\rangle-m|\Psi(x)\rangle=0, \label{DE}%
\end{equation}
where \ $\underline{\mathbf{\gamma}}^{\mathbf{a}}\in\mathbb{C(}4)$,
$\mathbf{a}=0,1,2,3$ is a set of \textit{constant} Dirac \textit{matrices}
satisfying
\begin{equation}
\ \underline{\mathbf{\gamma}}^{\mathbf{a}}\ \underline{\mathbf{\gamma}%
}^{\mathbf{b}}+\ \underline{\mathbf{\gamma}}^{\mathbf{b}}\ \underline
{\mathbf{\gamma}}^{\mathbf{a}}=2\eta^{\mathbf{ab}}. \label{DEC}%
\end{equation}

Due to the one-to-one correspondence between \textit{ideal }sections of
$\mathbb{C}\ell_{\mathrm{Spin}_{1,3}^{e}}^{l}(M,g)$, $\mathcal{C}%
\ell_{\mathrm{Spin}_{1,3}^{e}}^{l}(M,g)$ and of $S_{c}(M,g)$ as explained
above, we can \textit{translate} the Dirac Eq.(\ref{DE}) for a covariant
spinor field into an equation for a spinor field, which is a section of
$\mathbb{C}\ell_{\mathrm{Spin}_{1,3}^{e}}^{l}(M,g)$, and finally write an
equivalent equation for a \textit{DHSF }$\psi\in\sec\mathcal{C}\ell
_{\mathrm{Spin}_{1,3}^{e}}^{l}(M,g)$. In order to do that we introduce the
spin-Dirac operator.

\label{spin dirac}The \textit{spin-Dirac operator }acting on sections of
$\mathcal{C}\ell_{\mathrm{Spin}_{1,3}^{e}}^{l}(M,g)$ is the first order
differential operator \cite{lawmi}
\begin{equation}
{\mbox{\boldmath$\partial$}}^{s}=\theta^{\mathbf{a}}\mathbf{\nabla
}_{e_{\mathbf{a}}}^{s}, \label{DE02}%
\end{equation}
where $\{\theta^{\mathbf{a}}\}$ is a basis as defined in Eq.(\ref{DE0}).

\begin{remark}
It is crucial to keep in mind the distinction between the Dirac operator
${\mbox{\boldmath$\partial$}}$ (Eq.(\ref{12})) and the spin-Dirac operator
${\mbox{\boldmath$\partial$}}^{s}$ just defined.
\end{remark}

We now write Dirac equation in $\mathcal{C}\ell_{\mathrm{Spin}_{1,3}^{e}}%
^{l}(M,g)$, denoted \textit{DE}$\mathcal{C}\ell^{l}$. It is%
\begin{equation}
{\mbox{\boldmath$\partial$}}^{s}\mathbf{\psi E}^{\mathbf{21}}-m\mathbf{\psi
E}^{\mathbf{0}}-qA\mathbf{\psi}=0 \label{DHE01}%
\end{equation}
where $\mathbf{\psi}\in\sec\mathcal{C}\ell_{\mathrm{Spin}_{1,3}^{e}}^{l}(M,g)$
is a \textit{DHSF }and the $\mathbf{E}^{\mathbf{a}}\in\mathbb{R}_{1,3}$ are
such that $\mathbf{E}^{\mathbf{a}}\mathbf{E}^{\mathbf{b}}+\mathbf{E}%
^{\mathbf{b}}\mathbf{E}^{\mathbf{a}}=2\eta^{\mathbf{ab}}$. Multiplying
Eq.(\ref{DHE01}) on the right by the idempotent $\mathbf{f=}\frac{1}%
{2}(1+\mathbf{E}^{\mathbf{0}})\frac{1}{2}(1+i\mathbf{E}^{\mathbf{2}}%
\mathbf{E}^{\mathbf{1}})\in\mathbb{C\otimes R}_{1,3}$ we get after some simple
algebraic manipulations the following equation for the (complex) ideal left
spin-Clifford field $\Psi\mathbf{f=}\Psi\in\sec\mathbb{C}\ell_{\mathrm{Spin}%
_{1,3}^{e}}^{l}(M,g)$,%
\begin{equation}
i{\mbox{\boldmath$\partial$}}^{s}\Psi-m\Psi-qA\Psi=0. \label{DHE02}%
\end{equation}

Now we can easily show, that given any global trivializations $(U=M,\Theta)$
and $(U=M,\Phi),$ of $\mathcal{C}\ell(M,g)$ and $\mathcal{C}\ell
_{\mathrm{Spin}_{1,3}^{e}}^{l}(M,g),$ there exists matrix representations of
the $\{\theta^{\mathbf{a}}\}$ that are equal to the Dirac matrices
\ $\underline{\mathbf{\gamma}}^{\mathbf{a}}$ (appearing in Eq.(\ref{DE})). In
that way the correspondence between Eqs.(\ref{DE}), (\ref{DHE01}) and
(\ref{DHE02}) is proved.

\begin{remark}
We must emphasize at this point that we call Eq.(\ref{DHE01}) the
\textit{DE}$\mathcal{C}\ell^{l}$. It looks similar to the Dirac-Hestenes
equation (on Minkowski spacetime) discussed, e.g., in \cite{rod04}, \ but it
is indeed very different regarding its mathematical nature. It is an intrinsic
equation satisfied by a legitimate spinor field, namely a \textit{DHSF
}$\mathbf{\psi}\in\sec\mathcal{C}\ell_{\mathrm{Spin}_{1,3}^{e}}^{l}(M)$. The
question naturally arises: May we write an equation with the same mathematical
information of Eq.(\ref{DHE01}) but satisfied by objects living on the
Clifford bundle of an arbitrary Lorentzian spacetime, admitting a spin
structure? We now show that the answer to that question is yes.
\end{remark}

\subsection{The Dirac-Hestenes Equation}

We obtained above a Dirac equation, which we called \textit{DE}$C\ell^{l}$,
describing the motion of spinor fields represented by sections $\mathbf{\Psi}$
of $\mathcal{C}\ell_{\mathrm{Spin}_{1,3}^{e}}^{l}(M,g)$ in interaction with an
electromagnetic field $A\in\sec\mathcal{C}\ell(M,\mathtt{g}),$
\begin{equation}
{\mbox{\boldmath$\partial$}}^{s}\mathbf{\Psi E}^{\mathbf{21}}-qA\mathbf{\Psi
}=m\mathbf{\Psi E}^{\mathbf{0}}, \label{DE for Psi in Cl_left}%
\end{equation}
where ${\mbox{\boldmath$\partial$}}^{s}=\theta^{\mathbf{a}}\mathbf{\nabla
}_{e_{\mathbf{a}}}^{s}$, $\{\theta^{\mathbf{a}}\}$ is given by Eq.(\ref{DE0}),
$\mathbf{\nabla}_{e_{\mathbf{a}}}^{s}$ is the natural spinor covariant
derivative acting on $\sec\mathcal{C}\ell_{\mathrm{Spin}_{1,3}^{e}}^{l}(M,g)$
and $\{\mathbf{E}^{\mathbf{a}}\}\in\mathbb{R}^{1,3}\hookrightarrow
\mathbb{R}_{1,3}$ is such that $\mathbf{E}^{\mathbf{a}}\mathbf{E}^{\mathbf{b}%
}+\mathbf{E}^{\mathbf{b}}\mathbf{E}^{\mathbf{a}}=2\eta^{\mathbf{ab}}$. As we
already mentioned, although Eq.(\ref{DE for Psi in Cl_left}) is written in a
kind of Clifford bundle (i.e., $\mathcal{C}\ell_{\mathrm{Spin}_{1,3}^{e}}%
^{l}(M,g)$), it does not suffer from the inconsistency of representing spinors
as pure differential forms and, in fact, the object $\mathbf{\Psi}$ behaves as
it should under Lorentz transformations.

Of course, Eq.(\ref{DE for Psi in Cl_left}) is nothing more than a
\textit{rewriting} of the usual Dirac equation, where the role of the constant
gamma matrices is undertaken by the constant elements $\{\mathbf{E}%
^{\mathbf{a}}\}$ in $\mathbb{R}_{1,3}$ and by the set $\{\theta^{\mathbf{a}%
}\}$. In this way, Eq.(\ref{DE for Psi in Cl_left}) is \emph{not} a kind of
Dirac-Hestenes equation as discussed, e.g., in \cite{rod04}. It suffices to
say that (i) the state of the electron, represented by $\mathbf{\Psi}$, is not
a \textit{Clifford field} and (ii) the $\mathbf{E}^{a}$'s are just
\textit{constant} elements of $\mathbb{R}_{1,3}$ and not sections of 1-form
fields in $\mathcal{C}\ell(M,g)$. Nevertheless, as shown originally in
\cite{moro} Eq.(\ref{DE for Psi in Cl_left}) does lead to a multiform Dirac
equation once we carefully employ the theory of right and left actions on the
various Clifford bundles introduced above. It is that multiform equation that
we call the \textit{DHE}.

\subsubsection{Representatives of \emph{DHSF} on the Clifford Bundle}

Let $\{\mathbf{E}^{\mathbf{a}}\}$ be, as before, a fixed orthonormal basis of
$\mathbb{R}^{1,3}\hookrightarrow\mathbb{R}_{1,3}.$ Remember that these objects
are fundamental to the Dirac equation (\ref{DE for Psi in Cl_left}) in terms
of sections $\mathbf{\Psi}$ of $\mathcal{C}\ell_{\mathrm{Spin}_{1,3}^{e}}%
^{l}(M,g)$:%
\[
{\mbox{\boldmath$\partial$}}^{s}\mathbf{\Psi E}^{\mathbf{21}}-qA\mathbf{\Psi
}=m\mathbf{\Psi E}^{\mathbf{0}}.
\]
Let $_{s}\Xi\in\sec\mathbf{P}_{\mathrm{Spin}_{1,3}^{e}}(M)$ be a spin frame
and $\Xi$ its dual spin coframe on $M\ $and define the sections $\mathbf{1}%
_{\Xi}^{l},$ $\mathbf{1}_{\Xi}^{r}$ and $\theta_{\mathbf{a}}$, respectively on
$\mathcal{C}\ell_{\mathrm{Spin}_{1,3}^{e}}^{l}(M,g),$ $\mathcal{C}%
\ell_{\mathrm{Spin}_{1,3}^{e}}^{r}(M,g)$ and $\mathcal{C}\ell(M,g)$, as above.
Now we can use Eqs.(\ref{moro1}) and (\ref{transforms}) to write the above
equation in terms of sections of $\mathcal{C}\ell(M,g):$%
\begin{equation}
({\mbox{\boldmath$\partial$}}^{s}\mathbf{\Psi)1}_{\Xi}^{r}\text{
}\mathbf{\varepsilon}^{\mathbf{21}}\mathbf{1}_{\Xi}^{l}-qA\mathbf{\Psi
}=m\mathbf{\Psi1}_{\Xi}^{r}\text{\textbf{\ }}\mathbf{\varepsilon}^{\mathbf{0}%
}\mathbf{1}_{\Xi}^{l}.
\end{equation}
Right-multiplying by $\mathbf{1}_{\Xi}^{r}$ yields,%
\begin{equation}
\mathbf{\varepsilon}^{\mathbf{a}}(\mathbf{\nabla}_{e_{\mathbf{a}}}%
^{s}\mathbf{\Psi)1}_{\Xi}^{r}\mathbf{\varepsilon}^{\mathbf{21}}-qA\mathbf{\Psi
1}_{\Xi}^{r}=m\mathbf{\Psi1}_{\Xi}^{r}\mathbf{\varepsilon}^{\mathbf{0}}.
\end{equation}

It follows from Eq.(\ref{SCD11}) that
\begin{align}
(\mathbf{\nabla}_{e_{\mathbf{a}}}^{s}\mathbf{\Psi)1}_{\Xi}^{r}  &
=\mathbf{\nabla}_{e_{\mathbf{a}}}(\mathbf{\Psi1}_{\Xi}^{r})-\mathbf{\Psi
\nabla}_{e_{\mathbf{a}}}^{s}(\mathbf{1}_{\Xi}^{r})\nonumber\\
&  =\mathbf{\nabla}_{e_{\mathbf{a}}}(\mathbf{\Psi1}_{\Xi}^{r})+{\frac{1}{2}%
}\mathbf{\Psi1}_{\Xi}^{r}\omega_{e_{\mathbf{a}}}, \label{Leib 3}%
\end{align}
where Eq.(\ref{deriv of 1a}) was employed in the last step. Therefore%
\begin{equation}
\theta^{\mathbf{a}}\left[  \mathbf{\nabla}_{e_{\mathbf{a}}}(\mathbf{\Psi
1}_{\Xi}^{r})+{\frac{1}{2}}\mathbf{\Psi1}_{\Xi}^{r}\omega_{e_{\mathbf{a}}%
}\right]  \theta^{\mathbf{21}}-qA(\mathbf{\Psi1}_{\Xi}^{r})=m(\mathbf{\Psi
1}_{\Xi}^{r})\theta^{\mathbf{0}}.
\end{equation}
Thus it is natural to define, for each spin coframe $\Xi$, the Clifford field
$\psi_{\Xi}\in\sec\mathcal{C}\ell(M,g)$ by%
\begin{equation}
\mathit{\psi}_{\Xi}:=\mathbf{\Psi1}_{\Xi}^{r}. \label{definition of phi_Xi}%
\end{equation}
We then have%
\begin{equation}
\theta^{\mathbf{a}}\left[  \mathbf{\nabla}_{e_{\mathbf{a}}}\mathit{\psi}_{\Xi
}+{\frac{1}{2}}\mathit{\psi}_{\Xi}\omega_{e_{\mathbf{a}}}\right]
\theta^{\mathbf{21}}-qA\mathit{\psi}_{\Xi}=m\mathit{\psi}_{\Xi}\theta
^{\mathbf{0}}. \label{EDHprov}%
\end{equation}

A \textit{comment} about the nature of spinors is in order. As we
repeatedly\textit{\ }said in the previous sections, spinor fields should not
be ultimately regarded as fields of multiforms, for their behavior under
Lorentz transformations is not tensorial (they are able to distinguish between
$2\pi$ and $4\pi$ rotations). So, how can the identification above be correct?
The answer is that the definition in Eq.(\ref{definition of phi_Xi}) is
intrinsically spin-coframe dependent. Clearly, this is the price one ought to
pay if one wants to make sense of the procedure of representing spinors by
differential forms.

Note also that the covariant derivative acting on $\mathit{\psi}_{\Xi}$ in
Eq.(\ref{EDHprov}) is the tensorial covariant derivative $\mathbf{\nabla}_{V}$
on $\mathcal{C}\ell(M,g)$, as it should be. However, we see from the
expression above that $\mathbf{\nabla}_{V}$ acts on $\mathit{\psi}_{\Xi}$
together with the term ${\frac{1}{2}}\mathit{\psi}_{\Xi}\omega_{V}$.
Therefore, it is natural to define an \textit{effective covariant\ derivative}%
\ $\mathbf{\nabla}_{V}^{(s)}$ acting on $\mathit{\psi}_{\Xi}$ by%
\begin{equation}
\mathbf{\nabla}_{e_{\mathbf{a}}}^{(s)}\mathit{\psi}_{\Xi}:=\mathbf{\nabla
}_{e_{\mathbf{a}}}\mathit{\psi}_{\Xi}+{\frac{1}{2}}\mathit{\psi}_{\Xi}%
\omega_{e_{\mathbf{a}}}. \label{hahaha}%
\end{equation}
or, recalling Eq.(\ref{NCS00}),
\begin{equation}
\mathbf{\nabla}_{e_{\mathbf{a}}}^{(s)}\mathit{\psi}_{\Xi}=\partial
_{e_{\mathbf{a}}}(\mathit{\psi}_{\Xi})+{\frac{1}{2}}\omega_{e_{\mathbf{a}}%
}\mathit{\psi}_{\Xi}\text{,} \label{hahaha'}%
\end{equation}
which emulates the spinorial covariant derivative, as it should. We observe
moreover that if $\mathcal{C}\in\sec\mathcal{C}\ell(M,g)$ and if
$\mathit{\psi}_{\Xi}\in\sec\mathcal{C}\ell(M,g)$ is a representative of a
Dirac-Hestenes spinor field then%
\begin{equation}
\mathbf{\nabla}_{e_{\mathbf{a}}}^{(s)}\left(  \mathcal{C}\mathit{\psi}_{\Xi
}\right)  =\left(  \mathbf{\nabla}_{e_{\mathbf{a}}}\mathcal{C}\right)
\mathit{\psi}_{\Xi}+\mathcal{C(}\mathbf{\nabla}_{e_{\mathbf{a}}}%
^{(s)}\mathit{\psi}_{\Xi}). \label{HAHAHA}%
\end{equation}

With this notation, we finally have the Dirac-Hestenes equation for the
\emph{representative} Clifford field $\mathit{\psi}_{\Xi}\in\sec
\mathcal{C}\ell(M,\mathtt{g})$ of a \textit{DHSF} $\mathbf{\Psi}$ of
$\mathcal{C}\ell_{\mathrm{Spin}_{1,3}^{e}}^{l}(M,g)$ relative to the spin
coframe $\Xi$ on a Lorentzian spacetime:%

\begin{equation}
\theta^{\mathbf{a}}\mathbf{\nabla}_{\mathbf{e}_{\mathbf{a}}}^{(s)}%
\mathit{\psi}_{\Xi}\theta^{\mathbf{21}}-qA\mathit{\psi}_{\Xi}=m\mathit{\psi
}_{\Xi}\mathbf{\theta}^{\mathbf{0}}. \label{DHE in a RC spacetime}%
\end{equation}

Let us finally show that this formulation recovers the usual transformation
properties characteristic of the Hestenes' formalism as described, e.g., in
\cite{rod04}. Consider then, two spin coframes $\Xi,\Xi^{\prime}\in
\sec\mathbf{P}_{\mathrm{Spin}_{1,3}^{e}}(M),$ with $\Xi^{\prime}(x)=\Xi(x)u(x)
$, where $u(x)\in\mathrm{Spin}_{1,3}^{e}$ . It follows \cite{moro} that
$\mathit{\psi}_{\Xi^{\prime}}=\mathbf{\Psi1}_{\Xi^{\prime}}^{r}=\mathbf{\Psi
}u^{-1}\mathbf{1}_{\Xi}^{r}=\mathbf{\Psi1}_{\Xi}^{r}U^{-1}=\mathit{\psi}_{\Xi
}U^{-1}.$ Therefore, the various spin coframe dependent Clifford fields from
Eq.(\ref{DHE in a RC spacetime}) transform as%
\begin{align}
\theta_{\mathbf{a}}^{\prime}  &  =U\theta_{\mathbf{a}}U^{-1}%
,\label{transf dhsf}\\
\mathit{\psi}_{\Xi^{\prime}}  &  =\mathit{\psi}_{\Xi}U^{-1}.\nonumber
\end{align}
These are exactly the transformation rules one expects from fields satisfying
the Dirac-Hestenes equation.

\subsubsection{Passive Gauge Invariance of the DHE}

It is a trivial exercise to show that if
\[
\theta^{\prime\mathbf{a}}\mathbf{\nabla}_{e_{\mathbf{a}}^{\prime}}%
^{(s)}\mathit{\psi}_{\Xi^{\prime}}\theta^{\prime\mathbf{21}}-qA\mathit{\psi
}_{\Xi^{\prime}}=m\mathit{\psi}_{\Xi^{\prime}}\theta^{\prime\mathbf{0}}%
\]
then if the connection $\omega_{V}$ transforms as in
Eq.(\ref{connection transf}) then
\[
\theta^{\mathbf{a}}\mathbf{\nabla}_{e_{\mathbf{a}}}^{(s)}\mathit{\psi}_{\Xi
}\theta^{\mathbf{21}}-qA\mathit{\psi}_{\Xi}=m\mathit{\psi}_{\Xi}%
\theta^{\mathbf{0}}%
\]

This property will be referred as \textit{passive gauge invariance} of the
DHE. It shows that the fact that even if writing of the Dirac-Hestenes is, of
course, frame dependent, this fact does not imply in the selection of
\textit{any} preferred reference frame.

The concept of \textit{active} gauge invariance under local rotations of the
Dirac-Hestenes equation will be studied below.

\subsection{$[\nabla_{e_{\mathbf{a}}}^{s},\nabla_{e_{\mathbf{b}}}^{s}]\psi$}

Let $\psi\in\sec\mathcal{C}\ell^{(0)}(M,g)$ be a representative of a
Dirac-Hestenes spinor field in a spin frame $\Xi$ defining the orthonormal
basis $\{e_{\mathbf{a}}\}$ for $TM$ with $\{\theta^{\mathbf{a}}\}$,
$(\theta^{\mathbf{a}}\in\sec\bigwedge\nolimits^{1}(T^{\ast}M)\hookrightarrow
\mathcal{C\ell}(M,g))$ the corresponding dual basis and $\{\theta_{\mathbf{a}%
}\}$ the reciprocal basis of $\{\theta^{\mathbf{a}}\}$. Then, a trivial
calculation similar to the one that leads to Eq.(\ref{commut ident 1}) shows
that%
\begin{equation}
\lbrack\nabla_{e_{\mathbf{a}}}^{s},\nabla_{e_{\mathbf{b}}}^{s}]\psi=\frac
{1}{2}\mathfrak{R}(\theta_{\mathbf{a}}\wedge\theta_{\mathbf{b}})\psi
-(T_{\mathbf{ab}}^{\mathbf{c}}-\omega_{\mathbf{ab}}^{\mathbf{c}}%
+\omega_{\mathbf{ba}}^{\mathbf{c}})\mathbf{\nabla}_{e_{\mathbf{c}}}\psi,
\label{commut ident 2}%
\end{equation}
where $\mathfrak{R}(\theta_{\mathbf{a}}\wedge\theta_{\mathbf{b}})$ is the same
object as in Eq.(\ref{commut ident 1}). This shows the unifying power of our
formalism.\footnote{Also, compare that equation, e.g., with Eq.(6.4.54) of
Ramond%
\'{}%
s book \cite{ramond}, where some terms are missing.}

We are now prepared to discuss the meaning of local active Lorentz invariance
of Maxwell and Dirac-Hestenes Equations.

\section{Active Local Lorentz Invariance and Maxwell equations}

The action functional for the electromagnetic field $F$ generated by a current
$J$ in classical electromagnetic theory is (as it is well-known)%
\begin{equation}
\mathcal{A=}\int\nolimits_{U}F\wedge\star F-A\wedge\star J. \label{25}%
\end{equation}

Action (\ref{25}) is invariant under local active Lorentz transformations.
This statement is easy to see once we use the Clifford bundle formalism.
Indeed, taking into account that\footnote{Analogous argument applies to the
term $A\wedge\star J$.}
\begin{equation}
F\wedge\star F=(F\cdot F)\tau_{\mathtt{g}}, \label{26}%
\end{equation}
we see that if we perform an \textit{active }Lorentz transformation
\begin{equation}
F\mapsto\overset{R}{F}=RFR^{-1}, \label{26'}%
\end{equation}
where $R\in\sec\mathrm{Spin}_{1,3}^{e}(M)\hookrightarrow\sec\mathcal{C\ell
}^{(0)}\mathcal{(}M,g)$, we have
\begin{equation}
F\wedge\star F=\overset{R}{F}\wedge\star\overset{R}{F}. \label{27}%
\end{equation}
This follows since $\tau_{\mathtt{g}}=\theta^{5}$ commutes with even sections
of the Clifford bundle and $R\theta^{5}R^{-1}=\theta^{5}$.

Is $\overset{R}{F}$ a solution of Maxwell equations? The answer is in general
\textit{negative}. Indeed, if
\begin{equation}
dF=0,\qquad\delta F=-J, \label{28}%
\end{equation}
in general,
\begin{equation}
d(RFR^{-1})\neq0\text{, }\qquad\delta(RFR^{-1})\neq RJR^{-1}. \label{29}%
\end{equation}
\noindent This can be easily seen in the Clifford bundle formalism, since in
general,
\begin{equation}
{\mbox{\boldmath$\partial$}}(RFR^{-1})\neq R({\mbox{\boldmath$\partial$}}%
F)R^{-1}, \label{30}%
\end{equation}
because, in general, $\theta^{\mathbf{a}}R\neq R\theta^{\mathbf{a}}$.

So, in definitive, we showed that although Maxwell \textit{action} is
invariant under active local Lorentz transformations, Maxwell
\textit{equation} does not inherit that invariance. However, we can
\textit{modify} Maxwell equation in such way that we may propose that $F$ and
all $RFR^{-1}$ are gauge equivalent. In order for that to be the case, it is
necessary that the Dirac operator be ${\mbox{\boldmath$\partial$}}$ only in a
\textit{fiducial} gauge. In other gauges, we have (symbolically)
\begin{equation}
{\mbox{\boldmath$\partial$}}\left(  {}\right)  \mathbf{\mapsto}\overset
{R}{{\mbox{\boldmath$\partial$}}}=R{\mbox{\boldmath$\partial$}}\ \left(
{}\right)  R^{-1}. \label{31}%
\end{equation}
In this case, we always have
\begin{equation}
\overset{R}{{\mbox{\boldmath$\partial$}}}\overset{R}{F}=\overset{R}{J},
\label{32}%
\end{equation}
and we can say that distinct electromagnetic fields are also classified as
distinct equivalence classes, where $F$ and $\overset{R}{F}$ represent the
same field in different gauges. These different gauges arises from generalized
$G$-connections (Definition \ref{generalized connection}). Since this problem
appears also in an analogous way in the case of the Dirac equation, we discuss
in details, only this latter case. As we will see the suggestion that gauge
equivalent equations describe the same physical phenomena implies that the
different geometries on $(M,$\texttt{g}$)$ generated by equivalent gauge
connections, with different torsion and curvature tensors are
indistinguishable. The empirical use of a particular gauge is then a question
of \textit{convenience}.

\section{The Dirac-Hestenes Equation on a Riemann-Cartan Spacetime}

Let $(M,\mathtt{\mathbf{g}},\mathbf{\nabla},\tau_{\mathtt{\mathbf{g}}%
},\uparrow)$ be a general Riemann-Cartan spacetime. In this section we
investigate how to obtain a generalization of \ the Dirac-Hestenes equation
(introduced above for a Lorentzian spacetime) for a representative $\psi
_{\Xi_{u}}\in\sec\mathcal{C}\ell(M,g)$ of a Dirac-Hestenes spinor field
$\Psi\in\sec$ $\mathcal{C}\ell_{\mathrm{Spin}_{1,3}^{e}}^{l}(M,g).$ In order
to do that we first introduce a chart $(\varphi,U)$ from the maximal atlas of
$M$, with coordinate functions $\{x^{\mu}\}$. \ The associate coordinate basis
of $TU$ is denoted by $\{e_{\mu}=\frac{\partial}{\partial x^{\mu}}%
=\partial_{\mu}\}$ and we denote by $\{\gamma^{\mu}=dx^{\mu}\}$ its dual
basis. Moreover, we suppose that $\gamma^{\mu}\in\sec\bigwedge\nolimits^{1}%
(T^{\ast}M)\hookrightarrow\sec\mathcal{C}\ell(M,g)$. Also, let
$\{e_{\mathbf{a}}\}\in\sec\mathrm{P}_{\mathrm{SO}_{1,3}^{e}}(M) $ an
orthonormal frame and $\{\theta^{\mathbf{a}}\}$, $\theta^{\mathbf{a}}\in
\sec\bigwedge\nolimits^{1}(T^{\ast}M)\hookrightarrow\sec\mathcal{C}\ell(M,g)$
the corresponding dual basis. In what follows we are going to work in a
\textit{fixed} spin coframe $\Xi_{u}=(u,\{\theta^{\mathbf{a}}\})$ and so, in
order to simplify the notation we write simply $\psi$ instead of $\psi
_{\Xi_{u}}$.

Recall now that the Lagrangian density for a free Dirac-Hestenes spinor in
Minkowski spacetime $(M,\mathtt{\mathbf{\eta}},\mathring{D},\tau
_{\mathtt{\mathbf{\eta}}},\uparrow)$ \cite{FEMORO1} can be written in a global
Lorentz coordinate chart $\{\mathring{x}^{\mu}\}$ as%

\begin{align}
\mathcal{\mathring{L}}(\mathring{x},\psi,%
\mathring{{\boldmath\partial}}%
\psi)  &  =\mathfrak{\mathring{L}}(\mathring{x},\psi,%
\mathring{{\boldmath\partial}}%
\psi)d\mathring{x}^{0}\wedge d\mathring{x}^{1}\wedge d\mathring{x}^{2}\wedge
d\mathring{x}^{3}\nonumber\\
&  =\left[  (%
\mathring{{\boldmath\partial}}%
\psi\mathbf{i}\gamma_{3})\cdot\psi-m\psi\cdot\psi\right]  d\mathring{x}%
^{0}\wedge d\mathring{x}^{1}\wedge d\mathring{x}^{2}\wedge d\mathring{x}^{3},
\label{dhrc1}%
\end{align}
where $%
\mathring{{\boldmath\partial}}%
=d\mathring{x}^{\mu}\mathring{D}_{\frac{\partial}{\partial\mathring{x}^{\mu}}%
}$. The usual prescription of \textit{minimal coupling} between a given spinor
field and the (generalized) gravitational field,%
\[
d\mathring{x}^{\mu}\mapsto\theta^{\mathbf{a}}\text{, }\mathring{D}%
_{\frac{\partial}{\partial\mathring{x}^{\mu}}}\mapsto\nabla_{e_{\mathbf{a}}},
\]
suggests that we take the following Lagrangian for a (representative) $\psi$
of a Dirac-Hestenes spinor field in a Riemann-Cartan spacetime,%
\begin{align}
\mathcal{L}(x,\psi,{\mbox{\boldmath$\partial$}}^{(s)}\psi)  &  =\mathfrak{L}%
(x,\psi,{\mbox{\boldmath$\partial$}}^{s}\psi)dx^{0}\wedge dx^{1}\wedge
dx^{2}\wedge dx^{3}\nonumber\\
&  =\left[  ({\mbox{\boldmath$\partial$}}^{(s)}\psi\theta^{\mathbf{0}}%
\theta^{\mathbf{2}}\theta^{\mathbf{1}})\cdot\psi-m\psi\cdot\psi\right]
\sqrt{\left\vert \det\mathtt{\mathbf{g}}\right\vert }dx^{0}\wedge dx^{1}\wedge
dx^{2}\wedge dx^{3}, \label{dhrc2}%
\end{align}
where
\begin{equation}
{\mbox{\boldmath$\partial$}}^{(s)}{\psi}={\theta}^{\mathbf{a}}\nabla
_{e_{\mathbf{a}}}^{(s)}\psi={\theta}^{\mathbf{a}}\left(  \partial
_{e_{\mathbf{a}}}\psi+\frac{1}{2}\omega_{e_{a}}\psi\right)  \label{dhrc3}%
\end{equation}

Now, in order to use (almost directly) the Lagrangian formalism in
\textit{Minkowski} spacetime as presented in \cite{FEMORO1} for the present
problem we must recall that the term ${\mbox{\boldmath$\partial$}}_{x}%
\partial_{{\mbox{\boldmath$\partial$}}_{x}\psi}\mathfrak{\mathring{L}}$
$\ $appearing in the Euler-Lagrange equation \textit{(ELE)} for the
Dirac-Hestenes spinor field corresponds to $\frac{\partial}{\partial
\mathring{x}^{\mu}}\left(  \partial_{\frac{\partial\psi}{\partial\mathring
{x}^{\mu}}}\mathfrak{\mathring{L}}\right)  $ as can be proved by direct
calculation. If we take this observation into account we immediately obtain
that variation of the Lagrangian given by Eq.(\ref{dhrc2}) results in the
following \textit{ELE},%
\begin{equation}
\partial_{\psi}\mathfrak{L}-\partial_{\mu}\left(  \partial_{\partial_{\mu}%
\psi}\mathfrak{L}\right)  =0. \label{dhrc4}%
\end{equation}

But, since our Lagrangian is expressed in terms of the Pfaff derivatives
$\partial_{e_{\mathbf{a}}}\psi$ we must express $\partial_{\mu}\left(
\partial_{\partial_{\mu}\psi}\mathfrak{L}\right)  $ in terms of them. To do
that we first write
\begin{equation}
\theta^{\mathbf{a}}=h_{\mu}^{\mathbf{a}}dx^{\mu}. \label{dhrc5}%
\end{equation}
Then, since
\begin{equation}
\sqrt{\left\vert \det\mathtt{\mathbf{g}}\right\vert }=\frac{1}{\det(h_{\mu
}^{\mathbf{a}})}\equiv\mathfrak{h}^{-1} \label{dhrc6}%
\end{equation}
we have%
\begin{equation}
\mathfrak{L}(x,\psi,{\mbox{\boldmath$\partial$}}^{s}\psi)=\left[
({\mbox{\boldmath$\partial$}}^{s}\psi\theta^{\mathbf{0}}\theta^{\mathbf{2}%
}\theta^{\mathbf{1}})\cdot\psi-m\psi\cdot\psi\right]  \mathfrak{h}^{-1}.
\label{dhrc7}%
\end{equation}
Also,
\begin{equation}
\partial_{\partial_{\mu}\psi}\mathfrak{L=(}\partial_{\partial_{\mu}\psi
}\partial_{e_{\mathbf{a}}}\psi\mathfrak{)(}\partial_{\partial_{e_{\mathbf{a}}%
}\psi}\mathfrak{L)}=h_{\mathbf{a}}^{\mu}\partial_{\partial_{e_{\mathbf{a}}%
}\psi}\mathfrak{L} \label{dhrc8}%
\end{equation}
and Eq.(\ref{dhrc4}) becomes%
\begin{equation}
\partial_{\psi}\mathfrak{L-(\partial}_{\mu}h_{\mathbf{a}}^{\mu})\partial
_{\partial_{e_{\mathbf{a}}}\psi}\mathfrak{L-}\partial_{e_{\mathbf{a}}}\left(
\partial_{\partial_{e_{\mathbf{a}}}\psi}\mathfrak{L}\right)  =0. \label{dhrc9}%
\end{equation}
Now, we can verify that
\begin{equation}
\mathfrak{h}^{-1}\partial_{\mu}\mathfrak{h}=h_{\sigma}^{\mathbf{a}}%
\partial_{\mu}h_{\mathbf{a}}^{\sigma}, \label{dhrc10}%
\end{equation}
and taking into account that
\begin{equation}
\left[  e_{\mathbf{a}},e_{\mathbf{b}}\right]  =(h_{\mathbf{a}}^{\nu}%
\partial_{\mathbf{b}}h_{\nu}^{\mathbf{c}}-h_{\mathbf{b}}^{\nu}\partial
_{\mathbf{a}}h_{\nu}^{\mathbf{c}})e_{\mathbf{c}}=c_{\mathbf{ab}}^{\mathbf{c}%
}e_{\mathbf{c}} \label{dhrc11}%
\end{equation}
we can write
\begin{equation}
\partial_{\mu}h_{\mathbf{a}}^{\mu}=-c_{\mathbf{ab}}^{\mathbf{c}}%
+\mathfrak{h}^{-1}\partial_{e_{\mathbf{a}}}\mathfrak{h}, \label{dhrc12}%
\end{equation}
and Eq.(\ref{dhrc9}) becomes%
\begin{equation}
\partial_{\psi}\mathfrak{L}-[\partial_{e_{\mathbf{a}}}-c_{\mathbf{ab}%
}^{\mathbf{c}}+\mathfrak{h}^{-1}\partial_{e_{\mathbf{a}}}\mathfrak{h}]\left(
\partial_{\partial_{e_{\mathbf{a}}}\psi}\mathfrak{L}\right)  =0.
\label{dhrc13}%
\end{equation}
Now\footnote{The symbol $\langle$ $\rangle_{\psi}$ means projection on the
grades of $\psi$.},
\[
\partial_{\psi}\left(  {\theta}^{\mathbf{a}}\omega_{e_{\mathbf{a}}}\psi
\theta^{\mathbf{0}}\theta^{\mathbf{2}}\theta^{\mathbf{1}}\cdot\psi\right)
=\langle{\theta}^{\mathbf{a}}\omega_{e_{\mathbf{a}}}\psi\theta^{\mathbf{0}%
}\theta^{\mathbf{2}}\theta^{\mathbf{1}}+\omega_{e_{\mathbf{a}}}{\theta
}^{\mathbf{a}}\psi\theta^{\mathbf{0}}\theta^{\mathbf{2}}\theta^{\mathbf{1}%
}\rangle_{\psi},
\]
and then
\begin{align}
\mathfrak{h}\partial_{\psi}\mathfrak{L}  &  =\left(  {\theta}^{\mathbf{a}%
}\partial_{e_{\mathbf{a}}}\psi\theta^{\mathbf{0}}\theta^{\mathbf{2}}%
\theta^{\mathbf{1}}+\frac{1}{2}{\theta}^{\mathbf{a}}\omega_{e_{\mathbf{a}}%
}\psi\theta^{\mathbf{0}}\theta^{\mathbf{2}}\theta^{\mathbf{1}}+\frac{1}%
{2}\omega_{e_{\mathbf{a}}}{\theta}^{\mathbf{a}}\psi\theta^{\mathbf{0}}%
\theta^{\mathbf{2}}\theta^{\mathbf{1}}-2m\psi\right)  ,\nonumber\\
\partial_{\partial_{e_{\mathbf{a}}}\psi}\mathfrak{L}  &  =-\mathfrak{h}%
^{-1}\left(  {\theta}^{\mathbf{a}}\psi\theta^{\mathbf{0}}\theta^{\mathbf{2}%
}\theta^{\mathbf{1}}\right)  ,\nonumber\\
\partial_{e_{\mathbf{a}}}\left(  \partial_{\partial_{e_{\mathbf{a}}}\psi
}\mathfrak{L}\right)   &  =-\mathfrak{h}^{-1}\partial_{e_{\mathbf{a}}%
}\mathfrak{h}(\partial_{\partial_{e_{\mathbf{a}}}\psi}\mathfrak{L)}%
-\mathfrak{h}^{-1}{\theta}^{\mathbf{a}}\partial_{e_{\mathbf{a}}}\psi
\theta^{\mathbf{0}}\theta^{\mathbf{2}}\theta^{\mathbf{1}}, \label{dhrc14}%
\end{align}
and Eq.(\ref{dhrc13}) becomes
\begin{equation}
{\theta}^{\mathbf{a}}\nabla_{e_{\mathbf{a}}}^{(s)}\psi\theta^{\mathbf{0}%
}\theta^{\mathbf{2}}\theta^{\mathbf{1}}-\frac{1}{4}{\theta}^{\mathbf{a}}%
\omega_{e_{\mathbf{a}}}\psi\theta^{\mathbf{0}}\theta^{\mathbf{2}}%
\theta^{\mathbf{1}}+\frac{1}{4}{\theta}^{\mathbf{a}}\omega_{e_{\mathbf{a}}%
}\psi\theta^{\mathbf{0}}\theta^{\mathbf{2}}\theta^{\mathbf{1}}+\frac{1}%
{2}{\theta}^{\mathbf{a}}c_{\mathbf{ab}}^{\mathbf{b}}\psi-m\psi=0
\label{dhrc15a}%
\end{equation}
or, recalling that the components of the torsion tensor in an orthonormal
basis are given by%
\begin{equation}
T_{\mathbf{ab}}^{\mathbf{c}}=\omega_{\mathbf{ab}}^{\mathbf{c}}-\omega
_{\mathbf{ba}}^{\mathbf{c}}-c_{\mathbf{ab}}^{\mathbf{c}}, \label{dhrc15b}%
\end{equation}
and that, in particular $\omega_{\mathbf{ab}}^{\mathbf{b}}=\eta_{\mathbf{bb}%
}\omega_{\mathbf{a}}^{\mathbf{bb}}\ =0$, we have
\begin{align}
&  {\theta}^{\mathbf{a}}\nabla_{e_{\mathbf{a}}}^{(s)}\psi\theta^{\mathbf{0}%
}\theta^{\mathbf{2}}\theta^{\mathbf{1}}-\frac{1}{4}\omega_{e_{\mathbf{a}}%
}{\theta}^{\mathbf{a}}\psi\theta^{\mathbf{0}}\theta^{\mathbf{2}}%
\theta^{\mathbf{1}}-\frac{1}{2}c_{\mathbf{ab}}^{\mathbf{b}}{\theta
}^{\mathbf{a}}\psi\theta^{\mathbf{0}}\theta^{\mathbf{2}}\theta^{\mathbf{1}%
}-m\psi\nonumber\\
&  ={\theta}^{\mathbf{a}}\nabla_{e_{\mathbf{a}}}^{(s)}\psi\theta^{\mathbf{0}%
}\theta^{\mathbf{2}}\theta^{\mathbf{1}}-\frac{1}{4}{\theta}^{\mathbf{a}}%
\omega_{e_{\mathbf{a}}}\psi\theta^{\mathbf{0}}\theta^{\mathbf{2}}%
\theta^{\mathbf{1}}+\frac{1}{4}\omega_{e_{\mathbf{a}}}{\theta}^{\mathbf{a}%
}\psi\theta^{\mathbf{0}}\theta^{\mathbf{2}}\theta^{\mathbf{1}}-\frac{1}%
{2}c_{\mathbf{ab}}^{\mathbf{b}}{\theta}^{\mathbf{a}}\psi\theta^{\mathbf{0}%
}\theta^{\mathbf{2}}\theta^{\mathbf{1}}-m\psi\nonumber\\
&  ={\theta}^{\mathbf{a}}\nabla_{e_{\mathbf{a}}}^{(s)}\psi\theta^{\mathbf{0}%
}\theta^{\mathbf{2}}\theta^{\mathbf{1}}-\frac{1}{2}{\theta}^{\mathbf{a}%
}\lrcorner\omega_{e_{\mathbf{a}}}\psi\theta^{\mathbf{0}}\theta^{\mathbf{2}%
}\theta^{\mathbf{1}}-\frac{1}{2}c_{\mathbf{ab}}^{\mathbf{b}}{\theta
}^{\mathbf{a}}\psi\theta^{\mathbf{0}}\theta^{\mathbf{2}}\theta^{\mathbf{1}%
}-m\psi\label{dhrc16}\\
&  ={\theta}^{\mathbf{a}}\nabla_{e_{\mathbf{a}}}^{(s)}\psi\theta^{\mathbf{0}%
}\theta^{\mathbf{2}}\theta^{\mathbf{1}}+\frac{1}{2}T_{\mathbf{ab}}%
^{\mathbf{b}}{\theta}^{\mathbf{a}}\psi\theta^{\mathbf{0}}\theta^{\mathbf{2}%
}\theta^{\mathbf{1}}-m\psi=0\nonumber
\end{align}

Finally we write the Dirac-Hestenes equation in a general Riemann-Cartan
spacetime as%

\begin{equation}
{\mbox{\boldmath$\partial$}}^{(s)}\psi\theta^{\mathbf{2}}\theta^{\mathbf{1}%
}+\frac{1}{2}T\psi\theta^{\mathbf{0}}\theta^{\mathbf{2}}\theta^{\mathbf{1}%
}-m\psi\theta^{\mathbf{0}}=0, \label{dhrc17}%
\end{equation}
where%
\begin{equation}
T=T_{\mathbf{ab}}^{\mathbf{b}}{\theta}^{\mathbf{a}}. \label{dhrc18}%
\end{equation}
is called the \textit{torsion covector}. Note that in a Lorentzian manifold
$T=0$ and we come back to the Dirac-Hestenes equation given by
Eq.(\ref{EDHprov}). We observe moreover that the matrix representation of
Eq.(\ref{dhrc17}) coincides with an equation first proposed by Hehl and Datta
\cite{hehldatta}.

We observe yet that if we tried to get the equation of motion of a
Dirac-Hestenes spinor field in a Riemann-Cartan spacetime using the principle
of minimal coupling in the Dirac-Hestenes equation in Minkowski spacetime, we
would miss the term $\frac{1}{2}T\psi\theta^{\mathbf{2}}\theta^{\mathbf{1}}$
appearing in Eq.(\ref{dhrc17}). This would be very bad indeed, because in a
complete theory where the $\mathbf{\{}\theta^{\mathbf{a}}\mathbf{\}}$ and the
$\{\omega_{e_{\mathbf{a}}}\}$ must be dynamic fields it can be shown that
spinor fields generate torsion (details in \cite{hehldatta}).

\section{Active Lorentz Invariance of the Dirac-Hestenes Lagrangian}

In the proposed gauge theories of the gravitational field (see, e.g.,
\cite{mielke,ramond}) it is said that the Lagrangians and the corresponding
equations of motion of physical fields must be invariant under arbitrary local
Lorentz rotations. In this section we briefly investigate how to
mathematically implement such a hypothesis and what is its meaning for the
case of a Dirac-Hestenes spinor field on a Riemann-Cartan spacetime. The
Lagrangian we shall investigate is the one given by Eq.(\ref{dhrc7}), i.e.,%

\begin{equation}
\mathfrak{L}(x,\psi,{\mbox{\boldmath$\partial$}}^{(s)}\psi)=\left[
(\theta^{\mathbf{a}}\nabla_{e_{\mathbf{a}}}^{(s)}{}\psi\theta^{\mathbf{0}%
}\theta^{\mathbf{2}}\theta^{\mathbf{1}})\cdot\psi-m\psi\cdot\psi\right]
\mathfrak{h}^{-1} \tag{Dirac-Hestenes}%
\end{equation}

Observe that the Dirac-Hestenes Lagrangian has been written in a fixed
(passive gauge) individualized by a spin coframe $\Xi$ and we already know
that it is invariant under passive gauge transformations $\psi\mapsto\psi
U^{-1}$ ($UU^{-1}=1$, $U\in\sec\mathrm{Spin}_{1,3}^{e}(M)\hookrightarrow
\sec\mathcal{C\ell(}M,g)$ once the \ `connection' $2$-form $\omega_{V}$
transforms as given in Eq.(\ref{connection transf}), i.e.,
\begin{equation}
\frac{1}{2}\omega_{V}\mapsto U\frac{1}{2}\omega_{V}U^{-1}+(\mathbf{\nabla}%
_{V}U)U^{-1}. \label{passive connection}%
\end{equation}

Under an active rotation (gauge) transformation the fields transform in new
fields given by
\begin{align}
\psi &  \mapsto\psi^{\prime}=U\psi,\nonumber\\
\theta^{\mathbf{m}}  &  \mapsto\theta^{\prime\mathbf{m}}=U\theta^{\mathbf{m}%
}U^{-1}=\Lambda_{\mathbf{n}}^{\mathbf{m}}\theta^{\mathbf{n}},\nonumber\\
e_{\mathbf{m}}  &  \mapsto e_{\mathbf{m}}^{\prime}=(\Lambda^{-1})_{\mathbf{m}%
}^{\mathbf{n}}e_{\mathbf{n}}. \label{ari1}%
\end{align}

Now, according to the ideas of gauge theories, we must search for a new
connection $\nabla^{\prime s}$ such that the Lagrangian results invariant.
This will be the case if connections$\ \nabla^{s}$ and $\nabla^{\prime s}$ are
generalized $G$-connections (Definition \ref{generalized connection}), i.e.,
\begin{align}
\nabla_{e_{\mathbf{m}}^{\prime}}^{\prime(s)}(U\psi)  &  =U\nabla
_{e_{\mathbf{m}}}^{(s)}{}\psi,\nonumber\\
&  \text{or}\nonumber\\
\nabla_{e_{\mathbf{n}}}^{\prime(s)}(U\psi)  &  =\Lambda_{\mathbf{n}%
}^{\mathbf{m}}U\nabla_{e_{\mathbf{m}}}^{(s)}{}\psi. \label{ari2}%
\end{align}
Also, taking into account the structure of a representative of a spinor
covariant derivative in the Clifford bundle (see Eq.(\ref{NCS00})) we must
have we must have for the Pfaff derivative%
\begin{equation}
\partial_{\mathbf{e}_{n}}\mapsto\partial_{\mathbf{e}_{n}}^{\prime}%
=\Lambda_{\mathbf{n}}^{\mathbf{m}}\partial_{\mathbf{e}_{m}},
\end{equation}
and for the connection,%
\begin{align}
\omega_{e_{\mathbf{n}}}^{\prime}  &  =\Lambda_{\mathbf{n}}^{\mathbf{m}}\left(
U\omega_{e_{\mathbf{m}}}U^{-1}-2\mathbf{\partial}_{e_{\mathbf{m}}}%
(U)U^{-1}\right)  ,\nonumber\\
&  \text{or}\nonumber\\
\omega_{e_{\mathbf{m}}^{\prime}}^{\prime}  &  =U\omega_{e_{\mathbf{m}}}%
U^{-1}-2\mathbf{\partial}_{e_{\mathbf{m}}}(U)U^{-1}. \label{ari3}%
\end{align}

Write
\begin{align}
\omega_{e_{\mathbf{n}}}^{\prime}  &  =\frac{1}{2}\omega_{\mathbf{m}}%
^{\prime\mathbf{kl}}\theta_{\mathbf{k}}\wedge\theta_{\mathbf{l}}=\frac{1}%
{2}\omega_{\mathbf{m}}^{\prime\mathbf{kl}}\theta_{\mathbf{kl}}\in
\sec\mathcal{C\ell(}M,g),\nonumber\\
\omega_{e_{\mathbf{n}}}^{{}}  &  =\frac{1}{2}\omega_{\mathbf{m}}^{\mathbf{kl}%
}\theta_{\mathbf{k}}\wedge\theta_{\mathbf{l}}=\frac{1}{2}\omega_{\mathbf{m}%
}^{\mathbf{kl}}\theta_{\mathbf{kl}}\in\sec\mathcal{C\ell(}M,g),\nonumber\\
U  &  =\exp F\text{, }F=\frac{1}{2}F^{\mathbf{rs}}\theta_{\mathbf{rs}}\in
\sec\mathcal{C\ell(}M,g). \label{ari4}%
\end{align}
Recall that
\begin{align}
\omega_{\mathbf{n}}^{\mathbf{rs}}  &  =\eta^{\mathbf{ra}}\omega_{\mathbf{anb}%
}\eta^{\mathbf{sb}}=\omega_{\mathbf{nb}}^{\mathbf{r}}\eta^{\mathbf{sb}%
},\nonumber\\
\omega_{\mathbf{nk}}^{\mathbf{r}}  &  =\omega_{\mathbf{n}}^{\mathbf{rs}}%
\eta_{\mathbf{sk}}. \label{ari5}%
\end{align}

Then, from Eqs.(\ref{ari3}), (\ref{ari4}) and (\ref{ari5}) we get%

\begin{equation}
\omega_{\mathbf{nk}}^{\prime\mathbf{r}}=\Lambda_{\mathbf{q}}^{\mathbf{b}%
}\omega_{\mathbf{mb}}^{\mathbf{p}}\Lambda_{\mathbf{p}}^{\mathbf{r}}%
\Lambda_{\mathbf{k}}^{\mathbf{m}}-\eta_{\mathbf{sk}}\Lambda_{\mathbf{k}%
}^{\mathbf{m}}\partial_{e_{\mathbf{m}}}\left(  F^{\mathbf{rs}}\right)
\label{ari6}%
\end{equation}

Now, we recall that the components of the torsion tensors $\mathbf{T}$ and
$\mathbf{T}^{\prime}$ of the connections $\nabla$ and $\nabla^{\prime}$ in the
orthonormal basis $\{\mathbf{e}_{\mathbf{r}}\otimes\theta^{\mathbf{n}}%
\wedge\theta^{\mathbf{k}}\}$ are given by
\begin{align}
T_{\mathbf{nk}}^{\mathbf{r}}  &  =\omega_{\mathbf{nk}}^{\mathbf{r}}%
-\omega_{\mathbf{kn}}^{\mathbf{r}}-c_{\mathbf{nk}}^{\mathbf{r}},\nonumber\\
T_{\mathbf{nk}}^{\prime\mathbf{r}}  &  =\omega_{\mathbf{nk}}^{\prime
\mathbf{r}}-\omega_{\mathbf{kn}}^{\prime\mathbf{r}}-c_{\mathbf{nk}%
}^{\mathbf{r}}, \label{ari7}%
\end{align}
where $[e_{\mathbf{n}},e_{\mathbf{k}}]=c_{\mathbf{nk}}^{\mathbf{r}%
}e_{\mathbf{r}}$.

Let us suppose that we start with a torsion free connection $\nabla$. This
means that $c_{\mathbf{nk}}^{\mathbf{r}}=\omega_{\mathbf{nk}}^{\mathbf{r}%
}-\omega_{\mathbf{kn}}^{\mathbf{r}}$. Then,%
\begin{equation}
T_{\mathbf{nk}}^{\prime\mathbf{r}}=\Lambda_{\mathbf{n}}^{\mathbf{b}}%
\Lambda_{\mathbf{k}}^{\mathbf{m}}\Lambda_{\mathbf{p}}^{\mathbf{r}%
}c_{\mathbf{mb}}^{\mathbf{p}}-c_{\mathbf{nk}}^{\mathbf{r}}-\partial
_{e_{\mathbf{m}}}\left(  F^{\mathbf{rs}}\right)  \left[  \eta_{\mathbf{sk}%
}\Lambda_{\mathbf{n}}^{\mathbf{m}}-\eta_{\mathbf{sn}}\Lambda_{\mathbf{k}%
}^{\mathbf{m}}\right]  , \label{ari8}%
\end{equation}
and we see that $\mathbf{T}^{\prime}=0$ only for very particular gauge transformations.

We then arrive at the conclusion that to suppose the Dirac-Hestenes Lagragian
is invariant under active rotational gauge transformations implies in an
equivalence between torsion free and non torsion free connections. It is
always emphasized that\ in a theory where besides $\psi$, also the the tetrad
fields $\theta^{\mathbf{a}}$ and the connection $\omega$ are dynamic
variables, the torsion is not zero, because its source is the spin of the
$\psi$ field. Well, this is true in particular gauges, because as showed above
it seems that it is always possible to find gauges where the torsion is null.
Analogous conclusions are valid for the curvature tensors of the \ `gauge
equivalent connections', as the reader may verify.

\bigskip\appendix

\section{Principal Bundles and Associated Vector Bundles}

We recall in this and the next Appendices some of the main definitions and
concepts of the theory of principal bundles and their associated vector
bundles, including the theory of connections and generalized $G$-connections
in principal and vector bundles, which we shall need in order to introduce the
Clifford and spin-Clifford bundles over a Lorentzian manifold. Propositions
are in general presented without proofs, which can be found, e.g., in
\cite{choquet,koni,palais}.

\begin{definition}
\label{fiberbundle}A fiber bundle over $M$ with Lie group $G$ is denoted by
$(E,M,\mathbf{\pi},G,F)$. $E$ is a topological space called the total space of
the bundle, $\mathbf{\pi}:E\rightarrow M$ is a continuous surjective map,
called the canonical projection and $F$ is the typical fiber. The following
conditions must be satisfied:

a) $\mathbf{\pi}^{-1}(x)$, the fiber over $x$, is homeomorphic to $F$.

b) Let $\{U_{i},$ $i\in\mathfrak{I}\}$, where $\mathfrak{I}$ is an index set,
be a covering of $M$, such that:

\begin{itemize}
\item Locally a fiber bundle $E$ is trivial, i.e., it is diffeomorphic to a
product bundle, i.e., $\mathbf{\pi}^{-1}(U_{i})\simeq U_{i}\times F$ for all
$i\in\mathfrak{I}$.

\item The diffeomorphisms $\Phi_{i}:\mathbf{\pi}^{-1}(U_{i})\rightarrow
U_{i}\times F$ have the form
\begin{align}
\Phi_{i}(p)  &  =(\mathbf{\pi}(p),\phi_{i,x}(p))\label{fb4.1}\\[0.03in]
\left.  \phi_{i}\right\vert _{\pi^{-1}(x)}  &  \equiv\phi_{i,x}:\mathbf{\pi
}^{-1}(x)\rightarrow F\text{ is onto}%
\end{align}
The collection $\{(U_{i},\Phi_{i})\}$, $i\in\mathfrak{I}$, are said to be a
family of local trivializations for $E$.

\item The group $G$ acts on the typical fiber. Let $x\in U_{i}\cap U_{j}$.
Then,
\begin{equation}
\phi_{j,x}\circ\phi_{i,x}^{-1}:F\rightarrow F \label{fb4.3}%
\end{equation}
must coincide with the action of an element of $G$ for all $x\in U_{i}\cap
U_{j}$ and $i,j\in\mathfrak{I}$.

\item We call transition functions of the bundle the continuous induced
mappings
\begin{equation}
g_{ij}:U_{i}\cap U_{j}\rightarrow G\text{, where }g_{ij}(x)=\phi_{i,x}%
\circ\phi_{j,x}^{-1}.
\end{equation}

\end{itemize}
\end{definition}

For consistence of the theory the transition functions must satisfy the
cocycle condition
\begin{equation}
g_{ij}(x)g_{jk}(x)=g_{ik}(x).
\end{equation}

\begin{definition}
\label{principalbundle}\noindent$(P,M,\mathbf{\pi},G,F\equiv G)\equiv
(P,M,\mathbf{\pi},G)$ is called a principal fiber bundle\emph{\ (PFB)} if all
conditions in \ref{fiberbundle}\textbf{\ }are fulfilled and moreover, if there
is a right action of $G$ on elements $p\in P$, such that:

a) the mapping (defining the right action) $P\times G\ni$ $(p,g)\mapsto pg\in
P$ is continuous.

b) given $g,g^{\prime}\in G$ and $\forall p\in P$, $(pg)g^{\prime
}=p(gg^{\prime}).$

c) $\forall x\in M,\mathbf{\pi}^{-1}(x)$ is invariant under the action of $G$,
i.e., each element of $p\in\mathbf{\pi}^{-1}(x)$ is mapped into $pg\in
\mathbf{\pi}^{-1}(x)$, i.e., it is mapped into an element of the same fiber.

d) $G$ acts free and transitively on each fiber $\mathbf{\pi}^{-1}(x)$, which
means that all elements within $\mathbf{\pi}^{-1}(x)$ are obtained by the
action of all the elements of $G$ on any given element of the fiber
$\mathbf{\pi}^{-1}(x)$. This condition is, of course, necessary for the
identification of the typical fiber with $G$.\bigskip
\end{definition}

\begin{definition}
A bundle $(E,M,\mathbf{\pi}_{1},G=Gl(m,\mathcal{F}{}),F=\mathbf{V})$, where
$\mathcal{F}=\mathbb{R}$ or $\mathbb{C}$ (respectively the real and complex
fields), ${}Gl(m,\mathcal{F}{})$ is the linear group, and $\mathbf{V}$ is an
$m$-dimensional vector space over $\mathcal{F}\ {}$ is called a vector bundle.
\end{definition}

\begin{definition}
\textbf{\label{vectorbundle}}A vector bundle $(E,M,\mathbf{\pi},G,F)$ denoted
$E=P\times_{\rho}F$ is said to be associated to a \emph{PFB} bundle
$(P,M,\mathbf{\pi},G)$ by the linear representation $\rho$ of $G$ in
$F=\mathbf{V}$ (a linear space of finite dimension over an appropriate
field${}$, which is called the \emph{carrier space} of the representation) if
its transition functions are the images under $\rho$ of the corresponding
transition functions of the \emph{PFB} $(P,M,\mathbf{\pi},G)$. This means the
following: consider the following local trivializations of $P$ and $E$
respectively
\begin{align}
\Phi_{i}  &  :\mathbf{\pi}^{-1}(U_{i})\rightarrow U_{i}\times G,\quad
\label{fb4.6n}\\[0.03in]
\Xi_{i}  &  :\mathbf{\pi}_{1}^{-1}(U_{i})\rightarrow U_{i}\times
F,\quad\label{fb4.6}\\[0.03in]
\Xi_{i}(q)  &  =(\mathbf{\pi}_{1}(q),\chi_{i}(q))=(x,\chi_{i}%
(q)),\label{fb4.7}\\[0.03in]
\left.  \chi_{i}\right\vert _{\mathbf{\pi}_{1}^{-1}(x)}  &  \equiv\chi
_{i,x}:\mathbf{\pi}_{1}^{-1}(x)\rightarrow F, \label{fb4.8}%
\end{align}
where $\mathbf{\pi}_{1}:P\times_{\rho}F\rightarrow M$ is the projection of the
bundle associated to $(P,M,\mathbf{\pi},G)$. Then, for all $x\in U_{i}\cap
U_{j}$, $i,j\in\mathfrak{I}$, we have
\begin{equation}
\chi_{j,x}\circ\chi_{i,x}^{-1}=\rho(\phi_{j,x}\circ\phi_{i,x}^{-1}).
\label{fb4.9}%
\end{equation}
In addition, the fibers $\mathbf{\pi}^{-1}(x)$ are vector spaces isomorphic to
the representation space $V$.
\end{definition}

\begin{definition}
Let $(E,M,\mathbf{\pi},G,F)$ be a fiber bundle and $U\subset M$ an open set. A
local section of the fiber bundle $(E,M,\mathbf{\pi},G,F)$ on $U$ is a
mapping
\begin{equation}
s:U\rightarrow E\quad\text{such that}\quad\pi\circ s=Id_{U},
\end{equation}
If $U=M$ we say that $s$ is a \emph{global section}.
\end{definition}

\begin{remark}
There is a relation between sections and local trivializations for principal
bundles. Indeed, each local section $s,$ (on $U_{i}\subset M$) for a principal
bundle $(P,M,\mathbf{\pi},G)$ determines a local trivialization $\Phi
_{i}:\mathbf{\pi}^{-1}(U)\rightarrow U\times G,$ of $P$ by setting
\begin{equation}
\Phi_{i}^{-1}(x,g)=s(x)g=pg=R_{g}p.
\end{equation}
Conversely, $\Phi_{i}$ determines $s$ since
\begin{equation}
s(x)=\Phi_{i}^{-1}(x,e)\text{.} \label{fb4.23}%
\end{equation}

\end{remark}

\begin{proposition}
A principal bundle is trivial if and only if it has a global cross section.
\end{proposition}

\begin{proposition}
A vector bundle is trivial if and only if its associated principal bundle is trivial.
\end{proposition}

\begin{proposition}
\label{admit cs}Any fiber bundle $(E,M,\mathbf{\pi},G,F)$ such that $M$ is a
paracompact manifold and the fiber $F$ is a vector space admits a cross section.
\end{proposition}

\begin{remark}
Then, any vector bundle associated to a trivial principal bundle has non-zero
global sections. Note however that a vector bundle may admit a non-zero global
section even if it is not trivial. Indeed, as shown in the main text, any
Clifford bundle possesses a global identity section, and some spin-Clifford
bundles admits also identity sections once a trivialization is given.
\end{remark}

\begin{definition}
We say that the structure group $G^{\prime}$ of a fiber bundle
$(E,M,\mathbf{\pi},G,F)$ is reducible to $G$ if the bundle admits an
equivalent structure defined with a subgroup $G^{\prime}\ $of the structure
group $G$. More precisely, this means that the fiber bundle admits a family of
local trivializations such that the transition functions takes values in
$G^{\prime}$, i.e., $g_{ij}:U_{i}\cap U_{j}\rightarrow G^{\prime}$.
\end{definition}

\subsection{Frame Bundle}

The tangent bundle $TM$ to a differentiable $n$-dimensional manifold $M$ is an
associated bundle to a principal bundle called the frame bundle $F(M)=\bigcup
\nolimits_{x\in M}F_{x}M$, where $F_{x}M$ is the set of frames at $x\in M$.
Let $\{x^{i}\}$ be the coordinate functions associated to a local chart
$(U_{i},\varphi_{i})$ of the maximal atlas of $M$. Then, $T_{x}M $ $\ $has a
natural basis $\{\left.  \frac{\partial}{\partial x^{i}}\right\vert _{x}\}$ on
$U_{i}\subset M$. \ 

\begin{definition}
A frame at $T_{x}M$ is a set $\Sigma_{x}=\{\left.  e_{1}\right\vert
_{x},...,\left.  e_{n}\right\vert _{x}\}$ of linearly independent vectors such
that%
\begin{equation}
\left.  e_{i}\right\vert _{x}=F_{i}^{j}\left.  \frac{\partial}{\partial x^{j}%
}\right\vert _{x}, \label{fr1}%
\end{equation}
and where the matrix $(F_{i}^{j})$ with entries $A_{i}^{j}\in\mathbb{R}$,
belongs to the the real general linear group in $n$ dimensions
$Gl(n,\mathbb{R)}$. We write $(F_{i}^{j})$ $\in Gl(n,\mathbb{R)}$.
\end{definition}

A local trivialization $\phi_{i}:\pi^{-1}(U_{i})\mathbb{\rightarrow}%
U_{i}\times Gl(n,\mathbb{R)}$ is defined by
\begin{equation}
\phi_{i}(f)=(x,\Sigma_{x})\text{, }\pi(f)=x. \label{fr2}%
\end{equation}

The action of \ $a=(a_{i}^{j})\in Gl(n,\mathbb{R)}$ on a frame\ $f$ $\in F(U)
$ is given by $(f,a)\rightarrow fa$, where the new frame $fa\in F(U)$ is
defined by $\phi_{i}(fa)=(x,\Sigma_{x}^{\prime})$, $\mathbf{\pi}(f)=x$, and
\begin{align}
\Sigma_{x}^{\prime}  &  =\{\left.  e_{1}^{\prime}\right\vert _{x},...,\left.
e_{n}^{\prime}\right\vert _{x}\},\nonumber\\
\left.  e_{i}^{\prime}\right\vert _{x}  &  =\left.  e_{j}\right\vert _{x}%
a_{i}^{j}. \label{fr3}%
\end{align}
Conversely, given frames $\Sigma_{x}$ and $\Sigma_{x}^{\prime}$ there exists
$a=(a_{i}^{j})\in Gl(n,\mathbb{R)}$ such that Eq.(\ref{fr3}) is satisfied,
which means that $Gl(n,\mathbb{R)}$ acts on $F(M)$ actively.

Let $\{x^{i}\}$ and $\{\bar{x}^{i}\}$ be the coordinate functions associated
to the local chart $(U_{i},\varphi_{i})$ $(U_{i},\varphi_{i})$ and of the
maximal atlas of $M$. If $x\in U_{i}\cap U_{j}$ we have
\begin{align}
\left.  e_{i}\right\vert _{x}  &  =F_{i}^{j}\left.  \frac{\partial}{\partial
x^{j}}\right\vert _{x}=\bar{F}_{i}^{j}\left.  \frac{\partial}{\partial\bar
{x}^{j}}\right\vert _{x},\nonumber\\
(F_{i}^{j}),(\bar{F}_{i}^{j})  &  \in Gl(n,\mathbb{R)}. \label{fr4}%
\end{align}
Since $F_{i}^{j}=\bar{F}_{k}^{j}\left.  \left(  \frac{\partial x^{k}}%
{\partial\bar{x}^{i}}\right)  \right\vert _{x}$ we have that the transition
functions are%
\begin{equation}
g_{i}^{k}(x)=\left.  \left(  \frac{\partial x^{k}}{\partial\bar{x}^{i}%
}\right)  \right\vert _{x}\in Gl(n,\mathbb{R)}. \label{fr5}%
\end{equation}

\begin{remark}
Given $U\subset M$ we shall denote by $\Sigma\in\sec F(U)$ \ a section of
$F(U)$ $\subset F(M)$. This means that given a local trivialization $\phi
:\pi^{-1}(U)\mathbb{\rightarrow}U\times Gl(n,\mathbb{R)}$, $\phi
(\Sigma)=(x,\Sigma_{x})$, $\pi(f)=x$. Sometimes, we also use the sloppy
notation $\{e_{i}\}\in\sec F(U)$ or even $\{e_{i}\}\in\sec F(M)$ when the
context is clear.
\end{remark}

\subsection{Orthonormal Frame Bundle}

Suppose that the manifold $M$ is equipped with a metric field $g\in\sec
T^{2,0}M$ of signature $(p,q)$, $p+q=n$. Then, we can introduce
\textit{orthonormal} frames in each $T_{x}U$. In this case we denote an
orthonormal frame by $\Sigma_{x}=\{\left.  \mathbf{e}_{1}\right\vert
_{x},...,\left.  \mathbf{e}_{n}\right\vert _{x}\}$ and
\begin{align}
\left.  \mathbf{e}_{i}\right\vert _{x}  &  =h_{i}^{j}\left.  \frac{\partial
}{\partial x^{j}}\right\vert _{x},\label{fr6}\\
\left.  g(\left.  \mathbf{e}_{i}\right\vert _{x},\left.  \mathbf{e}%
_{j}\right\vert _{x})\right\vert _{x}  &  =\mathrm{diag}(1,1,...,1,-1,...,-1)
\end{align}
with $(h_{i}^{j})\in\mathrm{O}(p,q\mathbb{)}$, the \ real orthogonal group in
$n$ dimensions. In this case we say that the frame bundle has been
reduced\textit{\ }to the\textit{\ orthonormal frame bundle}, which will be
denoted by $\mathbf{P}_{\mathrm{O}(n)}(M)$. A section $\Sigma\in\sec$
$\mathbf{P}_{\mathrm{O}(n)}(U)$ is called a vierbein.

\begin{remark}
The principal bundle of orthonormal frames $\mathbf{P}_{\mathrm{SO}_{1,3}^{e}%
}(M)$ over a Lorentzian manifold modelling spacetime and its covering bundle
called spin bundle $\mathbf{P}_{\mathrm{Spin}_{1,3}^{e}}(M)$ discussed in
Section 2 play an important role in this article. Also, vector bundles
associated to these bundles are very important. Associated to $\mathbf{P}%
_{\mathrm{SO}_{1,3}^{e}}(M)$ we have the tensor bundle, the exterior bundle
and the Clifford bundle. Associated to $\mathbf{P}_{\mathrm{Spin}_{1,3}^{e}%
}(M)$ we have several spinor bundles, in particular the spin-Clifford bundle,
residence of the Dirac-Hestenes spinor fields. All those bundles and their
relations are studied in details in \cite{moro} and briefly discussed in
Section 2.
\end{remark}

Given two vector bundles $(E,M,\mathbf{\pi},G,\mathbf{V})$ and $(E^{\prime
},M^{\prime},\mathbf{\pi}^{\prime},G^{\prime},\mathbf{V}^{\prime})$ we have

\begin{definition}
The product bundle $E\times E^{\prime}$ is a fiber bundle whose basis space is
$M\times M^{\prime}$, the typical fiber is $\mathbf{V\oplus V}^{\prime}$, the
structural group of $E\times E^{\prime}$acts separately as $G$ and $G^{\prime
}$ in each one of the components of $\mathbf{V\oplus V}^{\prime}$ and the
projection $\mathbf{\pi\times\pi}^{\prime}$is such that $E\times E^{\prime
}\overset{\mathbf{\pi\times\pi}^{\prime}}{\rightarrow}M\times M^{\prime}$.
\end{definition}

\begin{definition}
\label{whitney sum}Given two vector bundles $(E,M,\mathbf{\pi},G,\mathbf{V}) $
and $(E^{\prime},M,\mathbf{\pi}^{\prime},G^{\prime},\mathbf{V}^{\prime})$ over
the same basis space, the Whitney sum bundle $E\oplus E^{\prime}$ is the
pullback of $E\times E^{\prime}$ by $h:M\rightarrow M\times M$, $h(p)=(p,p).$
\end{definition}

\begin{definition}
\label{tensor prod bundle}Given two vector bundles $(E,M,\mathbf{\pi
},G,\mathbf{V})$ and $(E^{\prime},M,\mathbf{\pi}^{\prime},G^{\prime
},\mathbf{V}^{\prime})$ over the same basis space, the tensor product bundle
$E\otimes E^{\prime}$ is the bundle obtained from $E$ and $E^{\prime}$ by
assigning the tensor product of fibers $\mathbf{\pi}_{x}^{-1}\otimes
\mathbf{\pi}_{x}^{\prime-1}$ for all $x\in M$.
\end{definition}

\begin{remark}
With the above definitions we can easily show that given three vector bundles,
say, $E,E^{\prime},E^{\prime\prime}$ we have%
\begin{equation}
E\oplus(E^{\prime}\otimes E^{\prime\prime})=(E\otimes E^{\prime}%
)\oplus(E\otimes E^{\prime\prime}) \label{DIST LAW}%
\end{equation}

\end{remark}

\section{Equivalent Definitions of a Connection in Principal Bundles}

To define the concept of a \emph{connection} on a \emph{PFB} $(P,M,\mathbf{\pi
},G)$, we recall that since $\dim(M)=m$, if $\dim(G)=n$, then $\dim(P)=n+m$.
Obviously, for all $x\in M$, $\mathbf{\pi}^{-1}(x)$ is an $n$-dimensional
submanifold of $P$ diffeomorphic to the structure group $G$ and $\mathbf{\pi}$
is a submersion, $\mathbf{\pi}^{-1}(x)$ is a closed submanifold of $P$ for all
$x\in M$.

The tangent space $T_{p}P$, $p\in\mathbf{\pi}^{-1}(x)$, is an $(n+m)$%
-dimensional vector space and the tangent space $V_{p}P\equiv T_{p}%
(\mathbf{\pi}^{-1}(x))$ to the fiber over $x$ at the same point $p\in
\mathbf{\pi}^{-1}(x)$ is an $n$-dimensional linear subspace of $T_{p}P$ called
the \emph{vertical subspace} of $T_{p}P$\footnote{Here we may be tempted to
realize that as it is possible to construct the vertical space for all $p\in
P$ then we can define a horizontal space as the complement of this space in
respect to $T_{p}P$. Unfortunately this is not so, because we need a smoothly
association of a horizontal space in every point. This is possible only by
means of a connection.}.

Now, roughly speaking a connection on $P$ is a rule that makes possible a
\emph{correspondence} between any two fibers along a curve $\sigma
:\mathbb{R}\supseteq{}I\rightarrow M,t\mapsto\sigma(t)$. If $p_{0}$ belongs to
the fiber over the point $\sigma(t_{0})\in\sigma$, we say that $p_{0}$ is
parallel translated along $\sigma$ by means of this \emph{correspondence}.

\begin{definition}
A horizontal lift of $\sigma$ is a curve $\hat{\sigma}:\mathbb{R}\supseteq
I{}\rightarrow P$ (described by the parallel transport of $p$).
\end{definition}

It is intuitive that such a transport takes place in $P$ along directions
specified by vectors in $T_{p}P$, which do not lie within the vertical space
$V_{p}P$. Since the tangent vectors to the paths of the basic manifold passing
through a given $x\in M$ span the entire tangent space $T_{x}M$, the
corresponding vectors \textbf{$Y$}$_{p}\in T_{p}P$ (in whose direction
parallel transport can generally take place in $P$) span a four-dimensional
linear subspace of $T_{p}P$ called the \emph{horizontal space} of $T_{p}P$ and
denoted by $H_{p}P$. Now, the mathematical concept of a connection can be
presented. This is done through three equivalent definitions given below which
encode rigorously the intuitive discussion given above. We have,\medskip

\begin{definition}
\noindent\textbf{\ \label{c1}} A connection on a \emph{PFB} $(P,M,\mathbf{\pi
},G)$ is an assignment to each $p\in P$ of a subspace $H_{p}P\subset T_{p}P$,
called the horizontal subspace for that connection, such that $H_{p}P$ depends
smoothly on $p$ and the following conditions hold:
\end{definition}

(i) $\mathbf{\pi}_{\ast}:H_{p}P\rightarrow T_{x}M$, $x=\mathbf{\pi}(p),$ is an isomorphism.

(ii) $H_{p}P$ depends smoothly on $p$.

(iii) ($R_{g})_{*}H_{p}P=H_{pg}P,\forall g\in G,$ $\forall p\in P$.

Here we denote by $\mathbf{\pi}_{\ast}$ the \emph{differential} of the mapping
$\mathbf{\pi}$ and by $(R_{g})_{\ast}$ the differential of the mapping
$R_{g}:P\rightarrow P$ (the right action) defined by $R_{g}(p)=pg$.

Since $x=\mathbf{\pi}(\hat{\sigma}(t))$ for any curve in $P$ such that
$\hat{\sigma}(t)\in\mathbf{\pi}^{-1}(x)$ and $\hat{\sigma}(0)=p_{0}$, we
conclude that $\mathbf{\pi}_{\ast}$ maps all vertical vectors in the zero
vector in $T_{x}M$, i.e., $\mathbf{\pi}_{\ast}(V_{p}P)=0$ and we
have\footnote{We also write $TP=HP\oplus VP.$},
\begin{equation}
T_{p}P=H_{p}P\oplus V_{p}P. \label{fb4.11}%
\end{equation}

Then every \textbf{$Y$}$_{p}\in T_{p}P$ can be written as
\begin{equation}
\mathbf{Y}=\mathbf{Y}_{p}^{h}+\mathbf{Y}_{p}^{v},\quad\quad\mathbf{Y}_{p}%
^{h}\in H_{p}P,\quad\quad\mathbf{Y}_{p}^{v}\in V_{p}P. \label{fb4.12}%
\end{equation}
Therefore, given a vector field $Y$ over $M$ it is possible to lift it to a
horizontal vector field over $P$, i.e., $\mathbf{\pi}_{\ast}($\textbf{$Y$%
}$_{p})=\mathbf{\pi}_{\ast}($\textbf{$Y$}$_{p}^{h})=Y_{x}\in T_{x}M$ for all
$p\in P$ with $\mathbf{\pi}(p)=x$. In this case, we call \textbf{$Y$}$_{p}^{h}
$ \emph{horizontal lift} of $Y_{x}$. We say moreover that \textbf{$Y $} is a
horizontal vector field over $P$ if \textbf{$Y$}$^{h}=$\textbf{$Y$}.

\begin{definition}
\noindent\label{c2} A \emph{connection} on a \emph{PFB} $(P,M,\mathbf{\pi},G)
$ is a mapping $\Gamma_{p}:T_{x}M\rightarrow T_{p}P$, such that $\forall p\in
P$ and $x=\mathbf{\pi}(p)$ the following conditions hold:
\end{definition}

(i) $\Gamma_{p}$ is linear.

(ii) $\mathbf{\pi}_{*}\circ\Gamma_{p}=Id_{T_{x}M}.$

(iii) the mapping $p\mapsto$ $\Gamma_{p}$ is differentiable.

(iv) $\Gamma_{R_{g}p}=(R_{g})_{*}\Gamma_{p}$, for all $g\in G$.

We need also the concept of parallel transport. It is given by

\begin{definition}
\noindent Let $\sigma:{}\ni I\rightarrow M,$ $t\mapsto\sigma(t)$ with
$x_{0}=\sigma(0)\in M$, be a curve in $M$ and let $p_{0}\in P$ such that
$\mathbf{\pi}(p_{0})=x_{0}$. The parallel transport of $p_{0}$ along $\sigma$
is given by the curve $\hat{\sigma}:{}\ni I\rightarrow P,t\mapsto\hat{\sigma
}(t)$ defined by
\begin{equation}
\frac{d}{dt}\hat{\sigma}(t)=\Gamma_{p}(\frac{d}{dt}\sigma(t)), \label{fb4.13}%
\end{equation}
with $p_{0}=\hat{\sigma}(0)$ and $\hat{\sigma}(t)=p_{\parallel t}$,
$\mathbf{\pi}(p_{\parallel t})=x.\medskip$
\end{definition}

In order to present yet a \emph{third} definition of a connection we need to
know more about the nature of the vertical space $V_{p}P$. For this, let
\textbf{$Y$}$\in T_{e}G=\mathfrak{G}$ be an element of the Lie algebra
$\mathfrak{G}$ of $G$. The vector \textbf{$Y$} is the tangent to the curve
produced by the exponential map%

\begin{equation}
\mathbf{Y}=\left.  \frac{d}{dt}\left(  \exp(t\mathbf{Y})\right)  \right\vert
_{t=0}.\label{fb4.14}%
\end{equation}
Then, for every $p\in P$ we can attach to each \textbf{$Y$}$\in T_{e}%
G=\mathfrak{G}$ a unique element \textbf{$Y$}$_{p}^{v}\in V_{p}P$ as follows:
let  $f:(-\varepsilon,\varepsilon)\rightarrow P$ be a curve in $P$ and
$\mathfrak{F}:P\rightarrow\mathbb{R}{}$ a smooth function and consider the
function $\mathfrak{F\circ}f(t)=\mathfrak{F}(p\exp t$\textbf{$Y$}$)$. Then we
have
\begin{equation}
\mathbf{Y}_{p}^{v}(\mathfrak{F})=\left.  \frac{d}{dt}\mathfrak{F}\left(
p\exp(t\mathbf{Y})\right)  \right\vert _{t=0}.\label{fb4.15}%
\end{equation}

By this construction we attach to each \textbf{$Y$}$\in T_{e}G=\mathfrak{G}$ a
unique vector field over $P$, called the fundamental field corresponding to
this element. We then have the canonical isomorphism
\begin{equation}
\mathbf{Y}_{p}^{v}\longleftrightarrow\mathbf{Y},\quad\mathbf{Y}_{p}^{v}\in
V_{p}P,\quad\mathbf{Y}\in T_{e}G=\mathfrak{G} \label{fb4.16}%
\end{equation}
from which we get
\begin{equation}
V_{p}P\simeq\mathfrak{G}. \label{fb4.17}%
\end{equation}

\begin{definition}
\noindent\label{c3} A connection on a \emph{PFB} $(P,M,\mathbf{\pi},G)$ is a
1-form field $\mathbf{\omega}$ on $P$ with values in the Lie algebra
$\mathfrak{G}$ $=T_{e}G$ such that $\forall p\in P$ we have,
\end{definition}

(i) $%
\boldsymbol{\omega}%
_{p}($\textbf{$Y$}$_{p}^{v})=$\textbf{$Y$} and \textbf{$Y$}$_{p}%
^{v}\longleftrightarrow$\textbf{$Y$}, where \textbf{$Y$}$_{p}^{v}\in V_{p}P$
and \textbf{$Y$}$\in T_{e}G=\mathfrak{G}$.

(ii) $%
\boldsymbol{\omega}%
_{p}$ depends smoothly on $p$.

(iii) $%
\boldsymbol{\omega}%
_{p}[(R_{g})_{\ast}$\textbf{$Y$}$_{p}]=(\mathrm{Ad}_{g^{-1}}\mathbf{\omega
}_{p})($\textbf{$Y$}$_{p})$, where $\mathrm{Ad}_{g^{-1}}\mathbf{\omega}%
_{p}=g^{-1}\mathbf{\omega}_{p}g$.

It follows that if $\{\mathcal{G}_{a}\}$ is a basis of $\mathfrak{G}$ and
$\{\theta^{i}\}$ is a basis for $T^{*}P$ then%

\begin{equation}%
\boldsymbol{\omega}%
_{p}=\omega_{p}^{a}\otimes\mathcal{G}_{a}=\omega_{i}^{a}(p)\theta_{p}%
^{i}\otimes\mathcal{G}_{a}, \label{fb4.18}%
\end{equation}
where $\omega^{a}$ are 1-forms on $P$.

Then the horizontal spaces can be defined by%

\begin{equation}
H_{p}P=\ker(%
\boldsymbol{\omega}%
_{p}), \label{fb4.19}%
\end{equation}
which shows the equivalence between the definitions.\medskip

\subsection{ The Connection on the Base Manifold}

\begin{definition}
Let $U\subset M$ and
\end{definition}

\begin{equation}
s:U\rightarrow\mathbf{\pi}^{-1}(U)\subset P,\quad\quad\mathbf{\pi}\circ
s=Id_{U}, \label{fb4.20}%
\end{equation}
be a local section of the \emph{PFB} $(P,M,\mathbf{\pi},G)$.

\begin{definition}
Let $%
\boldsymbol{\omega}%
$ be a connection on $P$. The 1-form $s^{\ast}%
\boldsymbol{\omega}%
$ (the pullback of $%
\boldsymbol{\omega}%
$ under $s$) by
\begin{equation}
(s^{\ast}%
\boldsymbol{\omega}%
\mathbf{)}_{x}(Y_{x})=%
\boldsymbol{\omega}%
_{s(x)}(s_{\ast}Y_{x}),\text{ }Y_{x}\in T_{x}M,\text{ }s_{\ast}Y_{x}\in
T_{p}P,\quad p=s(x),
\end{equation}
is called the \emph{local gauge potential}.
\end{definition}

It is quite clear that $s^{\ast}%
\boldsymbol{\omega}%
\in\sec T^{\ast}U\otimes\mathfrak{G}$. This object differs from the
\emph{gauge field} used by physicists by numerical constants (with units).
Conversely we have the following

\begin{proposition}
\noindent Given $\omega\in\sec T^{\ast}U\otimes\mathfrak{G}$ and a
differentiable section of $\mathbf{\pi}^{-1}(U)\subset P$, $U\subset M$, there
exists one and only one connection $%
\boldsymbol{\omega}%
$ on $\mathbf{\pi}^{-1}(U)$ such that $s^{\ast}%
\boldsymbol{\omega}%
=\omega$.\medskip
\end{proposition}

Consider now
\begin{equation}%
\begin{array}
[c]{c}%
\omega\in T^{\ast}U\otimes\mathfrak{G},\quad\omega=(\Phi^{-1}(x,e))^{\ast}%
\boldsymbol{\omega}%
=s^{\ast}%
\boldsymbol{\omega}%
,\quad s(x)=\Phi^{-1}(x,e),\\[2ex]%
\omega^{\prime}\in T^{\ast}U^{\prime}\otimes\mathfrak{G},\quad\omega^{\prime
}=(\Phi^{\prime-1}(x,e))^{\ast}%
\boldsymbol{\omega}%
=s^{\prime\ast}%
\boldsymbol{\omega}%
,\quad s^{\prime}(x)=\Phi^{\prime-1}(x,e).
\end{array}
\label{fb4.24}%
\end{equation}
Then we can write, for each $p\in P$ ($\mathbf{\pi}(p)=x$), parameterized by
the local trivializations $\Phi$ and $\Phi^{\prime}$ respectively as $(x,g)$
and $(x,g^{\prime})$ with $x\in U\cap U^{\prime}$, that
\begin{equation}
\mathbf{\omega}_{p}=g^{-1}dg+g^{-1}\omega_{x}g=g^{\prime-1}dg^{\prime
}+g^{\prime-1}\omega_{x}^{\prime}g^{\prime}. \label{fb4.25}%
\end{equation}
Now, if
\begin{equation}
g^{\prime}=hg, \label{fb4.26}%
\end{equation}
we immediately get from Eq.(\ref{fb4.25}) that
\begin{equation}
\omega_{x}^{\prime}=hdh^{-1}+h\omega_{x}h^{-1}, \label{fb4.27}%
\end{equation}
which can be called the \emph{transformation law} for the gauge
fields.\bigskip

\section{Exterior Covariant Derivatives.}

Let $%
{\displaystyle\bigwedge^{k}}
(P,\mathfrak{G})=%
{\displaystyle\bigwedge^{k}}
(T^{\ast}P)\otimes\mathfrak{G},0\leq k\leq n$, be the set of all $k$-form
fields over $P$ with values in the Lie algebra $\mathfrak{G}$ of the gauge
group $G$ (and, of course, the connection $\boldsymbol{\omega}\in\sec%
{\displaystyle\bigwedge^{1}}
(P,\mathfrak{G})$)$.$

\begin{definition}
For each $\boldsymbol{\varphi}\in\sec%
{\displaystyle\bigwedge^{k}}
(P,\mathfrak{G})$ we define the so called \emph{horizontal form }%
$\boldsymbol{\varphi}^{h}\in\sec%
{\displaystyle\bigwedge^{k}}
(P,\mathfrak{G})$ by
\begin{equation}
\boldsymbol{\varphi}_{p}^{h}(\mathbf{X}_{1},\mathbf{X}_{2},...,\mathbf{X}%
_{k})=\boldsymbol{\varphi}(\mathbf{X}_{1}^{h},\mathbf{X}_{2}^{h}%
,...,\mathbf{X}_{k}^{h}),
\end{equation}
where $\mathbf{X}_{i}\in T_{p}P$, $i=1,2,..,k$.
\end{definition}

Notice that $\boldsymbol{\varphi}_{p}^{h}(\mathbf{X}_{1},\mathbf{X}%
_{2},...,\mathbf{X}_{k})=0$ if one (or more) of the $\mathbf{X}_{i}\in T_{p}P$
are vertical.

\begin{definition}
$\boldsymbol{\varphi}\in\sec%
{\displaystyle\bigwedge^{k}}
(T^{\ast}P)\otimes\mathbf{V}$ (where $\mathbf{V}$ is a vector space) is said
to be horizontal if $\boldsymbol{\varphi}_{p}(\mathbf{X}_{1},\mathbf{X}%
_{2},...,\mathbf{X}_{k})=0$, if at least one (say) $\mathbf{X}_{i}\in T_{p}P$
is vertical.
\end{definition}

\begin{definition}
$\boldsymbol{\varphi}\in\sec%
{\displaystyle\bigwedge^{k}}
(T^{\ast}P)\otimes\mathbf{V}$ is said to be of type $(\rho,\mathbf{V})$ if
\ $\forall g\in G$ we have
\begin{equation}
R_{g}^{\ast}\boldsymbol{\varphi}=\rho(g^{-1})\boldsymbol{\varphi}.
\label{excnew}%
\end{equation}

\end{definition}

\begin{definition}
Let $\boldsymbol{\varphi}\in\sec%
{\displaystyle\bigwedge^{k}}
(T^{\ast}P)\otimes\mathbf{V}$ be horizontal. Then, $\boldsymbol{\varphi}$ is
said to be tensorial of type $(\rho,\mathbf{V})$.
\end{definition}

\begin{definition}
The exterior covariant derivative of $\boldsymbol{\varphi}\in\sec%
{\displaystyle\bigwedge^{k}}
(P,\mathfrak{G})$ in relation to the connection $\mathbf{\omega}$ is
\begin{equation}
D^{%
\boldsymbol{\omega}%
}\boldsymbol{\varphi}=(d\boldsymbol{\varphi})^{h}\in\sec%
{\displaystyle\bigwedge\nolimits^{k+1}}
(P,\mathfrak{G}), \label{ecd4.29}%
\end{equation}
where $D^{%
\boldsymbol{\omega}%
}\boldsymbol{\varphi}_{p}(\mathbf{X}_{1},\mathbf{X}_{2},...,\mathbf{X}%
_{k},\mathbf{X}_{k+1})=d\boldsymbol{\varphi}_{p}(\mathbf{X}_{1}^{h}%
,\mathbf{X}_{2}^{h},...,\mathbf{X}_{k}^{h},\mathbf{X}_{k+1}^{h})$. Notice that
$d\boldsymbol{\varphi}=d\boldsymbol{\varphi}^{a}\otimes\mathcal{G}_{a}$ where
$\boldsymbol{\varphi}^{a}\in\sec%
{\displaystyle\bigwedge^{k}}
(P)$, $a=1,2,...,n$.\bigskip
\end{definition}

\begin{definition}
\label{commutator}The commutato\emph{r} of $\boldsymbol{\varphi}\in\sec%
{\displaystyle\bigwedge^{i}}
(P,\mathfrak{G})$ and $\boldsymbol{\psi}\in\sec%
{\displaystyle\bigwedge^{j}}
(P,\mathfrak{G})$, $0\leq i,j\leq n$ by
$[\boldsymbol{\varphi},\boldsymbol{\psi}]\in\sec%
{\displaystyle\bigwedge^{i+j}}
(P,\mathfrak{G})$ such that if $\mathbf{X}_{1},...,\mathbf{X}_{i+j}\in\sec
TP$, then
\begin{equation}
\lbrack\boldsymbol{\varphi},\boldsymbol{\psi}](\mathbf{X}_{1},...,\mathbf{X}%
_{i+j})=\frac{1}{i!j!}\sum_{\sigma\in\mathcal{S}_{n}}(-1)^{\sigma
}[\boldsymbol{\varphi}(\mathbf{X}_{\iota(1)},...,\mathbf{X}_{\iota
(i)}),\boldsymbol{\psi}(\mathbf{X}_{\iota(i+1)},...,\mathbf{X}_{\iota(i+j)})],
\label{ecd4.30}%
\end{equation}
where $\mathcal{S}_{n}$ is the permutation group of $n$ elements and
$(-1)^{\sigma}=\pm1$ is the sign of the permutation. The brackets $[$ $,$ $] $
in the second member of Eq.(\ref{ecd4.30}) are the Lie brackets in
$\mathfrak{G}$.
\end{definition}

By writing
\begin{equation}%
\boldsymbol{\varphi}%
=%
\boldsymbol{\varphi}%
^{a}\otimes\mathcal{G}_{a},\quad%
\boldsymbol{\psi}%
=%
\boldsymbol{\psi}%
^{a}\otimes\mathcal{G}_{a},\quad%
\boldsymbol{\varphi}%
^{a}\in\sec%
{\displaystyle\bigwedge\nolimits^{i}}
(T^{\ast}P),\quad%
\boldsymbol{\psi}%
^{a}\in\sec%
{\displaystyle\bigwedge\nolimits^{j}}
(T^{\ast}P), \label{'4.31n}%
\end{equation}
we can write
\begin{equation}%
\begin{array}
[c]{rcl}%
\lbrack%
\boldsymbol{\varphi}%
,%
\boldsymbol{\psi}%
] & = &
\boldsymbol{\varphi}%
^{a}\wedge%
\boldsymbol{\psi}%
^{b}\otimes\lbrack\mathcal{G}_{a},\mathcal{G}_{b}]\\[1ex]
& = & f_{ab}^{c}(%
\boldsymbol{\varphi}%
^{a}\wedge%
\boldsymbol{\psi}%
^{b})\otimes\mathcal{G}_{c}%
\end{array}
\label{ecd4.32n}%
\end{equation}
where $f_{ab}^{c}$ are the structure constants of the Lie algebra ${}$.

With Eq.(\ref{ecd4.32n}) we can prove \emph{easily} the following important
properties involving commutators:
\begin{equation}
\lbrack\boldsymbol{\varphi},\boldsymbol{\psi}]=(-)^{1+ij}%
[\boldsymbol{\psi},\boldsymbol{\varphi}], \label{ecd4.33n}%
\end{equation}%
\begin{equation}
(-1)^{ik}%
[[\boldsymbol{\varphi},\boldsymbol{\psi}],\boldsymbol{\tau}]+(-1)^{ji}%
[[\boldsymbol{\psi},\boldsymbol{\tau}],\boldsymbol{\varphi}]+(-1)^{kj}%
[[\boldsymbol{\tau},\boldsymbol{\varphi}],\boldsymbol{\psi}]=0, \label{4.34n}%
\end{equation}%
\begin{equation}
d[\boldsymbol{\varphi},\boldsymbol{\psi}]=[d\boldsymbol{\varphi},\boldsymbol{\psi}]+(-1)^{i}%
[\boldsymbol{\varphi},d\boldsymbol{\psi}]. \label{ecd4.35n}%
\end{equation}
for $\boldsymbol{\varphi}\in\sec%
{\displaystyle\bigwedge^{i}}
(P,\mathfrak{G})$, $\boldsymbol{\psi}\in\sec%
{\displaystyle\bigwedge^{j}}
(P,\mathfrak{G})$, $\boldsymbol{\tau}\in\sec%
{\displaystyle\bigwedge^{k}}
(P,\mathfrak{G})$.

We shall also need the following identity
\begin{equation}
\lbrack%
\boldsymbol{\omega}%
,%
\boldsymbol{\omega}%
](\mathbf{\mathbf{X}_{1},\mathbf{X}_{2})=}2[%
\boldsymbol{\omega}%
(\mathbf{X}_{1}),%
\boldsymbol{\omega}%
(\mathbf{X}_{2})]. \label{ecd4.36n}%
\end{equation}
The proof of Eq.(\ref{ecd4.36n}) is given as follows:\newline(i) Recall that
\begin{equation}
\lbrack%
\boldsymbol{\omega}%
\mathbf{,}%
\boldsymbol{\omega}%
]=(%
\boldsymbol{\omega}%
^{a}\wedge%
\boldsymbol{\omega}%
^{b})\otimes\lbrack\mathcal{G}_{a},\mathcal{G}_{b}]. \label{ecd4.37n}%
\end{equation}
(ii) Let $\mathbf{X}_{1},\mathbf{X}_{2}\in\sec TP$ (i.e., $\mathbf{X}_{1}$ and
$\mathbf{X}_{2}$ are vector fields on $P$). Then,
\begin{equation}%
\begin{array}
[c]{rcl}%
\lbrack%
\boldsymbol{\omega}%
,%
\boldsymbol{\omega}%
](\mathbf{\mathbf{X}_{1},\mathbf{X}_{2})} & = & (%
\boldsymbol{\omega}%
^{a}({\mathbf{X}_{1}})%
\boldsymbol{\omega}%
^{b}(\mathbf{\mathbf{X}_{2}})-%
\boldsymbol{\omega}%
^{a}(\mathbf{\mathbf{X}_{2}})%
\boldsymbol{\omega}%
^{b}(\mathbf{\mathbf{X}_{1}}))[\mathcal{G}_{a},\mathcal{G}_{b}]\\[1ex]
& = & 2[%
\boldsymbol{\omega}%
(\mathbf{X}_{1}),%
\boldsymbol{\omega}%
(\mathbf{X}_{2})].
\end{array}
\label{ecd4.38n}%
\end{equation}

\begin{definition}
The curvature form of the connection $%
\boldsymbol{\omega}%
\in\sec%
{\displaystyle\bigwedge^{1}}
(P,\mathfrak{G})$ is $%
\boldsymbol{\Omega}%
^{%
\boldsymbol{\omega}%
}\in\sec%
{\displaystyle\bigwedge^{2}}
(P,\mathfrak{G})$ defined by
\begin{equation}%
\boldsymbol{\Omega}%
^{%
\boldsymbol{\omega}%
}=D^{%
\boldsymbol{\omega}%
}%
\boldsymbol{\omega}%
\mathbf{.} \label{ecd4.39n}%
\end{equation}

\end{definition}

\begin{definition}
The connection $%
\boldsymbol{\omega}%
$ is said to be flat if $%
\boldsymbol{\Omega}%
^{%
\boldsymbol{\omega}%
}=0$.
\end{definition}

\begin{proposition}
\noindent\
\begin{equation}
D^{%
\boldsymbol{\omega}%
}%
\boldsymbol{\omega}%
\mathbf{(\mathbf{X}_{1},\mathbf{X}_{2})}=d%
\boldsymbol{\omega}%
\mathbf{(X}_{1},\mathbf{X}_{2})+[%
\boldsymbol{\omega}%
(\mathbf{X}_{1}),%
\boldsymbol{\omega}%
(\mathbf{X}_{2})]. \label{ecd4.40n}%
\end{equation}

\end{proposition}

Eq.(\ref{ecd4.40n}) can be written using Eq.(\ref{ecd4.38n}) (and recalling
that $\mathbf{\omega}(\mathbf{X})=\omega^{a}(\mathbf{X})\mathcal{G}_{a}$).
Thus we have
\begin{equation}%
\boldsymbol{\Omega}%
^{%
\boldsymbol{\omega}%
}=D^{%
\boldsymbol{\omega}%
}%
\boldsymbol{\omega}%
=d%
\boldsymbol{\omega}%
+\frac{1}{2}[%
\boldsymbol{\omega}%
,%
\boldsymbol{\omega}%
]. \label{ecd4.41n}%
\end{equation}

\begin{proposition}
\emph{(Bianchi identity):}
\begin{equation}
D%
\boldsymbol{\Omega}%
^{%
\boldsymbol{\omega}%
}=0. \label{ecd4.42n}%
\end{equation}

\end{proposition}

\begin{proof}
\noindent\ (i) Let us calculate $d%
\boldsymbol{\Omega}%
^{%
\boldsymbol{\omega}%
}$. We have,
\begin{equation}
d%
\boldsymbol{\Omega}%
^{%
\boldsymbol{\omega}%
}=d\left(  d%
\boldsymbol{\omega}%
+\frac{1}{2}[%
\boldsymbol{\omega}%
,%
\boldsymbol{\omega}%
]\right)  . \label{ecd4.43n}%
\end{equation}
We now take into account that $d^{2}\boldsymbol{\omega}=0$ and that from the
properties of the commutators given by Eqs.(\ref{ecd4.33n}), (\ref{ecd4.43n})
and (\ref{ecd4.35n}) above, we have
\begin{align}
d[%
\boldsymbol{\omega}%
,%
\boldsymbol{\omega}%
]  &  =[d%
\boldsymbol{\omega}%
,%
\boldsymbol{\omega}%
]-[%
\boldsymbol{\omega}%
,d%
\boldsymbol{\omega}%
],\nonumber\\
\lbrack d%
\boldsymbol{\omega}%
,%
\boldsymbol{\omega}%
]  &  =-[%
\boldsymbol{\omega}%
,d%
\boldsymbol{\omega}%
],\nonumber\\
\lbrack\lbrack%
\boldsymbol{\omega}%
,%
\boldsymbol{\omega}%
],%
\boldsymbol{\omega}%
]  &  =0. \label{ecd4.44n}%
\end{align}
Using Eq.(\ref{ecd4.44n}) in Eq.(\ref{ecd4.43n}) gives
\begin{equation}
d%
\boldsymbol{\Omega}%
^{%
\boldsymbol{\omega}%
}=[d%
\boldsymbol{\omega}%
,%
\boldsymbol{\omega}%
]. \label{ecd4.45n}%
\end{equation}

(ii) In Eq.(\ref{ecd4.45n}) use Eq.(\ref{ecd4.41n}) and the last equation in
Eq.(\ref{ecd4.44n}) to obtain
\begin{equation}
d%
\boldsymbol{\Omega}%
^{%
\boldsymbol{\omega}%
}=[%
\boldsymbol{\Omega}%
^{%
\boldsymbol{\omega}%
},%
\boldsymbol{\omega}%
]. \label{ecd4.46n}%
\end{equation}

(iii) Use now the definition of the exterior covariant derivative
[Eq.(\ref{ecd4.30})] together with the fact that $%
\boldsymbol{\omega}%
\mathbf{(X^{h})}=0$, for all $\mathbf{X}\in T_{p}P$ to obtain
\[
D^{%
\boldsymbol{\omega}%
}%
\boldsymbol{\Omega}%
^{%
\boldsymbol{\omega}%
}=0.
\]
We can then write the very important formula (known as the Bianchi identity),
\begin{equation}
D^{%
\boldsymbol{\omega}%
}%
\boldsymbol{\Omega}%
^{%
\boldsymbol{\omega}%
}=d%
\boldsymbol{\Omega}%
^{%
\boldsymbol{\omega}%
}+[%
\boldsymbol{\omega}%
,%
\boldsymbol{\Omega}%
^{%
\boldsymbol{\omega}%
}]=0. \label{4.46nn}%
\end{equation}

\end{proof}

\subsection{ Local curvature in the Base Manifold $M$}

Let $(U,\Phi)$ be a local trivialization of $\mathbf{\pi}^{-1}(U)$ and $s$ the
associated cross section as defined above. Then, $s^{\ast}%
\boldsymbol{\Omega}%
^{%
\boldsymbol{\omega}%
}:=\Omega^{\omega}$ (the pullback of $%
\boldsymbol{\Omega}%
^{%
\boldsymbol{\omega}%
}$) is a well defined 2-form field on $U$ which takes values in the Lie
algebra $\mathfrak{G}.$ Let $\omega=s^{\ast}%
\boldsymbol{\omega}%
$ (see eq.(\ref{fb4.24})). If we recall that the differential operator $d$
commutes with the pullback, we immediately get
\begin{equation}
\Omega^{\omega}=s^{\ast}D^{%
\boldsymbol{\omega}%
}%
\boldsymbol{\omega}%
=d\omega+\frac{1}{2}\left[  \omega\mathbf{,}\omega\right]  . \label{ecd4.47n}%
\end{equation}
It is convenient to define the \textit{symbols}
\begin{align}
\mathbf{D}\omega &  :=s^{\ast}D^{%
\boldsymbol{\omega}%
}%
\boldsymbol{\omega}%
,\label{ecas}\\
\mathbf{D}\Omega^{\omega}  &  :=s^{\ast}D^{%
\boldsymbol{\omega}%
}%
\boldsymbol{\Omega}%
^{%
\boldsymbol{\omega}%
}%
\end{align}
and to write
\begin{equation}%
\begin{array}
[c]{rcl}%
\mathbf{D}\Omega^{\omega} & = & 0,\\[1ex]%
\mathbf{D}\Omega^{\omega} & = & d\Omega^{\omega}+\left[  \omega\mathbf{,}%
\Omega^{\omega}\right]  =0.
\end{array}
\label{base curv}%
\end{equation}
Eq.(\ref{base curv}) is also known as Bianchi identity.\medskip

\begin{remark}
\noindent\ In gauge theories (Yang-Mills theories) $\Omega^{\omega}$ is
(except for numerical factors with physical units) called a \emph{field
strength in the gauge} $\Phi$.\medskip
\end{remark}

\begin{remark}
\noindent\ When $G$ is a matrix group, as is the case in the presentation of
gauge theories by physicists, Definition \ref{commutator} of the commutator
$[\boldsymbol{\varphi},\boldsymbol{\psi}]\in\sec%
{\displaystyle\bigwedge^{i+j}}
(P,\mathfrak{G})$ ($\boldsymbol{\varphi}\in\sec%
{\displaystyle\bigwedge^{i}}
(P,\mathfrak{G}) $, $\boldsymbol{\psi}\in\sec%
{\displaystyle\bigwedge^{j}}
(P,\mathfrak{G})$) gives
\begin{equation}
\lbrack\boldsymbol{\varphi},\boldsymbol{\psi}]=\boldsymbol{\varphi}\wedge
\boldsymbol{\psi}-(-1)^{ij}\boldsymbol{\psi}\wedge\boldsymbol{\varphi},
\label{ecd4.49n}%
\end{equation}
where $\boldsymbol{\varphi}$ and $\boldsymbol{\psi}$ are considered as
matrices of forms with values in ${}\mathfrak{G}$ and
$\boldsymbol{\varphi}\wedge\boldsymbol{\psi}$ stands for the usual matrix
multiplication. Then, when $G$ is a matrix group, we can write
Eqs.(\ref{ecd4.41n}) and (\ref{ecd4.47n}) as
\begin{align}%
\boldsymbol{\Omega}%
^{%
\boldsymbol{\omega}%
}  &  =D^{%
\boldsymbol{\omega}%
}%
\boldsymbol{\omega}%
=d%
\boldsymbol{\omega}%
+%
\boldsymbol{\omega}%
\wedge%
\boldsymbol{\omega}%
\mathbf{,}\label{ecd4.50n}\\[0.01in]
\Omega^{\omega}  &  :=\mathbf{D}\omega=d\omega\mathbf{+}\omega\mathbf{\wedge
}\omega\mathbf{.} \label{ecd4.51n}%
\end{align}

\end{remark}

\subsection{ Transformation of the Field Strengths Under a Change of Gauge}

Consider two local trivializations $(U,\Phi)$ and $(U^{\prime},\Phi^{\prime})$
of $P$ such that $p\in\mathbf{\pi}^{-1}(U\cap U^{\prime})$ has $(x,g)$ and
$(x,g^{\prime})$ as images in $(U\cap U^{\prime})\times G$, where $x\in U\cap
U^{\prime}$. Let $s,s^{\prime}$ be the associated cross sections to $\Phi$ and
$\Phi^{\prime}$ respectively. By writing $s^{\prime\ast}%
\boldsymbol{\Omega}%
^{%
\boldsymbol{\omega}%
}=\Omega^{\omega\prime}$, we have the following relation for the local
curvature in the two different gauges such that $g^{\prime}=hg:$
\begin{equation}
\Omega^{\omega\prime}=h\Omega^{\omega}h^{-1},\quad\forall x\in U\cap
U^{\prime}. \label{ecd4.52n}%
\end{equation}

\bigskip

We now give the \emph{coordinate expressions} for the potential and field
strengths in the trivialization $\Phi$. Let $\langle x^{\mu}\rangle$ be a
local chart for $U\subset M$ and let $\left\{  \partial_{\mu}=\frac{\partial
}{\partial x^{\mu}}\right\}  $ and $\{dx^{\mu}\}$, $\mu=0,1,2,3$, be (dual)
bases of $TU$ and $T^{\ast}U$ respectively. Then,
\begin{align}
\omega &  =\omega^{a}\otimes\mathcal{G}_{a}=\omega_{\mu}^{a}dx^{\mu}%
\otimes\mathcal{G}_{a},\label{ecd4.53n}\\
\Omega^{\omega}  &  =(\Omega^{\omega})^{a}\otimes\mathcal{G}_{a}=\frac{1}%
{2}\Omega_{\mu\nu}^{a}dx^{\mu}\wedge dx^{\nu}\otimes\mathcal{G}_{a}.
\label{ecd4.54n}%
\end{align}
where $\omega_{\mu}^{a}$, $\Omega_{\mu\nu}^{a}:M\supset U\rightarrow
\mathbb{R}$ (or $\mathbb{C}$) and we get
\begin{equation}
\Omega_{\mu\nu}^{a}=\partial_{\mu}\omega_{\nu}^{a}-\partial_{\nu}\omega_{\mu
}^{a}+f_{bc}^{a}\omega_{\mu}^{b}\omega_{\nu}^{c}. \label{ecd4.55n}%
\end{equation}

The following objects appear frequently in the presentation of gauge theories
by physicists.
\begin{align}
(\Omega^{\omega})^{a}  &  =\frac{1}{2}\Omega_{\mu\nu}^{a}dx^{\mu}\wedge
dx^{\nu}=d{\omega}^{a}+\frac{1}{2}f_{bc}^{a}\omega^{b}\wedge\omega
^{c},\label{ecd4.56}\\
\Omega_{\mu\nu}^{\omega}  &  =\Omega_{\mu\nu}^{a}\mathcal{G}_{a}=\partial
_{\mu}\omega_{\nu}-\partial_{\nu}\omega_{\mu}+[\omega_{\mu},\omega_{\nu
}],\label{ecd4.57}\\
\omega_{\mu}  &  =\omega_{\mu}^{a}\mathcal{G}_{a}. \label{ecd4.58}%
\end{align}

We now give the local expression of Bianchi identity. Using
Eqs.(\ref{base curv}) and (\ref{ecd4.56}) we have
\begin{equation}
\mathbf{D}\Omega^{^{\omega}}:=\frac{1}{2}(\mathbf{D}\Omega^{^{\omega}}%
)_{\rho\mu\nu}dx^{\rho}\wedge dx^{\mu}\wedge dx^{\nu}=0. \label{ecd4.59}%
\end{equation}
By putting
\begin{equation}
(\mathbf{D}\Omega^{^{\omega}})_{\rho\mu\nu}:=\mathbf{D}_{\rho}\Omega_{\mu\nu
}^{^{\omega}} \label{ecd4.61}%
\end{equation}
we have
\begin{equation}
\mathbf{D}_{\rho}\Omega_{\mu\nu}^{^{\omega}}=\partial_{\rho}\Omega_{\mu\nu
}^{^{\omega}}+[\omega_{\rho},\Omega_{\mu\nu}^{^{\omega}}],
\end{equation}
and
\begin{equation}
\mathbf{D}_{\rho}\Omega_{\mu\nu}^{^{\omega}}+\mathbf{D}_{\mu}\Omega_{\nu\rho
}^{\omega}+\mathbf{D}_{\nu}\Omega_{\rho\mu}^{^{\omega}}=0. \label{ecd4.64}%
\end{equation}

Physicists call the operator
\begin{equation}
\mathbf{D}_{\rho}:=\partial_{\rho}+[\omega_{\rho},\;]. \label{4.65}%
\end{equation}
\textbf{\ }the \emph{covariant derivative}. The reason for this name will be
given below.

\subsection{Induced Connections\label{ind connection}}

Let $(P_{1},M_{1},\mathbf{\pi}_{1},G_{1})$ and $(P_{2},M_{2},\mathbf{\pi}%
_{2},G_{2})$ be two principal bundles and let $\mathcal{F}:P_{1}\rightarrow
P_{2}$ be a bundle homomorphism, i.e., $f$ is fiber preserving, induces a
diffeomorphism $f:M_{1}\rightarrow M_{2}$ and there exists a homomorphism
$\lambda:G_{1}\rightarrow G_{2}$ such that for $g_{1}\in G_{1}$, $p_{1}\in
P_{1}$ we have
\begin{equation}
\mathcal{F}(p_{1}g_{1})=R_{\lambda(g_{1})}\mathcal{F}(p_{1}) \label{ic1}%
\end{equation}

\begin{proposition}
Let $\mathcal{F}:P_{1}\rightarrow P_{2}$ be a bundle homomorphism. Then a
connection $%
\boldsymbol{\omega}%
_{1}$ on $P_{1}$determines a unique connection on $P_{2}.$
\end{proposition}

\begin{remark}
Let $(P,M,\mathbf{\pi}^{\prime},\mathrm{O}(p,q))=P_{\mathrm{O}(n)}(M)$ be the
orthonormal frame bundle, which is as explained above a reduction of the frame
bundle $F(M)$. Then, a connection on \ $P_{\mathrm{O}(n)}(M)$ determines a
unique connection on $F(M)$. This is a very important result that is usually
used implicitly.

\begin{proposition}
Let $F(M)$ be the frame bundle of a paracompact manifold $M$. Then, $F(M)$ can
be reduced to a principal bundle with structure group $\mathrm{O}(n)$, and to
each reduction there corresponds a Riemann metric field on $M.$
\end{proposition}
\end{remark}

\begin{remark}
If $M$ has dimension $4$, and we substitute $\mathrm{O}(n)\mapsto
\mathrm{SO}_{1,3}^{e}$ then to each reduction of $F(M)$ there corresponds \ a
Lorentzian metric field on $M.$
\end{remark}

\subsection{Linear Connections on a Manifold $M$}

\begin{definition}
\ A linear connection on a smooth manifold $M$ is a connection $%
\boldsymbol{\omega}%
\in\sec T^{\ast}F(M)\otimes gl(n,\mathbb{R)}$.
\end{definition}

\begin{remark}
Given a Riemannian (Lorentzian) manifold $(M,\mathtt{g})$ a connection on
$F(M)$ which is determined by a connection on the orthonormal frame bundle
$P_{\mathrm{O}(n)}(M)$ \emph{(}$P_{\mathrm{SO}_{1,3}^{e}}(M)$\emph{)} is
called a metric connection. After introducing the concept of covariant
derivatives on vector bundles, we can show that the covariant derivative of
the metric tensor with respect to a metric connection is null.
\end{remark}

Consider the mapping $\left.  f\right\vert _{p}:T_{x}M\rightarrow
\mathbb{R}^{n}$ (with $p=(x,\Sigma_{x})$ in a given trivialization) which
sends $\mathbf{v}\in T_{x}M$ into its components relative to the frame
$\Sigma_{x}=\{\left.  e_{1}\right\vert _{x},...,\left.  e_{n}\right\vert
_{x}\}$. Let $\{\left.  \theta^{j}\right\vert _{x}\}$ be the dual basis of
$\{\left.  e_{i}\right\vert _{x}\}$. Then,%
\begin{equation}
\left.  f\right\vert _{p}(\mathbf{v)=(}\left.  \theta^{j}\right\vert
_{x}(\mathbf{v))} \label{lc1}%
\end{equation}

\begin{definition}
The canonical soldering form of $M$ is the $1$-form $%
\boldsymbol{\theta}%
\in\sec T^{\ast}F(M)\otimes\mathbb{R}^{n}$ such that for any $v\in\sec
T_{p}F(M)$ such that $\mathbf{v=}\pi_{\ast}v$ we have%
\begin{align}
(%
\boldsymbol{\theta}%
\mathbf{(}v\mathbf{))}  &  \mathbf{:}=\left.
\boldsymbol{\theta}%
^{a}\right\vert _{p}(v\mathbf{)E}_{a}\nonumber\\
&  =\left.  \theta^{a}\right\vert _{x}(\mathbf{v)E}_{a}\mathbf{,} \label{sold}%
\end{align}
where $\{\mathbf{E}_{a}\}$ is the canonical basis of $\mathbb{R}^{n}$ and $\{%
\boldsymbol{\theta}%
^{a}\}$ is a basis of $T^{\ast}F(M)$, with $%
\boldsymbol{\theta}%
^{a}=\pi^{\ast}\theta^{a}$ \textbf{, }$\left.
\boldsymbol{\theta}%
^{a}\right\vert _{p}(v\mathbf{)=}\left.  \theta^{a}\right\vert _{x}%
(\mathbf{v)}$.
\end{definition}

\begin{definition}
The torsion form of a linear connection $%
\boldsymbol{\omega}%
\in\sec T^{\ast}F(M)\otimes gl(n,\mathbb{R)}$ is the $2$-form%
\[
D%
\boldsymbol{\theta}%
=%
\boldsymbol{\Theta}%
\in\sec\bigwedge\nolimits^{2}T^{\ast}F(M)\otimes\mathbb{R}^{n}.
\]

\end{definition}

As it is easy to verify, the soldering form $%
\boldsymbol{\theta}%
$ and the torsion $2$-form $%
\boldsymbol{\Theta}%
$ are tensorial of type $(\rho,\mathbb{R}^{n})$, where $\rho(g)=g$, $g\in
Gl(n,\mathbb{R)}$.

We can show, using the same techniques used in the calculation of $D^{%
\boldsymbol{\omega}%
}%
\boldsymbol{\omega}%
\mathbf{(\mathbf{X}_{1},\mathbf{X}_{2})}$ (Eq.(\ref{ecd4.40n})) that
\begin{equation}%
\boldsymbol{\Theta}%
=d%
\boldsymbol{\theta}%
+\mathbf{[}%
\boldsymbol{\omega}%
\mathbf{,}%
\boldsymbol{\theta}%
\mathbf{],} \label{lc2}%
\end{equation}
where $[$ , $]$ is the commutator product in the Lie algebra of the
\textit{affine} group $A(n,\mathbb{R})=Gl(n,\mathbb{R})\boxtimes\mathbb{R}%
^{n}$, where $\boxtimes$ means the \textit{semi-direct }product. Suppose that
$(\mathbf{E}_{a}^{b},\mathbf{E}_{c})$ is the canonical basis of
$a(n,\mathbb{R})$, the Lie algebra of $A(n,\mathbb{R})$. Recalling that%
\begin{align}%
\boldsymbol{\omega}%
(v\mathbf{)}  &  =%
\boldsymbol{\omega}%
_{b}^{a}(v)\mathbf{E}_{a}^{b},\label{lc3}\\%
\boldsymbol{\theta}%
\mathbf{(}v\mathbf{)}  &  =%
\boldsymbol{\theta}%
^{a}(v)\mathbf{E}_{a}, \label{lc4}%
\end{align}
we can show without difficulties that%
\begin{equation}
D^{%
\boldsymbol{\omega}%
}%
\boldsymbol{\Theta}%
=\mathbf{[}%
\boldsymbol{\Omega}%
\mathbf{,}%
\boldsymbol{\theta}%
\mathbf{]} \label{lc5}%
\end{equation}

\subsection{Torsion and Curvature on $M$}

Let $\{x^{i}\}$ be the coordinate functions associated to a local chart
$(U,\varphi)$ of the maximal atlas of $M$. Let $\Sigma\in\sec F(U)$ with
$e_{i}=F_{i}^{j}\frac{\partial}{\partial x^{j}}$ and $%
\boldsymbol{\theta}%
=%
\boldsymbol{\theta}%
^{a}\mathbf{E}_{a}$. Take $\mathbf{\pi}_{\ast}v=\mathbf{v}$. Then%
\begin{align}
(%
\boldsymbol{\theta}%
_{p}(v))  &  =\left.  f\right\vert _{p}(\mathbf{v)=}\left.  f\right\vert
_{p}\mathbf{(}dx^{j}(\mathbf{v)\partial}_{j}\mathbf{)=}\left.  f\right\vert
_{p}\mathbf{(}dx^{j}(\mathbf{v)(}F_{j}^{k}\mathbf{)}^{-1}\mathbf{e}%
_{k})\nonumber\\
&  =(\mathbf{(}F_{j}^{k}\mathbf{)}^{-1}dx^{j}(\pi_{\ast}v\mathbf{)).}
\label{lc6}%
\end{align}

With this result it is quite obvious that given any $\mathbf{w}\in
\mathbb{R}^{n}$, $%
\boldsymbol{\theta}%
$ determines a horizontal field $v_{\mathbf{w}}$ $\in\sec TF(M)$ by
\begin{equation}%
\boldsymbol{\theta}%
(v_{\mathbf{w}}(p))=\mathbf{w.} \label{lc7}%
\end{equation}

With these preliminaries we have the

\begin{proposition}
There is a bijective correspondence between sections of $T^{\ast}M\otimes
T_{s}^{r}M$ and sections of $T^{\ast}F(M)\otimes\mathbb{R}^{n_{q}}$, the space
of tensorial forms of the type $(\rho,\mathbb{R}^{n_{q}})$ in $F(M)$, with
$\rho$ and $q$ being determined \ by $T_{s}^{r}M$ .
\end{proposition}

Using the above proposition and recalling that the soldering form is tensorial
of type $(\rho(g),\mathbb{R}^{n})$, $\rho(g)=g,$ we see that it determines on
$M$ a vector valued differential $1$-form $\theta=e_{a}\otimes dx^{a}\in\sec
TM\otimes\bigwedge\nolimits^{1}(T^{\ast}M)$. Also, the torsion $%
\boldsymbol{\Theta}%
$ is tensorial of type $(\rho(g),\mathbb{R}^{n})$, $\rho(g)=g$ and thus define
a vector valued $2$-form on $M$, $\Theta=e_{a}\otimes\Theta^{a}\in\sec
TM\otimes\bigwedge\nolimits^{2}(T^{\ast}M)$. We can show from Eq.(\ref{lc2})
that given $u,w\in T_{p}F(M)$,%
\begin{equation}
\Theta^{a}(\mathbf{\pi}_{\ast}u,\mathbf{\pi}_{\ast}w)=d\theta^{a}(\mathbf{\pi
}_{\ast}u,\pi_{\ast}w)+\omega_{b}^{a}(\mathbf{\pi}_{\ast}u)\theta
^{b}(\mathbf{\pi}_{\ast}w)-\omega_{b}^{a}(\mathbf{\pi}_{\ast}w)\theta
^{b}(\mathbf{\pi}_{\ast}u). \label{lc8}%
\end{equation}

On the basis manifold this equation is often written:%
\begin{align}
\Theta &  =\mathbf{D}\theta=e_{a}\otimes(\mathbf{D}\theta^{a})\nonumber\\
&  =e_{a}\otimes(d\theta^{a}+\omega_{b}^{a}\wedge\theta^{b}), \label{lc10}%
\end{align}
where we recognize $\mathbf{D}\theta^{a}$ as the exterior covariant derivative
of index forms\footnote{Rigorously speaking, if Eq.(\ref{lc10}) is to agree
with Eq.(\ref{lc8}) we must have $\mathbf{\bar{\omega}}_{b}^{a}\wedge
\mathbf{\bar{\theta}}^{b}=\left(  \mathbf{\bar{\omega}}_{b}^{a}\otimes
\mathbf{\bar{\theta}}^{b}-\mathbf{\bar{\theta}}^{b}\otimes\mathbf{\bar{\omega
}}_{b}^{a}\right)  $. This is not the definition of the exterior product we
used in Section 2 because a factor $1/2$ is missing. However, this causes no
troubles in the calculations we did using the Clifford bundle formalism.}.

Also, the curvature $%
\boldsymbol{\Omega}%
^{%
\boldsymbol{\omega}%
}$ is tensorial of type $($\textrm{Ad}$,\mathbb{R}^{n^{2}})$. It then defines
$\Omega=e_{a}\otimes\theta^{b}\otimes\Omega_{b}^{a}\in\sec T_{1}^{1}%
M\otimes\bigwedge\nolimits^{2}(T^{\ast}M)$ which is easily find to be
\begin{align}
\Omega &  =e_{a}\otimes\theta^{b}\otimes\Omega_{b}^{a}\nonumber\\
&  =e_{a}\otimes\theta^{b}\otimes(d\omega_{b}^{a}+\omega_{c}^{a}\wedge
\omega_{b}^{c}), \label{lc11}%
\end{align}
where the $\Omega_{b}^{a}\in\sec\bigwedge\nolimits^{2}(T^{\ast}M)$ are the
curvature $2$-forms.
\begin{equation}
\Omega_{b}^{a}:=d\omega_{b}^{a}+\omega_{c}^{a}\wedge\omega_{b}^{c}
\label{LC12}%
\end{equation}
Note that sometimes the \textit{symbol} $\mathbf{D}\omega_{b}^{a}$ such that
$\Omega_{b}^{a}:=\mathbf{D}\omega_{b}^{a}$ is introduced in some texts. Of
course, the symbol $\mathbf{D}$ cannot be interpreted in this case as the
exterior covariant derivative of index forms. This is expected since $%
\boldsymbol{\omega}%
\in\sec%
{\displaystyle\bigwedge\nolimits^{1}}
(T^{\ast}P(M))\otimes gl(n,\mathbb{R})$ is \textit{not} tensorial.

\section{ Covariant Derivatives on Vector Bundles}

Consider \medskip a vector bundle $(E,M,\mathbf{\pi}_{1},G,\mathbf{V})$
denoted $E=P\times_{\rho}\mathbf{V}$ \emph{associated} to a \emph{PFB} bundle
$(P,M,\mathbf{\pi},G)$ by the linear representation $\rho$ of $G$ in the
vector space $\mathbf{V}$ over the field $\mathbb{F=R}$ or $\mathbb{C} $.
Also, let $\dim_{\mathbb{F}}\mathbf{V}=m$. Consider again the trivializations
of $P$ and $E$ given by Eqs.(\ref{fb4.6})-(\ref{fb4.8}). Then, we have
the\medskip

\begin{definition}
\label{def transp para}\noindent The \emph{parallel transport} of
$\mathbf{\Psi}_{\mathbf{0}}\in E$, $\mathbf{\pi}_{1}(\mathbf{\Psi}%
_{\mathbf{0}})=x_{0} $, along the curve $\sigma:{}\mathbb{R}\ni I\rightarrow
M$, $t\mapsto\sigma(t)$ from $x_{0}=\sigma(0)\in M$ to $x=\sigma(t)$ is the
element $\mathbf{\Psi}_{\mathbf{\parallel}t}\in E$ such that:
\end{definition}

(i) $\mathbf{\pi}_{1}(\mathbf{\Psi}_{\mathbf{\parallel}t})=x$,

(ii) $\chi_{i}(\mathbf{\Psi}_{\mathbf{\parallel}t})=\rho(\varphi
_{i}(p_{\parallel t})\circ\varphi_{i}^{-1}(p_{0}))\chi_{i}(\mathbf{\Psi
}_{\mathbf{0}})$.

(iii) $p_{\parallel t}\in P$ is the parallel transport of $p_{0}\in P$ along
$\sigma$ from $x_{0}$ to $x$ as defined in Eq.(\ref{fb4.13}) above.

\begin{definition}
\noindent Let $Y$ be a vector at $x_{0}$ tangent to the curve $\sigma$ (as
defined above). The covariant derivative of $\mathbf{\Psi\in}\sec E$ in the
direction of $Y$ is denoted $(D_{Y}^{E}\mathbf{\Psi})_{x_{0}}\mathbf{\in}\sec
E$ and
\begin{equation}
(D_{Y}^{E}\mathbf{\Psi})(x_{0})\equiv(D_{Y}^{E}\mathbf{\Psi})_{x_{0}}%
=\lim_{t\rightarrow0}\frac{1}{t}(\mathbf{\Psi}_{\mathbf{\parallel}t}%
^{0}-\mathbf{\Psi}_{\mathbf{0}}), \label{fb4.66}%
\end{equation}
where $\mathbf{\Psi}_{\mathbf{\parallel}t}^{0}$ is the \textquotedblleft
vector\textquotedblright\ $\mathbf{\Psi}_{t}\equiv\mathbf{\Psi(}%
\sigma\mathbf{(}t\mathbf{))}$\emph{\ }of a section $\mathbf{\Psi\in}\sec E$
parallel transported along $\sigma$ from $\sigma(t)$ to $x_{0}$, the only
requirement on $\sigma$ being
\begin{equation}
\left.  \frac{d}{dt}\sigma(t)\right\vert _{t=0}=Y. \label{fb4.67}%
\end{equation}

\end{definition}

In the local trivialization $(U_{i},\Xi_{i})$ of $E$ (see Eqs.(\ref{fb4.6}%
)-(\ref{fb4.8})) if $\Psi_{t}$ is the element in $\mathbf{V}$ representing
$\mathbf{\Psi}_{t}$, we have
\begin{equation}
\chi_{i}(\Psi_{\mathbf{\parallel}t}^{0})=\rho(g_{0}g_{t}^{-1})\chi
_{i\mid\sigma(t)}(\Psi_{t}). \label{fb4.68}%
\end{equation}
By choosing $p_{0}$ such that $g_{0}=e$ we can compute Eq.(\ref{fb4.66}):%

\begin{align}
(D_{Y}^{E}\mathbf{\Psi})_{x_{0}}  &  =\left.  \frac{d}{dt}\rho(g^{-1}%
(t)\Psi_{t})\right\vert _{t=0}\nonumber\\
&  =\left.  \frac{d\Psi_{t}}{dt}\right\vert _{t=0}-\left(  \left.
\rho^{\prime}(e)\frac{dg(t)}{dt}\right\vert _{t=0}\right)  (\Psi_{0}).
\label{fb4.70}%
\end{align}

This formula is trivially generalized for the covariant derivative in the
direction of an arbitrary vector field $Y\in\sec TM.$

With the aid of Eq.(\ref{fb4.70}) we can calculate, e.g., the covariant
derivative of $\mathbf{\Psi}\in\sec E$ in the direction of the vector field
$Y=\frac{\partial}{\partial x^{\mu}}\equiv\partial_{\mu}$. This covariant
derivative is denoted $D_{\partial_{\mu}}^{E}\!\mathbf{\Psi}$.

We need now to calculate $\left.  \frac{dg(t)}{dt}\right\vert _{t=0}$. In
order to do that, recall that if $\frac{d}{dt}$ is a tangent to the curve
$\sigma$ in $M$, then $s_{\ast}\left(  \frac{d}{dt}\right)  $ is a tangent to
$\hat{\sigma}$ the horizontal lift of $\sigma$, i.e., $s_{\ast}\left(
\frac{d}{dt}\right)  \in HP\subset TP.$ As defined before $s=\Phi_{i}%
^{-1}(x,e)$ is the cross section associated to the trivialization $\Phi_{i}$
of $P$ (see Eq.(\ref{fb4.6n})). Then, as $g$ is a mapping $U\rightarrow G$ we
can write
\begin{equation}
\left[  s_{\ast}(\frac{d}{dt})\right]  (g)=\frac{d}{dt}(g\circ\sigma).
\label{fb4.71}%
\end{equation}
To simplify the notation, introduce local coordinates $\langle x^{\mu
},g\rangle$ in $\pi^{-1}(U)$ and write $\sigma(t)=(x^{\mu}(t))$ and
$\hat{\sigma}(t)=(x^{\mu}(t),g(t))$. Then,
\begin{equation}
s_{\ast}\left(  \frac{d}{dt}\right)  =\dot{x}^{\mu}(t)\frac{\partial}{\partial
x^{\mu}}+\dot{g}(t)\frac{\partial}{\partial g}, \label{fb4.72}%
\end{equation}
in the local coordinate basis of $T(\mathbf{\pi}^{-1}(U))$. An expression like
the second member of Eq.(\ref{fb4.72}) defines in general a vector tangent to
$P$ but, according to its definition, $s_{\ast}\left(  \frac{d}{dt}\right)  $
is in fact horizontal. We must then impose that
\begin{equation}
s_{\ast}\left(  \frac{d}{dt}\right)  =\dot{x}^{\mu}(t)\frac{\partial}{\partial
x^{\mu}}+\dot{g}(t)\frac{\partial}{\partial g}=\alpha^{\mu}\left(
\frac{\partial}{\partial x^{\mu}}+\omega_{\mu}^{a}\mathcal{G}_{a}%
g\frac{\partial}{\partial g}\right)  , \label{fb4.73}%
\end{equation}
for some $\alpha^{\mu}$.

We used the fact that $\frac{\partial}{\partial x^{\mu}}+\omega_{\mu}%
^{a}\mathcal{G}_{a}g\frac{\partial}{\partial g}$ is a basis for $HP$, as can
easily be verified from the condition that $%
\boldsymbol{\omega}%
($\textbf{$Y$}$^{h})=0$, for all \textbf{$Y$}$\in HP$. We immediately get
that
\begin{equation}
\alpha^{\mu}=\dot{x}^{\mu}(t), \label{fb4.74}%
\end{equation}
and
\begin{align}
\frac{dg(t)}{dt}  &  =\dot{g}(t)=-\dot{x}^{\mu}(t)\omega_{\mu}^{a}%
\mathcal{G}_{a}g,\label{fb4.75}\\[0.03in]
\left.  \frac{dg(t)}{dt}\right\vert _{t=0}  &  =-\dot{x}^{\mu}(0)\omega_{\mu
}^{a}\mathcal{G}_{a}. \label{fb4.76}%
\end{align}
With this result we can rewrite Eq.(\ref{fb4.70}) as
\begin{equation}
(D_{Y}^{E}\mathbf{\Psi})_{x_{0}}=\left.  \frac{d\Psi_{t}}{dt}\right\vert
_{t=0}-\rho^{\prime}(e)\omega(Y)(\Psi_{0}),\quad\quad Y=\left.  \frac{d\sigma
}{dt}\right\vert _{t=0}, \label{fb4.77}%
\end{equation}
which generalizes trivially for the covariant derivative along a vector field
$Y\in\sec TM.\bigskip$

\begin{remark}
Many texts introduce the covariant derivative operator $D_{Y}^{E}$ acting on
sections of the vector bundle $E$ as follows:
\end{remark}

\begin{definition}
\label{dir conn def}A connection $D^{E}$ on $M$ is a mapping%
\begin{align}
D^{E}  &  :\sec TM\times\sec E\rightarrow\sec E,\nonumber\\
(X,\mathbf{\Psi)}  &  \mathbf{\mapsto}D_{Y}^{E}\mathbf{\Psi.} \label{cd1}%
\end{align}
such that $D_{Y}^{E}:\sec E\rightarrow\sec E$ satisfies the following
properties:%
\begin{equation}%
\begin{array}
[c]{cc}%
(i) & D_{X}^{E}(a\mathbf{\Psi)}=a(D_{X}^{E}\mathbf{\Psi),}\\
(ii) & D_{X}^{E}(\mathbf{\Psi+\Phi)}=D_{X}^{E}\mathbf{\Psi+}D_{X}%
^{E}\mathbf{\Phi,}\\
(iii) & D_{X}^{E}(f\mathbf{\Psi)}=X(f)+fD_{X}^{E}\mathbf{\Psi,}\\
(iv) & D_{X+Y}^{E}\mathbf{\Psi}=D_{X}^{E}\mathbf{\Psi+}D_{Y}^{E}\mathbf{\Psi
,}\\
(v) & D_{fX}^{E}\mathbf{\Psi}=fD_{X}^{E}\mathbf{\Psi.}%
\end{array}
\label{cd2}%
\end{equation}
$\forall$ $X,Y\in\sec TM$, $\mathbf{\Psi},\mathbf{\Phi\in}\sec E$\textbf{,
}$\forall a\in\mathbb{F=R}$ or $\mathbb{C}$ (the field of scalars entering the
definition of the vector space $\mathbf{V}$ of $E$) $\forall f\in C^{\infty
}(M)$, where $C^{\infty}(M)$ is the set of smooth functions with values in
$\mathbb{F}$.
\end{definition}

Of course, all properties in Eq.(\ref{cd2}) follows directly from
Eq.(\ref{fb4.77}). However, the point of view encoded in Definition
\ref{dir conn def} may be appealing to physicists. To see, first recall that
$E=P\times_{\rho}\mathbf{V}$. Recall that $\varrho$ stands for the
representation of $G$ in the vector space $\mathbf{V}$.

\begin{definition}
The dual bundle of $E$ is the bundle is the bundle $E^{\ast}=P\times
_{\rho^{\ast}}\mathbf{V}^{\ast}$, where $\mathbf{V}^{\ast}$ is the dual space
of $\mathbf{V}$ and $\varrho^{\ast}$ is the representation of $G$ in the
vector space $\mathbf{V}^{\ast}$.

\begin{example}
As examples, recall that the tangent bundle is $TM=F(M)\times_{\rho}%
\mathbb{R}^{n}$ where $\rho:Gl(n,\mathbb{R)\rightarrow}Gl(n,\mathbb{R)}$
denotes the standard representation. Then, $T^{\ast}M=F(M)\times_{\rho^{\ast}%
}(\mathbb{R}^{n})^{\ast}$ and the dual representation $\rho^{\ast\text{ }}%
$satisfies $\rho^{\ast}(g)=\rho(g^{-1})^{t}$. \ Also, the tensor bundle of
tensors of type $(r,s)$, the bundle of homogenous $k$-vectors and the bundle
of homogeneous $k$-forms are:%
\begin{align}
T_{s}^{r}M  &  =\bigotimes\nolimits_{s}^{r}TM=F(M)\times_{\bigotimes
\nolimits_{s}^{r}}(\bigotimes\nolimits_{s}^{r}\mathbb{R}^{n}),\nonumber\\
\bigwedge\nolimits^{k}(TM)  &  =F(M)\times_{\bigwedge\nolimits_{\rho}^{k}%
}\bigwedge\nolimits^{k}\mathbb{R}^{n},\nonumber\\
\bigwedge\nolimits^{k}(T^{\ast}M)  &  =F(M)\times_{\bigwedge\nolimits_{\rho
^{\ast}}^{k}}\bigwedge\nolimits^{k}\mathbb{R}^{n}, \label{cd3}%
\end{align}
where $\bigotimes\nolimits_{s}^{r}$, $\bigwedge\nolimits_{\rho}^{k}$ and
$\bigwedge\nolimits_{\rho^{\ast}}^{k}$ are the induced tensor product and
exterior powers representations.
\end{example}
\end{definition}

\begin{definition}
The bundle $E\otimes E^{\ast}$ is called the bundle of endomorphisms of $E$
and will be denoted by $\mathrm{End}(E)$.
\end{definition}

\begin{definition}
A connection $D^{E^{\ast}}acting$ on $E^{\ast}$ is defined by
\begin{equation}
(D_{X}^{E^{\ast}}\mathbf{\Upsilon}^{\ast})(\mathbf{\Psi})=X\left(
\mathbf{\Upsilon}^{\ast}(\mathbf{\Psi})\right)  -\mathbf{\Upsilon}^{\ast
}(D_{X}^{E}\mathbf{\Psi}), \label{CD3'}%
\end{equation}
$\forall\mathbf{\Upsilon}^{\ast}\in\sec E^{\ast}$, $\forall\mathbf{\Psi}%
\in\sec E$ and $\forall X\in\sec TM$.
\end{definition}

\begin{definition}
A connection $D^{E\otimes E^{\ast}}$ acting on sections of $E\otimes E^{\ast}$
is defined $\forall\mathbf{\Upsilon}^{\ast}\in\sec E^{\ast}$, $\forall
\mathbf{\Psi}\in\sec E$ and $\forall X\in\sec TM$ by%
\begin{equation}
D_{X}^{E\otimes E^{\ast}}\mathbf{\Upsilon}^{\ast}\otimes\mathbf{\Psi=}%
D_{X}^{E^{\ast}}\mathbf{\Upsilon}^{\ast}\otimes\mathbf{\Psi+\Upsilon}^{\ast
}\otimes D_{X}^{E}\mathbf{\Psi.} \label{CD3''}%
\end{equation}

\end{definition}

We shall abbreviate $D^{E\otimes E^{\ast}}$ by $D^{\mathrm{End}E}$.
Eq.(\ref{CD3''}) may be generalized in an obvious way in order to define a
connection on arbitrary tensor products of bundles $E\otimes E^{\prime}%
\otimes...\otimes E^{\prime...\prime}$. Finally, we recall for completeness
that given two bundles, say $E$ and $E^{\prime}$ and given connections $D^{E}$
and $D^{E^{\prime}}$there is an obvious connection $D^{E\oplus E^{\prime}}$
defined in the Whitney bundle $E\oplus E^{\prime}$ (recall definition
\ref{whitney sum}) It is given by
\begin{equation}
D_{X}^{E\oplus E^{\prime}}(\mathbf{\Psi}\oplus\mathbf{\Psi}^{\prime}%
)=D_{X}^{E}\mathbf{\Psi}\oplus D_{X}^{E^{\prime}}\mathbf{\Psi}^{\prime},
\label{CD}%
\end{equation}
$\forall\mathbf{\Psi}\in\sec E$, $\forall\mathbf{\Psi}^{\prime}\in\sec
E^{\prime}$ and $\forall X\in\sec TM$.

\subsection{Connections on $E$ over a Lorentzian Manifold}

In what follows we suppose that $(M,\mathtt{\mathbf{g}})$ is a Lorentzian
manifold. We recall that the manifold $M$ in a Lorentzian structure is
supposed paracompact. Then, according to Proposition \ref{admit cs} the
bundles $E,E^{\ast}$, $T_{s}^{r}M$ and $\mathrm{End}E$ admit global cross sections.

We then write for the covariant derivative of $\mathbf{\Psi\in}\sec E$ and
$X\in\sec TM,$%
\begin{equation}
D_{X}^{E}\mathbf{\Psi=}D_{X}^{0E}\mathbf{\Psi+}\mathcal{W}\mathbf{(}%
X\mathbf{)\Psi,} \label{cdn1}%
\end{equation}
where $\mathcal{W}\in\sec\mathrm{End}E\otimes T^{\ast}M$ will be called
\textit{connection} $1$-form (or \textit{potential)} for $D_{X}^{E}$ $\ $and
$D_{X}^{0E}$ is a well-defined connection on $E$, that we are going to determine.

Consider then a open set $U\subset M$ and a trivialization of $E$ in $U$. Such
a trivialization is said to be a \textit{choice of a gauge}.

Let $\{\mathbf{E}_{i}\}$ be the canonical basis of $\mathbf{V}$. Let $\left.
\mathbf{\Psi}\right\vert _{U}\in\sec$ $\left.  E\right\vert _{U}=\mathbf{\pi
}^{-1}(U)$. Consider the trivialization $\Xi:\mathbf{\pi}^{-1}(U)\rightarrow
U\times\mathbf{V}$, $\Xi(\mathbf{\Psi})=(\mathbf{\pi}(\mathbf{\Psi}%
),\chi(\mathbf{\Psi}))=(x,\chi(\mathbf{\Psi}))$.\ In this trivialization we
write
\begin{equation}
\left.  \mathbf{\Psi}\right\vert _{U}\mathbf{:=(}x\mathbf{,}\Psi(x)),
\end{equation}
$\Psi(x)\in\mathbf{V}$, $\forall x\in U$, with $\Psi:U\rightarrow\mathbf{V}$ a
smooth function. Let $\{s_{i}\}\in\sec\left.  E\right\vert _{U}$, $s_{i}%
=\chi^{-1}(\mathbf{E}_{i}),$ $i=1,2,...,m$ be a basis of sections of $\left.
E\right\vert _{U}$ and $\{e_{\mu}\}\in\sec F(U),\mu=0,1,2,3$ a basis for $TU$.
Let also $\{\varepsilon^{\nu}\}$, $\varepsilon^{\nu}\in\sec T^{\ast}U$, be the
dual basis of $\{e_{\mu}\}$ and $\{s^{\ast i}\}\in\sec\left.  E^{\ast
}\right\vert _{U}$, be a basis of sections of $\left.  E^{\ast}\right\vert
_{U}$ dual to the basis $\{s_{i}\}$.

We define the connection coefficients in the chosen gauge by
\begin{equation}
D_{e_{\mu}}^{E}s_{i}=\mathcal{W}_{\mu i}^{j}s_{j}. \label{cd4}%
\end{equation}
Then, if $\Psi=\Psi^{i}s_{i}$ and $X=X^{\mu}e_{\mu}$%
\begin{align}
D_{X}^{E}\Psi &  =X^{\mu}D_{e_{\mu}}^{E}(\Psi^{i}s_{i})\nonumber\\
&  =X^{\mu}\left[  e_{\mu}(\Psi^{i})+\mathcal{W}_{\mu j}^{i}\Psi^{j}\right]
s_{i}. \label{cd5}%
\end{align}

Now, let us concentrate on the term $X^{\mu}\mathcal{W}_{\mu j}^{i}\Psi
^{j}s_{i}$. It is, of course a new section $\mathbf{\digamma:=(}x,X^{\mu
}\mathcal{W}_{\mu j}^{i}\Psi^{j}s_{i})$ of $\left.  E\right\vert _{U}$ and
$X^{\mu}\mathcal{W}_{\mu j}^{i}\Psi^{j}s_{i}$ is linear in both $X$ and $\Psi.
$

This observation shows that $\mathcal{W}^{U}\in\sec(\mathrm{End}\left.
E\right\vert _{U})\otimes T^{\ast}U$, such that in the trivialization
introduced above is given by%
\begin{equation}
\mathcal{W}^{U}=\mathcal{W}_{\mu j}^{i}s_{j}\otimes s^{\ast i}\otimes
\varepsilon^{\mu} \label{cd6}%
\end{equation}
\textit{is} the representative of $\mathcal{W}$ in the chosen gauge.

Note that if $X\in\sec TU$ and $\mathbf{\Psi:=(}x\mathbf{,}\Psi(x))\in\sec$
$\left.  E\right\vert _{U}$ we have%
\begin{align}
\omega^{U}(X)  &  :=\omega_{X}^{U}=X^{\mu}\mathcal{W}_{\mu j}^{i}s_{j}\otimes
s^{\ast i},\nonumber\\
\omega_{X}^{U}(\Psi)  &  =X^{\mu}\mathcal{W}_{\mu j}^{i}\Psi^{i}s_{j}.
\label{CD7}%
\end{align}

We can then write%
\begin{equation}
D_{X}^{E}\Psi=X(\Psi)+\omega_{X}^{U}(\Psi), \label{cd9}%
\end{equation}
thereby identifying $D_{X}^{0E}\Psi=X(\Psi)$. In this case $D_{X}^{0E}$ is
called the standard \textit{flat} connection.

Now, we can state a very important result which has been used in Chapter 2 to
write the different decompositions of Riemann-Cartan connections.

\begin{proposition}
Let $\mathbf{D}^{0E}$ and \ $\mathbf{D}^{E}$ be arbitrary connections on $E$.
Then there exists $\mathcal{\bar{W}\in}\sec\mathrm{End}E\otimes T^{\ast}M$
such that for any $\mathbf{\Psi\in}\sec E$ and $X\in\sec TM,$%
\begin{equation}
\mathbf{D}_{X}^{E}\mathbf{\Psi=D}_{X}^{0E}\mathbf{\Psi+}\mathcal{\bar{W}%
}\mathbf{(}X\mathbf{)\Psi.} \label{cd10}%
\end{equation}

\end{proposition}

\subsection{Gauge Covariant Connections}

\begin{definition}
A connection $D^{E}$ on $E$ is said to be a $G$-connection if for any $g\in G
$ and any $\mathbf{\Psi\in}\sec E$ there exists a connection $\mathbf{D}%
^{\prime E}$ on $E$ such that for any $X\in\sec TM$
\begin{equation}
D_{X}^{\prime E}\left(  \rho(g)\mathbf{\Psi}\right)  =\rho(g)D_{X}%
^{E}\mathbf{\Psi.} \tag{d11}%
\end{equation}

\end{definition}

\begin{proposition}
If $D_{X}^{E}\mathbf{\Psi=}D_{X}^{0E}\mathbf{\Psi+}\mathcal{\bar{W}}%
\mathbf{(}X\mathbf{)\Psi}$ for $\mathbf{\Psi\in}\sec E$ and $X\in\sec TM$,
then $D_{X}^{\prime E}\mathbf{\Psi=}D_{X}^{0E}\mathbf{\Psi+}\mathcal{\bar{W}%
}^{\prime}\mathbf{(}X\mathbf{)\Psi}$ with
\begin{equation}
\mathcal{\bar{W}}^{\prime}\mathbf{(}X\mathbf{)=}g\mathcal{\bar{W}}%
\mathbf{(}X\mathbf{)}g^{-1}+gdg^{-1}. \label{d12}%
\end{equation}

\end{proposition}

Suppose that the vector bundle $E$ has the same structural group as the
orthonormal frame bundle $\mathbf{P}_{\mathrm{SO}_{1,3}^{e}}(M)$, which as we
know is a reduction of the frame bundle $F(M)$. In this case we give the

\begin{definition}
\label{generalized connection}A connection $\mathbf{D}^{E}$ on $E$ is said to
be a generalized $G$-connection if for any $g\in G$ and any $\mathbf{\Psi\in
}\sec E$ there exists a connection $\mathbf{D}^{\prime E}$ on $E$ such that
for any $X\in\sec TM$, $TM=$\textrm{P}$_{\mathrm{SO}_{1,3}^{e}}(M)\times
_{\rho^{TM}}\mathbb{R}^{4}$%
\begin{equation}
D_{X^{\prime}}^{\prime E}\left(  \rho(g)\mathbf{\Psi}\right)  =\rho
(g)D_{X}^{E}\mathbf{\Psi,} \label{d12'}%
\end{equation}
where $X^{\prime}=\rho^{TM}X\in\sec TM.$
\end{definition}

\subsection{Curvature Again}

\begin{definition}
Let $D^{E}$ be a $G$-connection on $E$. The curvature operator $\mathbf{R}%
^{E}\mathbf{\in}\sec\bigwedge\nolimits^{2}T^{\ast}M\otimes\mathrm{End}E$ of
$D^{E}$ is the mapping
\begin{align}
\mathbf{R}^{E}  &  \mathbf{:}\sec TM\otimes TM\otimes E\rightarrow
E,\label{cd13}\\
\mathbf{R}^{E}\mathbf{(}X,Y\mathbf{)\Psi}  &  =D_{X}^{E}D_{Y}^{E}\mathbf{\Psi
}-D_{Y}^{E}D_{X}^{E}\mathbf{\Psi}-D_{[X,Y]}^{E}\mathbf{\Psi}\nonumber
\end{align}%
\begin{equation}
\mathbf{R}^{E}\mathbf{(}X,Y\mathbf{)}=D_{X}^{E}D_{Y}^{E}-D_{Y}^{E}D_{X}%
^{E}-D_{[X,Y]}^{E}, \tag{d14}%
\end{equation}
for any $\mathbf{\Psi\in}\sec E$ and $X,Y\in\sec TM$.
\end{definition}

If \ $X=\partial_{\mu},Y=\partial_{\nu}\in\sec TU$ are coordinate basis
vectors associated to the coordinate functions $\{x^{\mu}\}$ we have%
\begin{equation}
\mathbf{R}^{E}(\partial_{\mu},\partial_{\nu}):=\mathbf{R}_{\mu\nu}^{E}=\left[
D_{\partial_{\mu}}^{E},D_{\partial_{\nu}}^{E}\right]  \label{d14}%
\end{equation}

In a local basis \{$s_{i}\otimes s^{\ast j}\}$ of $\mathrm{End}E$ we have
under the local trivialization used above
\begin{align}
\mathbf{R}_{\mu\nu}^{E}  &  =\mathbf{R}_{\mu\nu b}^{a}s_{a}\otimes s^{\ast
b},\nonumber\\
\mathbf{R}_{\mu\nu b}^{a}  &  =\partial_{\mu}\mathcal{W}_{\nu b}^{a}%
-\partial_{\nu}\mathcal{W}_{\mu b}^{a}+\mathcal{W}_{\mu c}^{a}\mathcal{W}_{\nu
b}^{c}-\mathcal{W}_{\nu c}^{a}\mathcal{W}_{\mu b}^{c} \label{d15}%
\end{align}

Eq.(\ref{d15}) can also be written%
\begin{equation}
\mathbf{R}_{\mu\nu}^{E}=\partial_{\mu}\mathcal{W}_{\nu}-\partial_{\nu
}\mathcal{W}_{\mu}+\left[  \mathcal{W}_{\mu},\mathcal{W}_{\nu}\right]
\label{d16}%
\end{equation}

\subsection{Exterior Covariant Derivative Again}

\begin{definition}
Consider $\Psi\mathbf{\otimes}A_{r}$ $\in\sec E\otimes\bigwedge\nolimits^{r}%
T^{\ast}M$ and \ $B_{s}$ $\in\bigwedge\nolimits^{s}T^{\ast}M$ . We define
$(\Psi\mathbf{\otimes}A_{r})\otimes_{\wedge}B_{s}$ by
\begin{equation}
(\Psi\mathbf{\otimes}A_{r})\otimes_{\wedge}B_{s}=\Psi\mathbf{\otimes(}%
A_{r}\wedge B_{s}) \label{d17}%
\end{equation}

\end{definition}

\begin{definition}
Let $\Psi\mathbf{\otimes}A_{r}$ $\in\sec E\otimes\bigwedge\nolimits^{r}%
T^{\ast}M$ and $\mathbf{\Pi\otimes}B_{s}$ $\in\sec\mathrm{End}E\otimes
\bigwedge\nolimits^{s}T^{\ast}M.$ We define $(\Pi\mathbf{\otimes}B_{s}%
)\otimes_{\wedge}(\Psi\mathbf{\otimes}A_{r})$ \textbf{\ }$by$%
\begin{equation}
(\Pi\mathbf{\otimes}B_{s})\otimes_{\wedge}(\Psi\mathbf{\otimes}A_{r}%
)=\Pi\mathbf{(}\Psi\mathbf{)\otimes}(B_{s}\wedge A_{r}) \label{d17'}%
\end{equation}

\end{definition}

\begin{definition}
Given a connection D$^{E}$ acting on $E,$ the exterior covariant derivative
\textbf{d}$^{D^{E}}$ acting on sections of $E\otimes\bigwedge\nolimits^{r}%
T^{\ast}M$ and the exterior covariant derivative \textbf{d}$^{D^{\mathrm{EndE}%
}}$ sections of $\mathrm{End}E\otimes\bigwedge\nolimits^{s}T^{\ast}M$
$\ (r,s=0,1,2,3,4)$ are given by

(i) if $\mathbf{\Psi\in}\sec E$ then for any $X,Y\in\sec TM$%
\begin{equation}
\mathbf{d}^{D^{E}}\Psi\mathbf{(}X\mathbf{)=}D^{E}\Psi\mathbf{,} \label{d18}%
\end{equation}

(ii) For any $\Psi\mathbf{\otimes}A_{r}$ $\in\sec E\otimes\bigwedge
\nolimits^{r}T^{\ast}M$
\begin{equation}
\mathbf{d}^{D^{E}}\left(  \Psi\mathbf{\otimes}A_{r}\right)  =\mathbf{d}%
^{D^{E}}\Psi\mathbf{\otimes\wedge}A_{r}+\Psi\mathbf{\otimes}dA_{r},
\label{d19}%
\end{equation}

(iii) For any $\mathbf{\Pi\otimes}B_{s}$ $\in\sec\mathrm{End}E\otimes
\bigwedge\nolimits^{s}T^{\ast}M$%
\begin{equation}
\mathbf{d}^{D^{\mathrm{EndE}}}(\Pi\mathbf{\otimes}B_{s})=\mathbf{d}%
^{D^{\mathrm{EndE}}}\Pi\mathbf{\otimes}_{\wedge}B_{s}+\Pi\mathbf{\otimes
}dB_{s}, \label{d20}%
\end{equation}

\end{definition}

\begin{proposition}
Consider the bundle product $\mathfrak{E=}$ ($\mathrm{End}E\otimes
\bigwedge\nolimits^{s}T^{\ast}M)\otimes_{\wedge}E\otimes\bigwedge
\nolimits^{r}(T^{\ast}M)$. Let $\mathbf{\Pi}=\Pi\mathbf{\otimes}B_{s}$
$\in\sec\mathrm{End}E\otimes\bigwedge\nolimits^{s}T^{\ast}M$ and
$\mathbf{\Psi}=\Psi\mathbf{\otimes}A_{r}$ $\in\sec E\otimes\bigwedge
\nolimits^{r}T^{\ast}M$. Then the exterior covariant derivative $\mathbf{d}%
^{D^{\mathfrak{E}}}$\textbf{\ }acting on sections of $\mathfrak{E}$ satisfies%
\begin{equation}
\mathbf{d}^{D^{\mathfrak{E}}}(\mathbf{\Pi}\otimes_{\wedge}\mathbf{\Psi
})=(\mathbf{d}^{D^{\mathrm{EndE}}}\mathbf{\Pi)}\otimes_{\wedge}\mathbf{\Psi
+(}-1\mathbf{)}^{s}\mathbf{\Pi\otimes}_{\wedge}\mathbf{d}^{D^{E}}%
\mathbf{\Psi.} \label{d21}%
\end{equation}

\end{proposition}

\begin{exercise}
We can now show several interesting results, which make contact with results
obtained earlier when we analyzed the connections and curvatures on principal
bundles and which allowed us sometimes the use of sloppy notations in the main text:

(i) Suppose that the bundle admits a flat connection $D^{0E}.$ We put
$\mathbf{d}^{D^{0E}}=d$. Then, if \ $\chi\in\sec E\otimes\bigwedge
\nolimits^{r}T^{\ast}M$ \ we have%
\[
\mathbf{d}^{D^{0E}}\chi=\mathbf{d}\chi\mathbf{+}\mathcal{W}\mathbf{\otimes
}_{\wedge}\mathbf{\chi}%
\]

(ii) If $\chi\in\sec E\otimes\bigwedge\nolimits^{r}T^{\ast}M$ $\ $we have%
\begin{equation}
(\mathbf{d}^{D^{E}})^{2}\chi=\mathbf{R}^{E}\mathbf{\otimes}_{\wedge}\chi.
\label{d22}%
\end{equation}

(iii) If $\chi\in\sec E\otimes\bigwedge\nolimits^{r}T^{\ast}M$ $\ $we have%
\begin{equation}
(\mathbf{d}^{D^{E}})^{3}\chi=\mathbf{R}^{E}\mathbf{\otimes}_{\wedge}%
\mathbf{d}^{D^{E}}\chi\label{D23}%
\end{equation}

(iii) Suppose that the bundle admits a flat connection $D^{0E}.$ We put
$\mathbf{d}^{D^{0E}}=d$. Then, if

(iv) If $\mathbf{\Pi}$ $\in\sec\mathrm{End}E\otimes\bigwedge\nolimits^{s}%
T^{\ast}M$
\begin{equation}
\mathbf{d}^{D^{\mathrm{End}E}}\mathbf{\Pi=d\Pi+[}\mathcal{W}\mathbf{,\Pi]}
\label{d24}%
\end{equation}

(v)%
\begin{equation}
\mathbf{d}^{D^{\mathrm{End}E}}\mathbf{R}^{E}=0 \label{d25}%
\end{equation}

(vi)
\begin{equation}
\mathbf{R}^{E}=d\mathcal{W}+\mathcal{W}\mathbf{\mathbf{\otimes}_{\wedge}%
}\mathcal{W} \label{d26}%
\end{equation}

\end{exercise}

\begin{remark}
Note that $\mathbf{R}^{E}\neq\mathbf{d}^{D^{\mathrm{End}E}}\mathcal{W}%
\mathbf{.}$
\end{remark}

We hope that the material presented in the Appendix be enough to permit our
reader to follow the more difficult parts of the text and in particular to see
the reason for or use of many eventual sloppy notations.\medskip

\textbf{Acknowledegement}: R. da Rocha is grateful to CAPES for financial support.

\end{document}